\documentclass[prd,nofootinbib,
amsmath,amssymb,aps,
reprint,
floatfix
]{revtex4-2}

\usepackage[colorlinks=true
,urlcolor=blue
,anchorcolor=blue
,citecolor=blue
,filecolor=blue
,linkcolor=blue
,menucolor=blue
,linktocpage=true
,pdfproducer=medialab
,pdfa=true
]{hyperref}

\usepackage{hyperref}
\usepackage[T1]{fontenc}
\usepackage{graphicx}
\usepackage{comment}
\usepackage{bm}
\usepackage{braket}
\usepackage{cancel}
\usepackage{lmodern}
\usepackage{ulem}

\newcommand\aap{A\&A}                
\newcommand\apjl{ApJ}                
\newcommand\apjs{ApJS}               
\newcommand\jcap{J.~Cosmology Astropart. Phys.} 
\newcommand\mnras{MNRAS}             
\newcommand\sovast{Soviet~Ast.}      

\newcommand{\eq}[1]{\begin{equation}\begin{split} #1 \end{split}\end{equation}}

\newcommand{\lr}[1]{\left( #1 \right)}

\newcommand{\zf}{z_\mathrm{f}}

\newcommand{\m}{\mathrm}

\begin{document}
\title{Constraint on the early-formed dark matter halos using the free-free emission in the Planck foreground analysis}
\author{Katsuya T. Abe}
\email{abe_kt@nagoya-u.jp}
\author{Teppei Minoda}
\author{Hiroyuki Tashiro}
\affiliation{Division of Particle and Astrophysical Science,
Graduate School of Science, Nagoya University,
Chikusa, Nagoya 464-8602, Japan}

\begin{abstract}
We provide a new constraint on the small-scale density fluctuations,
evaluating the diffuse background free-free emission
from dark matter halos in the dark ages. 
If there exists a large amplitude of the matter density fluctuations on small scales,
the excess enhances the early formation of dark matter halos.
When the virial temperature is sufficiently high, the gas in a halo is heated up and ionized by thermal collision.
The heated ionized gas emits photons by the free-free process.
We would observe the sum of these photons
as the diffuse background free-free emission.
Assuming the analytical dark matter halo model including the gas density and temperature profile, 
we calculate the intensity of the diffuse background free-free emission from early-formed dark matter halos in the microwave frequency range.
Comparing with the recent foreground analysis on cosmic microwave background,
we obtain the constraint on the excess of the density fluctuations on small scales.
Our constraint corresponds to
$P_\zeta \lesssim 10^{-7}$ for
$k \simeq 1-100~\mathrm{Mpc}^{-1}$ with assuming the delta-function-type curvature power spectrum.
Therefore, our constraint is
the most stringent constraint
on the perturbations below $1~\rm Mpc$ scales.
\end{abstract}

\maketitle

\section{Introduction}\label{sec: introduction}
The primordial perturbations
are the initial seeds of various cosmological structures:
the temperature anisotropies of
the Cosmic Microwave Background (CMB), 
the Large-Scale Structure (LSS),
clusters of galaxies, galaxies, and so on.
Therefore, the statistical properties of
the primordial perturbations are well studied.
In particular, the power spectrum
of the primordial perturbations
is one of the important statistics
to understand the cosmological structure formation.
It is thoroughly measured on large length scales
through the observations of the CMB on
$10^{-3}~\m{Mpc}^{-1} \lesssim k
\lesssim 10^{-1}~\m{Mpc}^{-1}$
\cite{2020A&A...641A...6P}
and the LSS (e.g., galaxy clustering, weak lensing,
Ly-$\alpha$ forest and so on)
on $k \sim 1~\m{Mpc}^{-1}$
\cite{2006ApJS..163...80M,2019JCAP...07..017C,2021A&A...646A.129J,
2021PhRvD.103d3522C,2021ApJS..253....3H,2021arXiv210513549D}.

These observations have revealed that
the amplitude of the power spectrum is
in order of $P_\zeta \sim 10^{-9}$ on those scales
and is almost scale-invariant.
These statistical natures are consistent with
the primordial perturbations predicted in the inflationary paradigm
\cite{1982PhLB..117..175S,1982PhRvL..49.1110G}.
On the other hand,
on smaller scales with $k > 1~\m{Mpc}^{-1}$,
the primordial perturbations are not known well. 
Many studies suggest that
the excess power on small scales can be induced
by a lot of mechanisms in the early universe:
multi-field inflation~\cite{2009JCAP...02..014A},
phase transition
\cite{1997PhRvL..78..791S, 2001MNRAS.324..977B},
early matter-dominated era
\cite{2014JHEP...04..138B},
fast-rolling scalar field domination
\cite{2018PhRvD..98f3504R},
and so on.
Therefore, the investigation of
the primordial perturbations on small scales can
improve our knowledge of the primordial universe.

One of the powerful probes for the small-scale perturbations 
is the CMB distortion~\cite{2012MNRAS.419.1294C,2014PTEP.2014fB107T,2014MNRAS.438.2065C}. 
The small-scale primordial perturbations 
are drown out by the Silk damping,
however, instead, the distortion of the CMB energy spectrum would be created. 
Therefore, measurements of the CMB distortion
lead us to further understanding of the primordial perturbations
\cite{1994ApJ...430L...5H,
2012ApJ...758...76C,2012PhRvL.109b1302P, 2012PhRvD..86b3514D,2013MNRAS.434.1619C}.
For example, Ref.~\cite{2012ApJ...758...76C}
has pointed out that
the CMB distortion measurement by COBE/FIRAS
\cite{1996ApJ...473..576F,1994ApJ...420..439M}
provides the constraint in order of $P_\zeta \lesssim 10^{-5}$
for the wave number range, $k \approx 1-10^4~\rm Mpc^{-1}$,
and next-generation CMB measurements like PIXIE
\cite{2019arXiv190901593C}
can be a probe of the level of the scale-invariant spectrum,
$P_\zeta \lesssim 10^{-8}$.
Another hopeful measurement is future redshifted 21-cm observations
which are expected to explore the density perturbations
on smaller scales than the Silk scale~\cite{2004PhRvL..92u1301L}.
Since the 21-cm signal comes from neutral hydrogen,
the measurements of the spatial fluctuations
in the redshifted 21-cm signal can trace
the density fluctuations before the epoch of reionization.
In order to measure the fluctuations of
21-cm signals from high redshifts,
the radio interferometer telescope,
the Square Kilometre Array (SKA), is now constructing.

The large amplitude of the perturbations on small scales
promotes the structure formation 
and produces substantial collapsed objects in the early universe.
One of such collapsed objects
is a primordial black hole~(PBH)~\cite{1967SvA....10..602Z}.
PBHs are considered to be responsible for
a non-negligible fraction of dark matter~(DM) compositions
\cite{1971MNRAS.152...75H, 1975Natur.253..251C}.
Furthermore, they are attracting attention,
because of the recent detection of gravitational wave~(GW) events 
via LIGO and Virgo collaboration~\cite{2016PhRvL.116f1102A},
as sources of GWs.
The abundance of PBHs has been studied in
a wide range of the PBH mass scale~\cite{2020arXiv200212778C}.
Since the PBH mass scale corresponds to
the scale of the overdense region collapsing to PBHs,
the constraint on the PBH abundance can be converted 
to the limit on the perturbations
in the wide range of the scale~\cite{2018JCAP...01..007E}.
However, the PBH formation requires
large amplitude of curvature power spectrum as
$P_\zeta \sim 10^{-2}-10^{-1}$.
Therefore, their constraint on the perturbations
is relatively not so tight.

The density fluctuations are induced by the primordial perturbations
and grow gravitationally well after the matter-radiation equality.
In the growth, the overdense region on small scales
finally collapses into the nonlinear structures called
ultra-compact minihalos or minihalos in higher redshifts
than the standard hierarchical structure formation
due to the scale-invariant primordial perturbations~\cite{2009ApJ...707..979R}.
With the self-annihilating DM model,
the abundance of these early-formed DM halos
is constrained by the present gamma-ray observations. 
Accordingly, the limit on the primordial perturbations
can be obtained with the formation scenario
\cite{2009PhRvL.103u1301S,2010PhRvD..82h3527J,2012PhRvD..85l5027B}.
However, the constraint on the primordial perturbations
highly depends on the parameters of the self-annihilating DM.
The constraint on the early-formed DM halo abundance,
which is independent of the DM model
can be obtained from redshifted 21-cm observations.
Depending on their mass, early-formed DM halos
can host abundant neutral hydrogen gas. 
Therefore, redshifted 21-cm observations
can probe the abundance of early-formed DM halos
\cite{2002ApJ...572L.123I}.
Several studies have shown that
SKA can provide the stringent limit
on the abundance of early-formed collapsed objects
and the amplitude of the primordial perturbations
\cite{2014JCAP...03..001S,2014PhRvD..90h3003S,
2018JCAP...02..053S,2020MNRAS.494.4334F}.

In this paper, we discuss the diffuse background free-free emission
in the CMB frequency range from early-formed DM halos.
In the formation of DM halos, baryonic gas is collapsed into them.
During the collapse, the gas is heated up by the gravitational potential 
and ionized by the collisions between gas particles
if the halo virial temperature is high enough,
as $T_\m{vir} \gtrsim 10^4~\m{K}$.
Such hot ionized gas can emit photons via free-free emission,
and they would be observed as
the diffuse background free-free emission.

The diffuse background free-free emission has
already been identified as one of the important foreground components in the CMB analysis~\cite{2016A&A...594A..10P,2020A&A...641A...4P}.
Although most of
the observed free-free emission is
believed to be Galactic origin,
a lot of cosmological free-free emitters can be considered,
for example, the intergalactic medium (IGM)~\cite{2004ApJ...606L...5C},
the galaxy groups and clusters after the reionization~\cite{2011MNRAS.410.2353P},
and the structure formation during the reionization~\cite{2019MNRAS.486.3617L}.
The estimated amplitude is roughly 10\% level of
the observed free-free emission component.

Here, we focus on the early-formed DM halos with
$M_\m{halo} \sim 10^7-10^{13} M_\odot$ and
estimate the contribution to the
observed free-free emission component
in the CMB frequency range.
These mass scales roughly correspond to the perturbation scales $k \sim 1-100~\m{Mpc}^{-1}$
and there is a possible window
for the excess of the primordial perturbations 
to produce DM halos much earlier than in the standard hierarchical structure formation scenario.
Comparing our estimation of the diffuse background free-free emission
with the observed free-free emission component
\cite{2016A&A...594A..10P,2020A&A...641A...4P},
we obtain the constraint on the primordial density fluctuations
on small scales.

The rest of this paper is organized as follows. 
In section~\ref{sec: calculation}, we provide
the halo model describing 
the density, temperature, and ionization fraction of gas
by assuming 
the hydrostatic equilibrium,
and the collisional-ionization equilibrium with isothermal gas. 
Then, we calculate the
intensity of the free-free emission from individual halos
for different mass $M_\m{halo}$ and formation redshift $z_\m{f}$.
Next, in section~\ref{section:all-sky},
considering the halo formation history,
we formulate the diffuse background intensity
which is the sum of the free-free emission
from early-formed halos.
The results are shown in section~\ref{section:results}.
We also discuss the application of our results to the constraint on the primordial curvature perturbations and obtain the limit by comparing
the intensity of the observed free-free emission
in the CMB frequency range.
We conclude in section~\ref{section:conclusion}.
Throughout this letter,
we take the flat $\Lambda$CDM model with:
$(\Omega_{\rm m},\Omega_{\rm b},h,n_{\rm{s}},\sigma_{8})$
=$(0.32,0.049,0.67,0.97,0.81)$~\cite{Planck2018_cospara}.

\section{Free-free emission from the individual halo}
\label{sec: calculation}
In this section, we estimate
the intensity of the free-free emission from individual halos.
It
depends on
the number density and temperature profiles of ionized gas.
First, we construct the simple analytical model of
ionized gas in a DM halo with mass $M_\m{halo}$.

\subsection{gas structure of DM halo}
\label{sec:internal}

Suppose that a DM halo forms with a mass of $M_\m{halo}$ in a redshift $z_{\rm f}$.
We assume that, after the virialization,
the DM density profile is given
by the NFW profile~\cite{NFWprofile1997},
\eq{\rho_{\mathrm{NFW}}(r)=
\frac{\rho_s}{\frac{r}{r_{s}}\left(1+\frac{r}{r_{s}}\right)^{2}},
}
where $r_s$ is the scale radius, which is related to the virial radius
\eq{
R_{\m{ vir}}
(M_\m{halo})=
\lr{
\frac{3M_\m{halo}}
{4\pi\Delta_c 
\rho_{\m{m}}(z_\m{f})}}^{1/3},
}
with the matter density $\rho_{\m{m}}(z)$ and
the concentration parameter,~$c_s\equiv R_{\m{ vir}}/r_{s}$.

The scale density $\rho_s$ is given by 
\eq{
\rho_s
= \frac{M_\m{halo}}{4 \pi r_s^3  F(c_s)}
= \frac{c_s^{3}}{3 F(c_s)} \Delta_c 
\rho_{\m{m}}(z_\m{f}),
}
where $\Delta_c=18\pi^2$ is the spherical overdensity of the halo,
and the function~$F(x)$ is described as
\eq{
F(x)=\ln (1+x)- \cfrac{x}{1+x}.
}

The concentration parameter $c_s$
varies depending on the mass and redshift of the halo.
Although
no simulation covers
both the mass and redshift range
we are interested in this paper,
Ref.~\cite{2020arXiv200714720I} has performed N-body simulations
and suggested the mass-concentration relation
from the simulation results
including DM halos with mass $M_\m{halo}>10^9 ~M_\odot$
in the redshift range of $0< z < 14$.
This relation is based on the analytic form
given in Ref.~\cite{2019ApJ...871..168D},
and in this model, the concentration parameter depends on
the spectral index of the primordial curvature power spectrum.
Because this concentration parameter becomes larger
as the spectral index increases,
we conservatively take the concentration parameter
for the scale-invariant case
to reduce the impact of the concentration parameter on our final results.

In the halo formation, the gas in a DM halo
is heated to the virial temperature by the virial shock.
The virial temperature $T_{\m{vir}}$ is given by~\cite{2013ASSL..396..177J}
\eq{\label{eq: temp_vir}
T_{\text {vir }}(M_\m{halo},z_\m{f})
=&1.98 \times 10^{4}
\left(\cfrac{\mu}{0.6}\right)
\left(\cfrac{M_\m{halo}}{10^{8} h^{-1} M_{\odot}}\right)^{2/3} \\
&\times \left[\cfrac{\Omega_\m{m}}{\Omega_\m{m}(z_\m{f})}
\frac{\Delta_{c}}{18 \pi^{2}}\right]^{1 / 3}
\left(\cfrac{1+z_\m{f}}{10}\right)~\mathrm{K},
} 
where $\mu$ is the mean molecular weight (we set $\mu=0.6$),
$\Omega_\m{m}(z)$ is the matter density parameter
at a redshift $z$ after the matter-dominant era,
\eq{
\Omega_{\m{m}}(z)=\frac{\Omega_{\m{m}}(1+z)^{3}}
{\Omega_{\m{m}}(1+z)^{3}+\Omega_{\Lambda}}~,
}
with $\Omega_\m{m} = \Omega_\m{m}(z=0)$,
$k_\m{B}$ is the Boltzmann constant, and $m_\m{p}$ is mass of proton.

To determine the gas density profile in a DM halo,
we make the assumption that the gas in the DM halo
has an isothermal profile 
with the virial temperature $T_{\rm vir}$
and is in the hydrostatic equilibrium between the gas
pressure and the DM potential of the NFW profile.
Under these conditions, the gas number density profile
is given in the analytical form~\cite{1998ApJ...497..555M},
\eq{\label{eq: gas_dens_pro_makino}
n_{\rm gas}(r)=n_{\rm gas, 0} \exp \left[-\frac{\mu m_{\mathrm{p}}}{2 k_{\mathrm{B}} T_{\mathrm{vir}}}\left(V_{\mathrm{esc}}^{2}(0)-V_{\mathrm{esc}}^{2}(r)\right)\right].
}
Here, 
$V_{\m{esc}}$ is the escape velocity written by
\eq{
V_{\mathrm{esc}}^{2}(r)
=& 2 \int_{r}^{\infty} \frac{G M\left(r^{\prime}\right)}{r^{\prime 2}} d r^{\prime} \\
=&\frac{2G M_\m{halo}}{R_{\mathrm{vir}}}
\frac{F(c_s x_r)+c_s x_r /(1+c_s x_r)}{x_r F(c_s)},
}
where $x_r=r/R_{\m{vir}}$,
and $M(r) =\int^{r}_0 4\pi r'^2 \rho_\m{gas}(r')~dr'$.

The number density at the central core, $n_{\m{gas,0}}$,
in Eq.~\eqref{eq: gas_dens_pro_makino}
is obtained by fixing the total gas mass in the sphere with $R_{\m{vir}}$ to $M_{\m{gas}}=M_\m{halo}\Omega_{\m{b}}/\Omega_{\m{m}}$,
\eq{
n_{\m{gas, 0}}(z)=\frac{\left(\Delta_{c} / 3\right)
c_s^{3} e^{A}}{\int_{0}^{c_s}(1+t)^{A / t} t^{2} d t}
\left(\frac{\Omega_\m{b}}{\Omega_\m{m}}\right)
\frac{ \rho_{\m{cri}}(z)}{m_{\rm p}},
}
where $A \equiv
V_{\mathrm{esc}}^{2}(0) R_{\m{vir}} / G M_\m{halo}
=2 c_s / F(c_s)$.

In this work, we assume that DM halos
do not host any stars in themselves for simplicity.
In the absence of ionization photon sources including stars, 
collisional ionization is a possible process
to ionize the gas in a DM halo.
Since the collisional ionization and recombination timescales are sufficiently
shorter than the cosmological timescale,
we may assume that halo gas is in the steady-state, by balancing 
between the collisional ionization and the case-B recombination.
In such a case,
the ionization fraction in a DM halo
with the temperature $T_{\rm halo}$
is determined by
\eq{\label{eq: ionization_fraction_equilibrium}
x_\m{e}(T_{\m{halo}}) = 
\frac{\tilde{C}_{\m{coll}}}{\tilde{C}_{\m{coll}}+\tilde{A}_{\m{rec}}},
}
where $\tilde{C}_{\m{coll}}=C_{\m{coll}}/(1-x_{\m{e}})$,
and $C_{\m{coll}}$ is the electron-collisional ionization rate
which is given by the fitting formula~\cite{doi:10.1063/1.555700},
\eq{\label{eq: ion_rate}
C_{\m{coll}}\approx5.85 \times 10^{-9}~T_{4}^{1/2}
\mathrm{e}^{-T_{\mathrm{H}}/T_{\m{halo}}}
n_{\mathrm{gas}}(1-x_{\mathrm{e}})~\mathrm{cm}^{3}/\mathrm{s}.
}
Here, $T_{\m{H}}\approx1.58\times 10^5~\m{K}$
and $T_{4} \equiv T_{\m{halo}}/10^4~\m{K}$.
In Eq.~\eqref{eq: ion_rate}, $\tilde{A}_{\m{rec}}=A_{\m{rec}}/x_{\m{e}}$,
and $A_{\m{rec}}$ is the recombination rate obtained by
\eq{\label{eq: rec_rate}
A_{\m{rec}}= -\alpha_{\mathrm{H}}n_{\mathrm{gas}}x_{\mathrm{e}},
}
where $\alpha_{\m{H}}$ is the case-B recombination coefficient for hydrogen
and fitted by Ref.~\cite{1991A&A...251..680P} as
\eq{
\alpha_{\mathrm{H}}=1.14\times 10^{-13}
\frac{a T_{4}^{b}}{1+c T_4^{d}} ~\mathrm{cm}^{3}/\mathrm{~s}
}
with the fitting parameters,
$a = 4.309$, $b = -0.6166$, $c =  0.6703$, and $d = 0.5300$.
In our model, the ionization fraction is only controlled by the gas temperature. 
When the gas temperature is larger than $10^4 $~K,
the ionization fraction goes up close to the unity.

Although the gas becomes hot by the virial shock
at the formation of the DM halo,
the gas will be cooled down by the cooling mechanism related to baryon physics.
Assuming the cooling timescale $t_\m{cool}$, 
the temperature is expressed at a given redshift~$z$,
\eq{\label{eq: temp_halo}
T_{\m{halo}}(M_\m{halo},z,\zf) = T_{\m{vir}}(M_\m{halo},\zf)~\m{exp}\lr{-\frac{\Delta_t(z,\zf)}{t_\m{cool}}},
}
where $\Delta_t$ is the cosmic time duration in the redshift range $[z,\zf]$
in the matter dominated epoch, 
\eq{
\Delta_t \approx \frac{2}{3}\lr{\frac{1}{H(z)}-\frac{1}{H(\zf)}},
}
with the Hubble parameter~$H(z)$.

The main cooling mechanism depends on the redshift and the gas properties
including the gas density, temperature, and ionization fraction.
One of the main cooling mechanisms is the free-free emission cooling.
The timescale might be defined by
\eq{
t_{\m{ff}}(M,z) = \frac{3/2 n_{\m{gas}}T_{\m{halo}}}{\int d\nu \epsilon_{\nu}^{\m{ff}}}.
\label{eq:ff_cool}
}
where $\epsilon_{\nu}^{\m{ff}}$ is the efficiency of free-free emission which is defined in the following subsection, especially in Eq.~\eqref{eq: brems_emit_rate}.
When the gas is ionized (even if partially),
the Compton cooling by CMB photons also can be the main cooling mechanism.
In this case, the cooling timescale depends on only the redshift,
\eq{
t_{\rm Comp}(z) =\frac{3m_{\m{e}} c}{4\sigma_{\mathrm{T}}\rho_{\gamma}}f_{\m{cool}}
\simeq 1.4\times 10^7f_{\m{cool}}\lr{\frac{1+z}{20}}^{-4}\mathrm{yr},
\label{eq:Comp_cool}
}
where $m_{\m{e}}$ is the mass of electron, $\rho_{\gamma}$ is the CMB energy density,
and $f_{\m{cool}}\equiv(1+x_e)/2x_e$ represents the dependence on the ionized fraction. Here we assume that the baryon is composed of only hydrogen atoms.

Through the paper, we adopt the cooling timescale for the smaller value of Eqs.~\eqref{eq:ff_cool} and \eqref{eq:Comp_cool},
that is, $t_\m{cool}= \min (t_{\m{ff}},t_{\rm Comp})$.
Because of the cooling, even if the gas temperature
is high enough to ionize the gas at the formation time,
the temperature decreases, and finally
the gas becomes neutral by following
Eq.~\eqref{eq: ionization_fraction_equilibrium}.

As we mentioned above, we assume that stars are not formed in DM halos, 
and we consider only the Compton cooling as the cooling process.
However, when the virial temperature is larger than $10^4$~K,
the atomic cooling becomes effective
and leads to further collapsing of the gas and the star formation.
Stars could heat up and 
ionize the gas in such massive DM halos substantially.
While that can lead to enhance the free-free emission,
the timescale and the efficiency of the star formation are still uncertain.
Therefore, we do not include these effects.

\subsection{free-free emission}
When the gas in DM halos has a high temperature enough 
for the collisional ionization, 
the ionized gas can emit radiation through thermal free-free emission.
The emissivity of the thermal free-free emission at 
a frequency $\nu$ is given by~\cite{Radiative_process_in_Astrophysics}
\begin{align}\label{eq: brems_emit_rate}
\epsilon_{\nu}^{\m{ff}}
=& \frac{2^3  e^6}{3 m_\m{e} c^3}
\left(\frac{2 \pi}{3 m_\m{e} k_{\rm B} T_\m{halo}}\right)^{1/2} \nonumber \\
&\times x_{\rm e}^2{ n^2_{\rm gas}}
\exp({- h_{\m{p}} \nu/k_{\rm B} T_\m{halo}})
\bar{g}_{\m{ff}}~,
\end{align}
where $h_{\m{p}}$ is the Planck constant.
In this paper, we are interested in the CMB frequency range or below them.
Since these frequencies are much smaller than
the halo temperature for the free-free emission,
$h_{\m{p}} \nu \ll k_{\rm B} T_{\rm halo}$,
we take $\exp(- h_{\m{p}} \nu/k_{\rm B} T_\m{halo}) \approx 1$.
In Eq.~\eqref{eq: brems_emit_rate},
$\bar{g}_{\m{f f}}$ is a {velocity-averaged Gaunt factor}.
Ref.~\cite{2011piim.book.....D} has provided
the fitting formula for the $\bar{g}_{\m{f f}}$,
\eq{\label{eq: gaunt_factor}
\bar{g}_{\m{f f}}=\log \left\{\exp \left[5.960-\sqrt{3} / \pi \log \left(\nu_{9}T_{4}^{-3 / 2}\right)\right]+\mathrm{e}\right\},
}
where $\nu_{9}\equiv \nu/(1~\m{GHz})$,
and $\m{e}$ is the Napier's constant.
Since, in our gas model,
the number density has radial dependence,
the emissivity also depends on the radius in a DM halo.

Now we calculate
the mean intensity of the free-free emission
from a DM halo.
When the optical depth of the free-free absorption is negligible,
the mean intensity of the free-free emission
induced by an individual halo is obtained by 
\begin{align}
\label{eq: Inu_w/_impact_param}
I_{\nu}^{\m{ind}}(z,z_{\rm f},M_\m{halo}) = 
\frac{\int \epsilon_{\nu}^{\m{ff}}dV}{ S_{\m{halo}}},
\end{align}
where $S_{\m{halo}}$ is the physical cross section of a DM halo on the sky,
$S_{\m{halo}} = \pi R_{\m{vir}}^2$.

Since the gas profile depends on the redshift of the emission~$z$,
the formation redshift~$z_{\rm f}$,
and the mass of the DM halos~$M_\m{halo}$,
Eq.~\eqref{eq: Inu_w/_impact_param} depends on these parameters.

\section{Diffuse background free-free emission}
\label{section:all-sky}

Now we consider the diffuse background intensity
emitted from DM halos with mass~$M_\m{halo}$
residing spherical shell between from $z$ to $z+dz$.
This
intensity is obtained by integrating the contributions
from the DM halos formed until redshift~$z$,
\eq{\label{eq: dI_sky_dz}
&dI_\m{\nu}^{\m{ff}}(z,M_\m{halo})
=\lr{\int^{\infty}_{z} f_{\rm sky} I_{\nu}^{\m{ind}}
\frac{dn_{\m{halo}}^{\m{com}}}{{~d} z_f}d\zf}\frac{dV_{\rm com}}{dz} dz~.
}
Here, we omit the dependencies of
$f_{\rm sky}$ and $I_{\nu}^{\m{ind}}$ on $(z,z_{\rm f},M_\m{halo})$.
In Eq.~\eqref{eq: dI_sky_dz},
$V_{\rm com}$ is the total comoving volume out to redshift~$z$,
and $n_\m{halo}^{\rm com}$ is the comoving number density of halos
with mass $M_\m{halo}$ at a redshift~$z$, which we discuss later.
Additionally, 
$f_{\rm sky}$ in Eq.~\eqref{eq: dI_sky_dz}
is the sky fraction of the DM halos
formed at $z_{\rm f}$ with mass $M_\m{halo}$
and given by
\eq{\label{eq: sky_z}
f_{\rm sky } = \frac{\Omega_{\m{halo}}}{4\pi},
}
where the solid angle~$\Omega_{\m{halo}}(z,z_{\rm f}, M_\m{halo})$
of a DM halo is given by
$\Omega_{\m{halo}} = \pi{((1+z)R_{\m{vir}})^2}/{\chi^2}$
with the comoving distance $\chi$, from $z=0$ to $z$.

Taking into account the redshift effect due to the cosmological expansion,
the diffuse background intensity
at the observed frequency~$\nu_{\m{obs}}$ is calculated by
\eq{\label{eq: Inu_full}
I_{\m{obs}} (\nu_{\m{obs}})
=\int^{\infty}_{z_{\m{cut}}}~dz~\frac{1}{(1+z)^3}
\frac{dI_\m{\nu_{\m{em}}}^{\m{ff}}(z,M_\m{halo})}{dz},
}
where $\nu_{\m{em}}=(1+z)\nu_{\m{obs}}$, and $z_{\m{cut}}$ is the lower limit of the redshift. We are interested in the free-free emission from DM halos created by the excess fluctuations from the scale-invariant spectrum.
Therefore, we set $z_{\rm cut}$ to the redshift at which the mass variance due to the scale-invariant spectrum reaches the critical density contrast $\delta_c = 1.69$, so that
$\delta_ c = \sigma_{\rm i}(z, M_{\rm halo})$, where $\sigma_{\rm i} (z, M)$ is the mass variance calculated from the scale-invariant spectrum at $z$.

\subsection{DM halo number density}
In order to calculate the number density of DM halos in~Eq.~\eqref{eq: dI_sky_dz},
we assume that the density fluctuations obey Gaussian statistics.
We represent the mass variance with mass $M_{\rm halo}$ 
at the present epoch as $\sigma_0(M_{\rm halo})$.
The mass variance at a redshift~$z$ is 
$\sigma(z, M_{\rm halo}) = D(z) \sigma_0(M_{\rm halo})$
with the linear growth rate $D(z)$ 
which is normalized as $D(0)=1$ at the present.
Using the spherical collapse model,
at a redshift $z_\m{f}$ in the matter dominated epoch, 
the mass fraction of DM halos with mass $>M_{\rm halo}$ is given by
\eq{
f_{\rm coll}(z_\m{f}, M_{\rm halo}) = {\m{erfc}}\lr{\frac{\nu_\m{c}}{\sqrt{2}}},
}
where $\nu_\m{c}\equiv \delta_\mathrm{c}/\sigma (z,M_\mathrm{halo})$.

The mass distribution of DM halos
depends on the shape of the mass variance as a function of $M_{\m{halo}}$.
However in this work, instead of specifying the shape of the mass variance,
we simply assume that
the number density of DM halos with mass $M_{\rm halo}$ is given by

\eq{\label{eq: num_halo_com}
n^{\rm com}_{\rm halo} (z_\m{f}, M_{\rm halo})
= \Omega_\m{m} \frac{\rho_{\rm crit}}{M_{\rm halo}}
f_{\rm coll}(z_\m{f}, M_{\rm halo}).
}
Since the mass fraction, $\sigma(M_{\m{halo}})$ includes the contribution from not only dark matter halos with $M_{\rm halo}$ but also with $M>M_{\rm halo}$, this evaluation leads to overestimate the number density of $M_{\rm halo}$.
In other words, we neglect the DM halos with $M>M_{\rm halo}$ in the above treatment.
However, the
free-free intensity is larger for massive halos than that for lighter mass.
We discuss this point in the next section.
Therefore, the obtained intensity is
the lowest signals of the free-free emission 
for the given mass variance with a single mass $M_{\rm halo}$.
Consequently, the limit on the primordial curvature perturbations would be estimated conservatively.

Accordingly, the derivative of the number density
with respect to the formation redshift, $z_{\rm f}$ is 
\eq{\label{eq: num_den_press}
\frac{dn^{\m{com}}_{\m{halo}}}{d\zf}
= \sqrt{\cfrac{2}{\pi}}~
\frac{\Omega_\m{m} \rho_{\rm crit}}{M_{\m{halo}}}~
\nu_{c,0} \exp\lr{-\frac{\nu_c^2}{2}},
}
where $\nu_{c,0}=\delta_\m{c}/\sigma_0(M_{\m{halo}})$.
Through this halo number density,
the diffuse background intensity
from early-formed DM halos
depends on the density fluctuations,
whose statistical property we represent by two parameters
($M_{\rm halo},~\sigma_0(M_{\m{halo}})$).

In the next section, we demonstrate that
the observation of the all-sky averaged free-free emission
is a powerful probe of the density fluctuations on small scales, 
calculating the amplitude of
the diffuse background intensity
with different sets of two parameters~($M_{\rm halo},~\sigma_0(M_{\m{halo}})$).

\section{Results and Discussion}
\label{section:results}
\begin{figure}
    \centering
    \includegraphics[width=1.0\hsize]{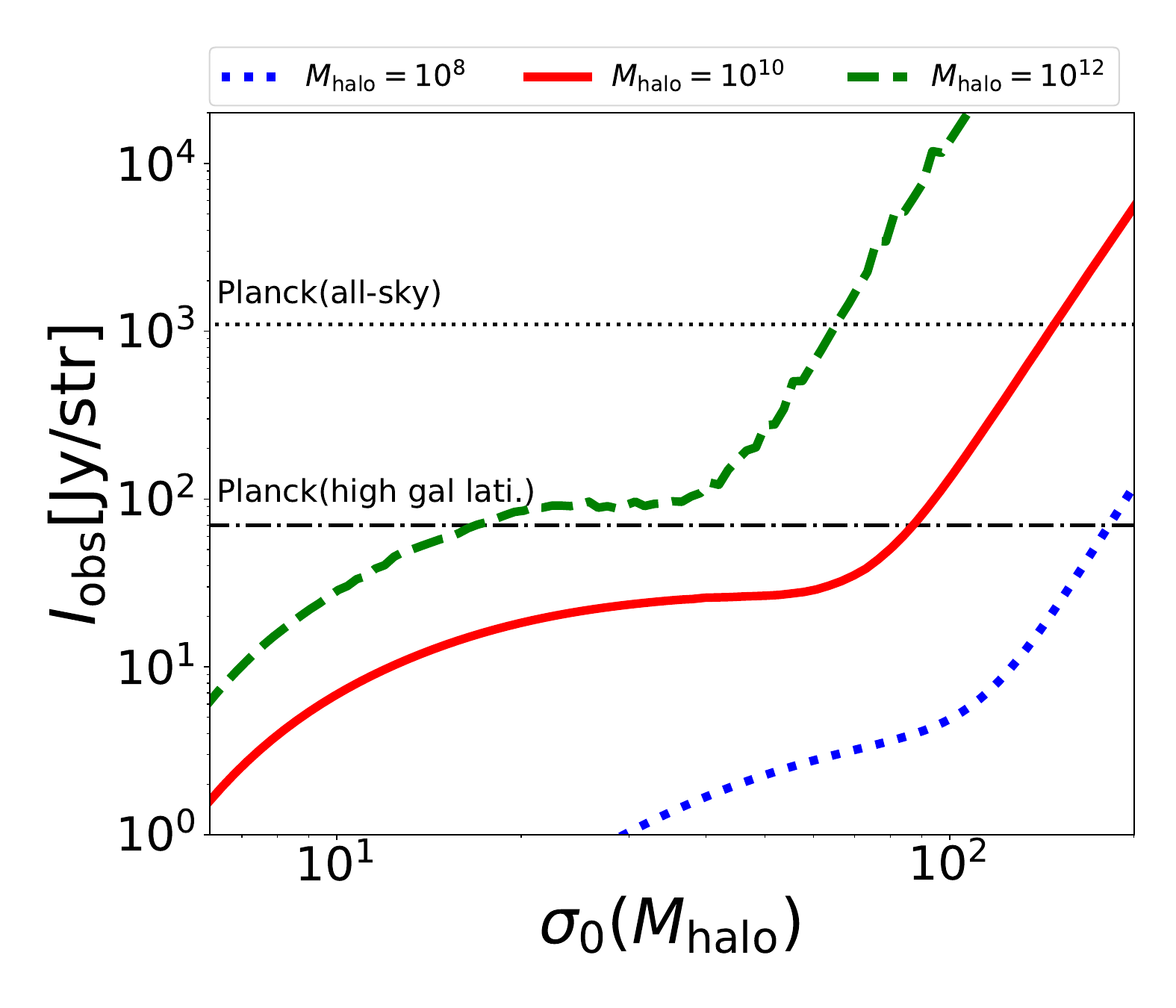}
    \caption{
    The diffuse background intensity in Eq.~\eqref{eq: Inu_full} with different parameter sets $(M_{\m{halo}},\sigma_0(M_{\m{halo}}))$ at $\nu=70\m{GHz}$. The blue dotted line shows the intensity as a function of $\sigma_0$ with $M_{\m{halo}}=10^8M_{\odot}$.
    The red solid and green dashed lines show the ones with $M_{\m{halo}}=10^{10},~10^{12}M_{\odot}$ respectively. The thin black dotted line shows
    the intensity of the all-sky averaged free-free emission observed
    by the Planck satellite, and the thin black dash-dotted line shows
    the intensity averaged only in high galactic latitude region.
    See Fig.~\ref{fig: free_planck} for the detail.}
    \label{fig: Iobs_sigma0}
\end{figure}

In Fig.~\ref{fig: Iobs_sigma0}, we show 
the dependence of the diffuse background intensity
from early-formed DM halos at $\nu = 70~\rm GHz$ on the mass variance~$\sigma_0(M)$ for different $M=M_{\rm halo}$.
It is worth mentioning that, in the standard $\Lambda$CDM model consistent with the Planck data,
the mass variance are $\sigma_0(M) =12.0,~6.5$ and $3.5$ for $M = 10^{10},~10^{11}$ and $10^{12} M_\odot$, respectively.
When $\sigma_0(M)$ becomes large, the formation of DM halos with mass~$M$ would start in higher redshifts, and
the number density of DM halos emitting the free-free emission would increase.
Therefore, the large $\sigma_0(M)$ enhances the amplitude of
the diffuse background intensity
in these mass scales.
Fig.~\ref{fig: Iobs_sigma0} tells us that
more massive halos emit the stronger free-free intensity than less massive halos.
This reason can be interpreted that
the massive halos have higher virial temperature
at their formation time,
and they survive longer until the cooling down and the neutralization of the baryon gas.

\begin{figure}
    \centering
    \includegraphics[width=1.0\hsize]{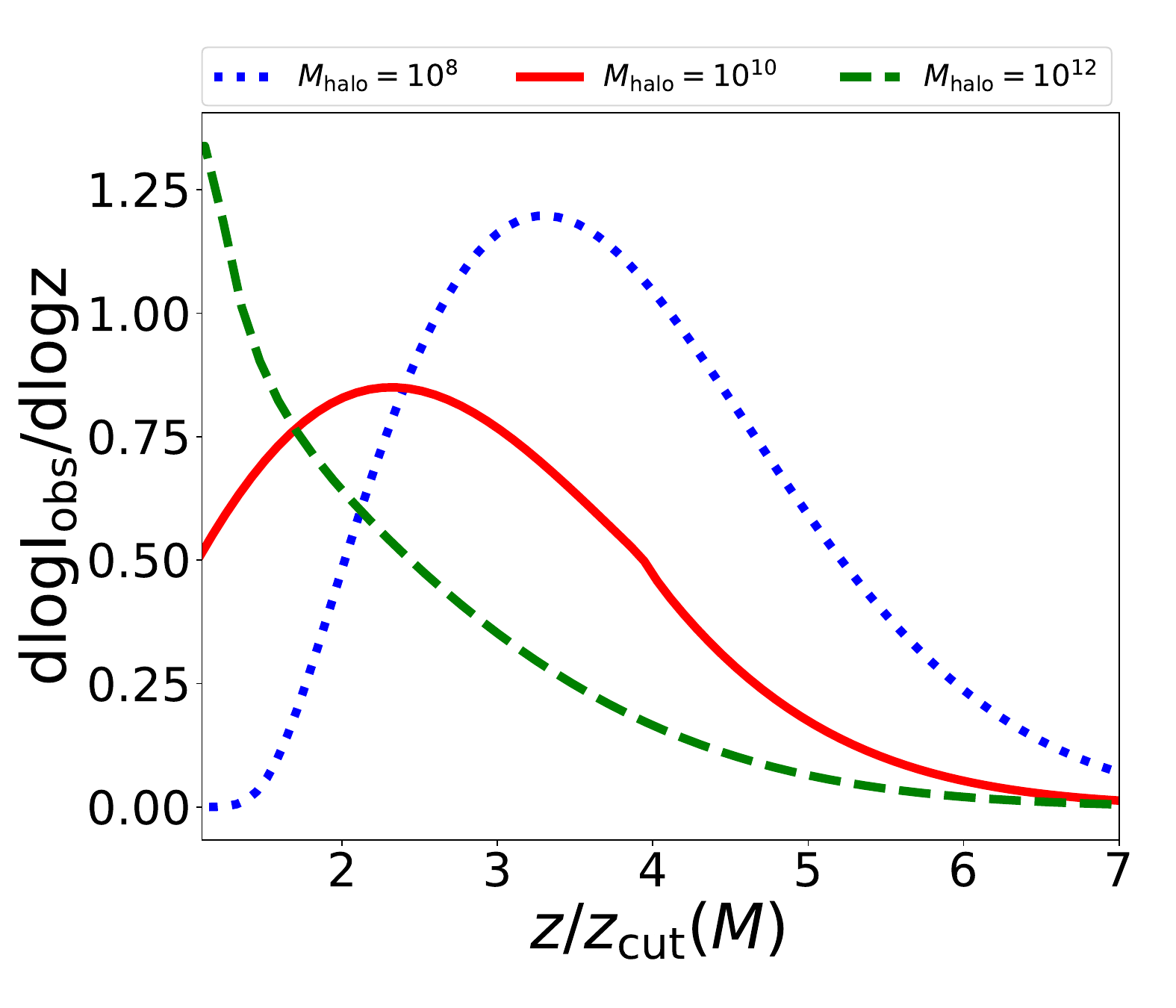}
    \caption{The redshift distribution of the diffuse background intensity induced by early-formed halos with mass and mass variance $(M_{\m{halo}}[M_{\odot}],\sigma_0)=(10^{8},24.8), (10^{10},19.5), (10^{12},14.2)$.
    The color and line styles are the same as in Fig.~\ref{fig: Iobs_sigma0}.
    }
    \label{fig: dIobs_dz}
\end{figure}

Additionally, Fig.~\ref{fig: dIobs_dz}
shows the redshift distribution of the diffuse background intensity with at each redshift for $M_{\m{halo}} = 10^8,10^{10},$ and $10^{12} M_\odot$.
In this figure, we fix the mass variance as $\sigma_0= 24.8, 19.5,$ and $14.2$, respectively.

\begin{figure}
    \centering
    \includegraphics[width=1.0\hsize]{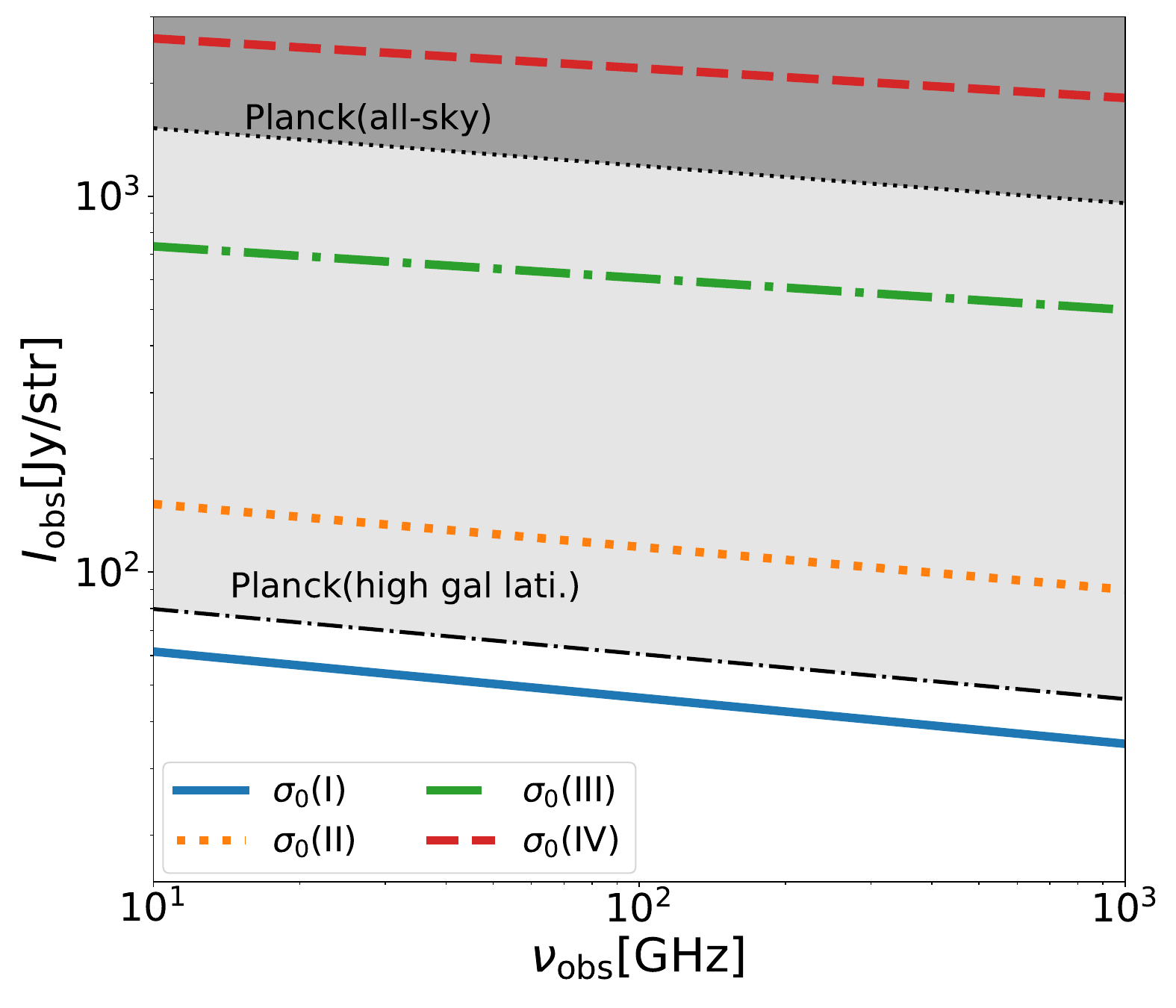}
    \caption{
    The diffuse background intensity
    induced by early-formed halos with mass $M_{\m{halo}}=10^{12}M_{\odot}$. The thick lines with different color and line types represent
    the diffuse background intensity
    with different amplitude of $\sigma_0$.
    From bottom to top, the lines are for $\sigma_0({\rm i})=(12.7,41.7, 59.2, 73.0)$ with (i=I,II,III,IV).
    The thin black dotted line and the thin black dash-dotted line are the same as in Fig.~\ref{fig: Iobs_sigma0}.
    }
    \label{fig: Iobs_nu_M12}
\end{figure}

Fig.~\ref{fig: Iobs_nu_M12} presents the frequency dependence of
the diffuse background intensity
in the CMB frequency range.
These frequencies are
smaller than $k_{\rm B} T_{\rm halo}/h_{\m{p}}$.
Therefore, the frequency dependence comes from only the Gaunt factor of Eq.~\eqref{eq: gaunt_factor}.
The dependence is the same as in the
free-free emission component
in the CMB data analysis~\cite{2016A&A...594A..10P}.

\begin{figure}
    \centering
    \includegraphics[width=1.0\hsize]{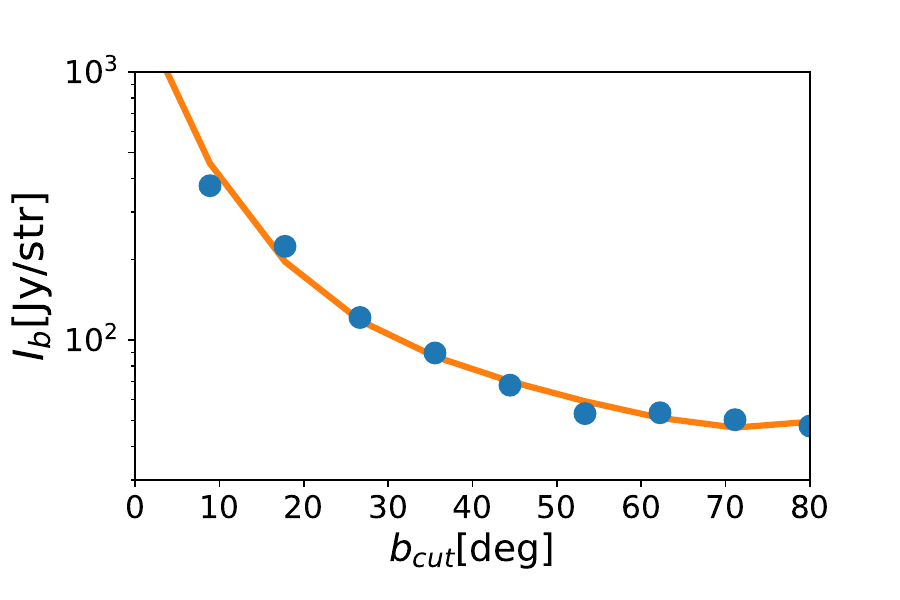}
    \caption{
    The averaged intensity of the free-free emission
    from the Planck foreground map. The horizontal axis represents the absolute galactic latitude $|b_{\rm cut}|$
    for masking the Galactic disk component. The orange line is the fitting curve for illustrative purposes.
    } 
    \label{fig: free_planck}
\end{figure}

In the Planck CMB analysis, the free-free emission component
has been separated from other diffuse foreground emissions.
The obtained intensity map of the free-free emission
tells us that the emission is strongly anisotropic.
At low galactic latitude, the emission is strong, while the emission becomes weak as the galactic latitude increases.
From the Planck data,
the all-sky averaged free-free emission component
is identified as $I_\nu \approx 10^3~\rm Jy/str$ at the frequency $\nu =70~\rm GHz$~\cite{2016A&A...594A..10P,2020A&A...641A...4P}.
To avoid the contamination from the Galactic disk, 
we also calculate
the intensity of the observed free-free emission component
by masking the region at galactic latitude $|b|<b_{\rm cut}$.
Fig.~\ref{fig: free_planck} shows the dependence of the averaged intensity on the cutoff galactic latitude $b_{\rm cut}$ at $\nu = 70~$GHz.
The averaged intensity becomes small as $b_{\rm cut}$ increases, and we have found that
when we set $b_{\rm cut} > 60 ~$deg., so that the region with $|b| <b_{\rm cut}$ is masked, the resultant averaged intensity observed by Planck drops down roughly to
$70$~Jy/str~(in the brightness temperature,
$T_{b} \sim  0.7 \times(\nu/70~{\rm GHz})^{-2}{\rm \mu K}$).
For comparison, we plot
the intensity of the all-sky averaged free-free emission
$I_\nu \approx 10^3~\rm Jy/str$,
and only in high galactic latitude,
$I_\nu \approx 70~\rm Jy/str$ 
as black dotted and dash-dotted lines,
respectively,
in Figs.~\ref{fig: Iobs_sigma0} and~\ref{fig: Iobs_nu_M12}.

\begin{figure}
    \centering
    \includegraphics[width=1.0\hsize]{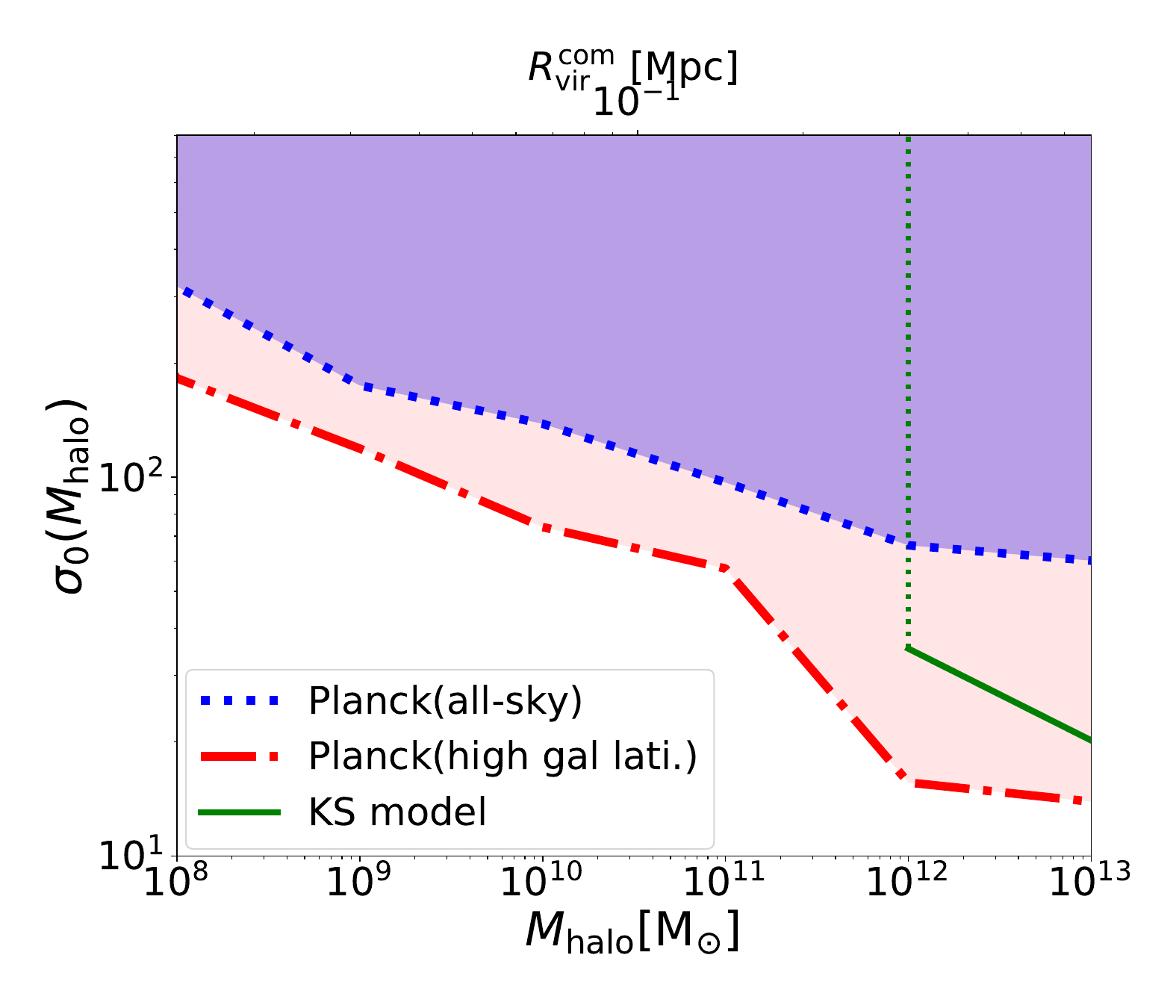}
    \caption{The constraint on $\sigma_{0}(M_{\m{halo}})$
    from the observed free-free emission.
    The blue dotted line represents the limit from the Planck all-sky observation. The red dash-dotted line shows the one by the Planck observation only in the high galactic latitude region.
    The green solid line shows the same constraint as the red dash-dotted line but with the KS model for the gas profile in the DM halos.
    Below $M_{\rm halo} < 10^{12}~M_\odot$, the signals in the KS model are suppressed too strongly to obtain the constraint.}
    \label{fig: sigma_M_limit}
\end{figure}

Since the diffuse background intensity
from early-formed DM halos cannot exceed
the intensity of the observed free-free emission,
we can obtain the constraint on $\sigma_0 (M_\m{halo})$.
Fig.~\ref{fig: sigma_M_limit} represents
the upper limit on
the allowed $\sigma_0(M_\m{halo})$ for each mass~$M_\m{halo}$.
The parameter sets in the colored region show the excluded ones where the diffuse background intensity is larger than the measured signal.
It is noted that the dependence of the signal amplitude on the DM halo mass $M_{\rm halo}$
is weak in the mass range between 
$10^{12}~M_{\odot}$ and $10^{13}~M_{\odot}$.
This is because
the free-free emission from individual halos
gets higher as the halo mass increases,
while the number density of DM halos decreases for a fixed $\sigma_0$.
For halos with mass $10^{11}~M_\odot \lesssim M_{\rm halo} \lesssim 10^{12}~M_\odot$,
the diffuse background intensity weakly depends on the mass variance $20 \lesssim \sigma_0 \lesssim 50$, as shown in Fig.~\ref{fig: Iobs_sigma0}.
Therefore the constraint on $\sigma_0 (M_\mathrm{halo})$ strongly changes on these mass scales.
Finally, in the lightest mass range as $M_{\rm halo} \lesssim 10^{9}~M_\odot$, the virial temperature is not high enough to ionize the gas inside DM halos.
Therefore, the signal strength decreases, and resultantly, the constraint on $\sigma_0(M_{\m{halo}})$ quickly becomes weak.

We find that
the fitting formula of the limitation of $\sigma_0(M_{\m{halo}})$ by
the intensity of the observed free-free emission
averaged in all-sky
is
\eq{\label{eq: fit_constraint_sigma0_M}
\log_{10}\sigma_{0}<-0.0061x^3+0.21x^2-2.56x+12.43,
}

where $x=\log_{10}M_{\m{halo}}$.
We mention that this fitting function can be adopted in the mass region $M_{\m{halo}}=(10^8M_{\odot},10^{13}M_{\odot})$.

We have obtained the constraint on the mass variance on $M_{\rm halo}$.
It is also worth discussing this constraint in terms of the primordial curvature perturbations.
The mass variance is calculated from the power spectrum of the curvature perturbations $\mathcal{P}_{\zeta }(k)$ by
\begin{align}
\sigma_0^2 (M) = \int d\log k ~\cfrac{4}{25}~\cfrac{k^4}{\Omega_\mathrm{m}^2 H_0^4} ~\mathcal{P}_{\zeta }(k) T^2(k)  W_k^2(kR_{M}(M)),
\label{eq:sigma_P}
\end{align}
where $W_k(x)$ is the
Fourier component of the top-hat window function,
and $R_M(M)$ is the comoving scale in which the mass $M$
is enclosed with the background matter density $\Omega_{\rm m} \rho_{\rm cri}$.
Besides, we adopt the transfer function for the matter density fluctuations $T(k)$ as
\cite{2008cosm.book.....W}
\begin{align}
T(k) = \cfrac{45}{2} \cfrac{\Omega_\mathrm{m}^2 H_0^2}{\Omega_\mathrm{r} k^2} \left(-\cfrac{7}{2}+\gamma_\mathrm{E} + \ln{\left(\cfrac{4\sqrt{\Omega_\mathrm{r}}k}{\sqrt{3}\Omega_\mathrm{m} H_0}\right)} \right)~,
\label{eq:transfer}
\end{align}
with the Euler-Mascheroni constant $\gamma_\mathrm{E} \simeq 0.577$.
Since we would like to constrain the excess of the fluctuations from the scale-invariant spectrum on small scales, we implicitly assume the blue-tilted power spectrum on these scales. 
In this case, 
$\mathcal{P}_{\zeta }(k) W_k^2(kR)$ in Eq.~\eqref{eq:sigma_P} has a maximum value at $k=2\pi/R$.
Therefore, Eq.~\eqref{eq:sigma_P} can be approximated to
\eq{\label{eq: sigma0_M_P_zeta}
\sigma_0^2 (M) \approx  \left. \cfrac{4}{25}~\cfrac{k^4}{\Omega_\mathrm{m}^2 H_0^4}~ \mathcal{P}_{\zeta }(k) T^2(k) \right|_{k= k_{\rm p}},
}
where $k_{\rm p}=2\pi /R_{M}(M_{\rm halo})$ and we adopt the sharp-$k$ filter, $k=k_{\rm{p}}$.

\begin{figure}
    \centering
    \includegraphics[width=1.0\hsize]{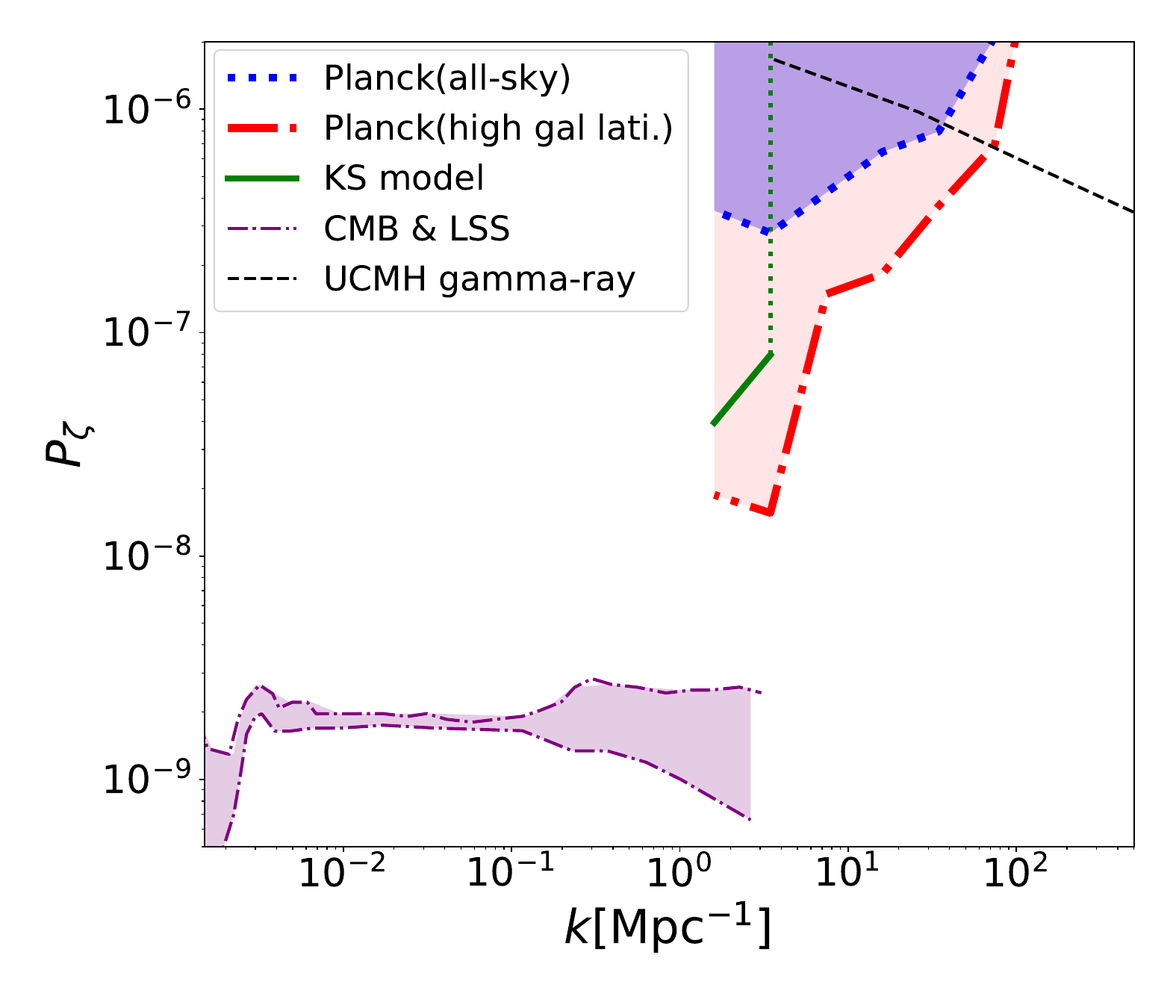}
    \caption{
    The constraint on the primordial curvature power spectrum $P_{\zeta}(k)$ through non-detection of
    the diffuse background intensity
    induced by early-formed DM halos with their mass of $M_{\m{halo}}$. The three thick lines: red dash-dotted, green solid, and blue dotted lines show the constraints by the same as in Fig.~\ref{fig: sigma_M_limit}. The purple dash-dotted line and shaded region represent the allowed parameter region obtained from the CMB and LSS observation~\cite{2009JCAP...07..011N,2010JCAP...01..016N,2011MNRAS.413.1717B}. The black dashed line shows the one by the theoretical prediction of the gamma-ray induced by the UCMH with the self-annihilating DM model~\cite{2012PhRvD..85l5027B}.}
    \label{fig: Pzeta_k}
\end{figure}

Using Eq.~\eqref{eq: sigma0_M_P_zeta}, we rewrite our constraint on $\sigma_0(M_{\rm halo})$ in Fig.~\ref{fig: sigma_M_limit} to the limit on the amplitude of the delta-function type primordial curvature power spectrum, $\cal P_{\zeta }$ as a function of different $k$. 
Fig.~\ref{fig: Pzeta_k} is a summary of our constraint on the amplitude of the curvature perturbations.
On scales $1~{\rm Mpc}^{-1} < k < 100~{\rm Mpc}^{-1}$,
our constraint is stronger than the previous ones.

At the last of this section, we discuss the dependence of our results on the gas profile. 
The diffuse background intensity
is proportional to the square of the number density of gas particle as shown in Eq.~\eqref{eq: brems_emit_rate}.
We have evaluated the diffuse background intensity
with the gas profile based on the hydrostatic equilibrium and the isothermal condition.
In order to estimate the model uncertainty of the gas profile, we also calculate with the different gas profile model~(KS model) suggested in Ref.~\cite{2002MNRAS.336.1256K}.
In the KS model, without the isothermal condition, the gas density and temperature profiles are obtained from the hydrostatic equilibrium and the condition that the shape of the gas density profile becomes the same as the one of DM in the outer region of the DM halo.
Calculating the diffuse background intensity
with this model and
we find out that the signals become smaller than in the isothermal case.
Although it depends on halo mass,
the extent of the signal suppression is a factor of two 
in the mass range between $10^{12} M_\odot < M < 10^{13} M_\odot$.
Comparing 
the free-free emission at the high galactic latitude, 
we obtain the constraint in the case of the KS model
and plot it in Figs.~\ref{fig: sigma_M_limit} and \ref{fig: Pzeta_k}. 
Because of the signal suppression,
the constraint in the KS model is weaker than in the isothermal case as well.

\section{Conclusion}
\label{section:conclusion}

In this work, we have estimated
the diffuse background free-free emission
due to the early formation of DM halos. 
If there are a large amplitude of density fluctuations on small scales,
DM halos form efficiently after the matter-radiation equality. 
Inside DM halos with high virial temperature, the gas is ionized by the thermal collision and can emit photons by the free-free process.

First, we construct the gas model in DM halos, assuming isothermal gas with the hydrostatic equilibrium 
and calculate the free-free emission from individual halos.
In our model, the gas is heated up to the virial temperature at the halo formation
and cooled down by the Compton scattering with CMB photons.
The ionization of gas is achieved by
thermal collisional ionization.
The gas with a large virial temperature is ionized and can emit radiation through free-free emission. However since the Thomson scattering cooled down the gas,
the gas quickly becomes neutral and terminates the free-free emission.
The intensity is completely determined by the halo mass and the
formation redshift.

Next, assuming the excess of density fluctuations from the scale-invariant power spectrum on small scales,
we have computed the diffuse background intensity.

Since it depends on the formation history of DM halos,
the intensity
is sensitive to the density fluctuations.
We have found that the large mass variance promotes
the DM halo formation at high redshifts,
and increases the diffuse background intensity.

By comparing
the intensity of the free-free emission observed
by the Planck satellite,
we have put an upper limit on the mass variance at the present epoch, $\sigma_0(M)$.
The observed free-free emission is highly anisotropic and
the intensity averaged only in high galactic latitude region
is lower than that of the all-sky averaged free-free emission.
We have shown that the comparison with
the emission at high galactic latitude
makes the constraint stronger.
The obtained constraint on the mass variance from
intensity averaged only in high galactic latitude region
is $\sigma_0 < 20$ in the mass range
$10^{12} M_\odot < M < 10^{13} M_\odot$.
This constraint allows us to provide the limit on the amplitude of the delta-function-type primordial curvature power spectrum,
$P_\zeta(k) \lesssim 10^{-7}$
for $1~\mathrm{Mpc}^{-1} \lesssim k \lesssim 100~\mathrm{Mpc}^{-1}$.

We have demonstrated that the measurements of
the diffuse background free-free emission
can probe the abundance of early-formed DM halos.
However, there are several theoretical uncertainties related to
the nonlinear structure formation in our estimation.
One of such uncertainties is the baryon gas profile in DM halos. 
In order to investigate this uncertainty, we calculate
the diffuse background intensity
with a different gas profile model as in Ref.~\cite{2002MNRAS.336.1256K}.
We have found that the difference of the gas profile models provides the suppression of the signal by a factor of two.

One way to improve the theoretical prediction of
the diffuse background free-free emission
is to perform
numerical simulations of the early structure formation including baryon physics.
Not only the gas profile mentioned above but also 
the star formation in early-formed DM halos can affect
the diffuse background intensity
because radiation from stars can heat up and ionize
baryon gas inside halos.
In addition, the further improvement of the constraint
is expected by combining the other observational probes,
21-cm line observations,
thermal and kinetic Sunyaev-Zel'dovich (SZ) effect,
and so on.
We leave them for our future work.

\acknowledgements
This work is supported by Japan Society for the Promotion of Science~(JSPS) KAKENHI Grants
No.~JP20J22260 (K.T.A),
and No.~JP21K03533 (H.T),
and supported by JSPS Overseas Research Fellowships (T.M).

\bibliography{article}

\begin{thebibliography}{61}%
\makeatletter
\providecommand \@ifxundefined [1]{%
 \@ifx{#1\undefined}
}%
\providecommand \@ifnum [1]{%
 \ifnum #1\expandafter \@firstoftwo
 \else \expandafter \@secondoftwo
 \fi
}%
\providecommand \@ifx [1]{%
 \ifx #1\expandafter \@firstoftwo
 \else \expandafter \@secondoftwo
 \fi
}%
\providecommand \natexlab [1]{#1}%
\providecommand \enquote  [1]{``#1''}%
\providecommand \bibnamefont  [1]{#1}%
\providecommand \bibfnamefont [1]{#1}%
\providecommand \citenamefont [1]{#1}%
\providecommand \href@noop [0]{\@secondoftwo}%
\providecommand \href [0]{\begingroup \@sanitize@url \@href}%
\providecommand \@href[1]{\@@startlink{#1}\@@href}%
\providecommand \@@href[1]{\endgroup#1\@@endlink}%
\providecommand \@sanitize@url [0]{\catcode `\\12\catcode `\$12\catcode
  `\&12\catcode `\#12\catcode `\^12\catcode `\_12\catcode `\%12\relax}%
\providecommand \@@startlink[1]{}%
\providecommand \@@endlink[0]{}%
\providecommand \url  [0]{\begingroup\@sanitize@url \@url }%
\providecommand \@url [1]{\endgroup\@href {#1}{\urlprefix }}%
\providecommand \urlprefix  [0]{URL }%
\providecommand \Eprint [0]{\href }%
\providecommand \doibase [0]{https://doi.org/}%
\providecommand \selectlanguage [0]{\@gobble}%
\providecommand \bibinfo  [0]{\@secondoftwo}%
\providecommand \bibfield  [0]{\@secondoftwo}%
\providecommand \translation [1]{[#1]}%
\providecommand \BibitemOpen [0]{}%
\providecommand \bibitemStop [0]{}%
\providecommand \bibitemNoStop [0]{.\EOS\space}%
\providecommand \EOS [0]{\spacefactor3000\relax}%
\providecommand \BibitemShut  [1]{\csname bibitem#1\endcsname}%
\let\auto@bib@innerbib\@empty
\bibitem [{\citenamefont {{Planck Collaboration}}\ \emph
  {et~al.}(2020{\natexlab{a}})\citenamefont {{Planck Collaboration}},
  \citenamefont {{Aghanim}}, \citenamefont {{Akrami}}, \citenamefont
  {{Ashdown}}, \citenamefont {{Aumont}}, \citenamefont {{Baccigalupi}},
  \citenamefont {{Ballardini}}, \citenamefont {{Banday}}, \citenamefont
  {{Barreiro}}, \citenamefont {{Bartolo}}, \citenamefont {{Basak}},
  \citenamefont {{Battye}}, \citenamefont {{Benabed}}, \citenamefont
  {{Bernard}}, \citenamefont {{Bersanelli}}, \citenamefont {{Bielewicz}},
  \citenamefont {{Bock}}, \citenamefont {{Bond}}, \citenamefont {{Borrill}},
  \citenamefont {{Bouchet}}, \citenamefont {{Boulanger}}, \citenamefont
  {{Bucher}}, \citenamefont {{Burigana}}, \citenamefont {{Butler}},
  \citenamefont {{Calabrese}}, \citenamefont {{Cardoso}}, \citenamefont
  {{Carron}}, \citenamefont {{Challinor}}, \citenamefont {{Chiang}},
  \citenamefont {{Chluba}}, \citenamefont {{Colombo}}, \citenamefont
  {{Combet}}, \citenamefont {{Contreras}}, \citenamefont {{Crill}},
  \citenamefont {{Cuttaia}}, \citenamefont {{de Bernardis}}, \citenamefont {{de
  Zotti}}, \citenamefont {{Delabrouille}}, \citenamefont {{Delouis}},
  \citenamefont {{Di Valentino}}, \citenamefont {{Diego}}, \citenamefont
  {{Dor{\'e}}}, \citenamefont {{Douspis}}, \citenamefont {{Ducout}},
  \citenamefont {{Dupac}}, \citenamefont {{Dusini}}, \citenamefont
  {{Efstathiou}}, \citenamefont {{Elsner}}, \citenamefont {{En{\ss}lin}},
  \citenamefont {{Eriksen}}, \citenamefont {{Fantaye}}, \citenamefont
  {{Farhang}}, \citenamefont {{Fergusson}}, \citenamefont {{Fernandez-Cobos}},
  \citenamefont {{Finelli}}, \citenamefont {{Forastieri}}, \citenamefont
  {{Frailis}}, \citenamefont {{Fraisse}}, \citenamefont {{Franceschi}},
  \citenamefont {{Frolov}}, \citenamefont {{Galeotta}}, \citenamefont
  {{Galli}}, \citenamefont {{Ganga}}, \citenamefont {{G{\'e}nova-Santos}},
  \citenamefont {{Gerbino}}, \citenamefont {{Ghosh}}, \citenamefont
  {{Gonz{\'a}lez-Nuevo}}, \citenamefont {{G{\'o}rski}}, \citenamefont
  {{Gratton}}, \citenamefont {{Gruppuso}}, \citenamefont {{Gudmundsson}},
  \citenamefont {{Hamann}}, \citenamefont {{Handley}}, \citenamefont
  {{Hansen}}, \citenamefont {{Herranz}}, \citenamefont {{Hildebrandt}},
  \citenamefont {{Hivon}}, \citenamefont {{Huang}}, \citenamefont {{Jaffe}},
  \citenamefont {{Jones}}, \citenamefont {{Karakci}}, \citenamefont
  {{Keih{\"a}nen}}, \citenamefont {{Keskitalo}}, \citenamefont {{Kiiveri}},
  \citenamefont {{Kim}}, \citenamefont {{Kisner}}, \citenamefont {{Knox}},
  \citenamefont {{Krachmalnicoff}}, \citenamefont {{Kunz}}, \citenamefont
  {{Kurki-Suonio}}, \citenamefont {{Lagache}}, \citenamefont {{Lamarre}},
  \citenamefont {{Lasenby}}, \citenamefont {{Lattanzi}}, \citenamefont
  {{Lawrence}}, \citenamefont {{Le Jeune}}, \citenamefont {{Lemos}},
  \citenamefont {{Lesgourgues}}, \citenamefont {{Levrier}}, \citenamefont
  {{Lewis}}, \citenamefont {{Liguori}}, \citenamefont {{Lilje}}, \citenamefont
  {{Lilley}}, \citenamefont {{Lindholm}}, \citenamefont {{L{\'o}pez-Caniego}},
  \citenamefont {{Lubin}}, \citenamefont {{Ma}}, \citenamefont
  {{Mac{\'\i}as-P{\'e}rez}}, \citenamefont {{Maggio}}, \citenamefont {{Maino}},
  \citenamefont {{Mandolesi}}, \citenamefont {{Mangilli}}, \citenamefont
  {{Marcos-Caballero}}, \citenamefont {{Maris}}, \citenamefont {{Martin}},
  \citenamefont {{Martinelli}}, \citenamefont {{Mart{\'\i}nez-Gonz{\'a}lez}},
  \citenamefont {{Matarrese}}, \citenamefont {{Mauri}}, \citenamefont
  {{McEwen}}, \citenamefont {{Meinhold}}, \citenamefont {{Melchiorri}},
  \citenamefont {{Mennella}}, \citenamefont {{Migliaccio}}, \citenamefont
  {{Millea}}, \citenamefont {{Mitra}}, \citenamefont {{Miville-Desch{\^e}nes}},
  \citenamefont {{Molinari}}, \citenamefont {{Montier}}, \citenamefont
  {{Morgante}}, \citenamefont {{Moss}}, \citenamefont {{Natoli}}, \citenamefont
  {{N{\o}rgaard-Nielsen}}, \citenamefont {{Pagano}}, \citenamefont
  {{Paoletti}}, \citenamefont {{Partridge}}, \citenamefont {{Patanchon}},
  \citenamefont {{Peiris}}, \citenamefont {{Perrotta}}, \citenamefont
  {{Pettorino}}, \citenamefont {{Piacentini}}, \citenamefont {{Polastri}},
  \citenamefont {{Polenta}}, \citenamefont {{Puget}}, \citenamefont {{Rachen}},
  \citenamefont {{Reinecke}}, \citenamefont {{Remazeilles}}, \citenamefont
  {{Renzi}}, \citenamefont {{Rocha}}, \citenamefont {{Rosset}}, \citenamefont
  {{Roudier}}, \citenamefont {{Rubi{\~n}o-Mart{\'\i}n}}, \citenamefont
  {{Ruiz-Granados}}, \citenamefont {{Salvati}}, \citenamefont {{Sandri}},
  \citenamefont {{Savelainen}}, \citenamefont {{Scott}}, \citenamefont
  {{Shellard}}, \citenamefont {{Sirignano}}, \citenamefont {{Sirri}},
  \citenamefont {{Spencer}}, \citenamefont {{Sunyaev}}, \citenamefont
  {{Suur-Uski}}, \citenamefont {{Tauber}}, \citenamefont {{Tavagnacco}},
  \citenamefont {{Tenti}}, \citenamefont {{Toffolatti}}, \citenamefont
  {{Tomasi}}, \citenamefont {{Trombetti}}, \citenamefont {{Valenziano}},
  \citenamefont {{Valiviita}}, \citenamefont {{Van Tent}}, \citenamefont
  {{Vibert}}, \citenamefont {{Vielva}}, \citenamefont {{Villa}}, \citenamefont
  {{Vittorio}}, \citenamefont {{Wandelt}}, \citenamefont {{Wehus}},
  \citenamefont {{White}}, \citenamefont {{White}}, \citenamefont {{Zacchei}},\
  and\ \citenamefont {{Zonca}}}]{2020A&A...641A...6P}%
  \BibitemOpen
  \bibfield  {author} {\bibinfo {author} {\bibnamefont {{Planck
  Collaboration}}}, \bibinfo {author} {\bibfnamefont {N.}~\bibnamefont
  {{Aghanim}}}, \bibinfo {author} {\bibfnamefont {Y.}~\bibnamefont {{Akrami}}},
  \bibinfo {author} {\bibfnamefont {M.}~\bibnamefont {{Ashdown}}}, \bibinfo
  {author} {\bibfnamefont {J.}~\bibnamefont {{Aumont}}}, \bibinfo {author}
  {\bibfnamefont {C.}~\bibnamefont {{Baccigalupi}}}, \bibinfo {author}
  {\bibfnamefont {M.}~\bibnamefont {{Ballardini}}}, \bibinfo {author}
  {\bibfnamefont {A.~J.}\ \bibnamefont {{Banday}}}, \bibinfo {author}
  {\bibfnamefont {R.~B.}\ \bibnamefont {{Barreiro}}}, \bibinfo {author}
  {\bibfnamefont {N.}~\bibnamefont {{Bartolo}}}, \bibinfo {author}
  {\bibfnamefont {S.}~\bibnamefont {{Basak}}}, \bibinfo {author} {\bibfnamefont
  {R.}~\bibnamefont {{Battye}}}, \bibinfo {author} {\bibfnamefont
  {K.}~\bibnamefont {{Benabed}}}, \bibinfo {author} {\bibfnamefont {J.~P.}\
  \bibnamefont {{Bernard}}}, \bibinfo {author} {\bibfnamefont {M.}~\bibnamefont
  {{Bersanelli}}}, \bibinfo {author} {\bibfnamefont {P.}~\bibnamefont
  {{Bielewicz}}}, \bibinfo {author} {\bibfnamefont {J.~J.}\ \bibnamefont
  {{Bock}}}, \bibinfo {author} {\bibfnamefont {J.~R.}\ \bibnamefont {{Bond}}},
  \bibinfo {author} {\bibfnamefont {J.}~\bibnamefont {{Borrill}}}, \bibinfo
  {author} {\bibfnamefont {F.~R.}\ \bibnamefont {{Bouchet}}}, \bibinfo {author}
  {\bibfnamefont {F.}~\bibnamefont {{Boulanger}}}, \bibinfo {author}
  {\bibfnamefont {M.}~\bibnamefont {{Bucher}}}, \bibinfo {author}
  {\bibfnamefont {C.}~\bibnamefont {{Burigana}}}, \bibinfo {author}
  {\bibfnamefont {R.~C.}\ \bibnamefont {{Butler}}}, \bibinfo {author}
  {\bibfnamefont {E.}~\bibnamefont {{Calabrese}}}, \bibinfo {author}
  {\bibfnamefont {J.~F.}\ \bibnamefont {{Cardoso}}}, \bibinfo {author}
  {\bibfnamefont {J.}~\bibnamefont {{Carron}}}, \bibinfo {author}
  {\bibfnamefont {A.}~\bibnamefont {{Challinor}}}, \bibinfo {author}
  {\bibfnamefont {H.~C.}\ \bibnamefont {{Chiang}}}, \bibinfo {author}
  {\bibfnamefont {J.}~\bibnamefont {{Chluba}}}, \bibinfo {author}
  {\bibfnamefont {L.~P.~L.}\ \bibnamefont {{Colombo}}}, \bibinfo {author}
  {\bibfnamefont {C.}~\bibnamefont {{Combet}}}, \bibinfo {author}
  {\bibfnamefont {D.}~\bibnamefont {{Contreras}}}, \bibinfo {author}
  {\bibfnamefont {B.~P.}\ \bibnamefont {{Crill}}}, \bibinfo {author}
  {\bibfnamefont {F.}~\bibnamefont {{Cuttaia}}}, \bibinfo {author}
  {\bibfnamefont {P.}~\bibnamefont {{de Bernardis}}}, \bibinfo {author}
  {\bibfnamefont {G.}~\bibnamefont {{de Zotti}}}, \bibinfo {author}
  {\bibfnamefont {J.}~\bibnamefont {{Delabrouille}}}, \bibinfo {author}
  {\bibfnamefont {J.~M.}\ \bibnamefont {{Delouis}}}, \bibinfo {author}
  {\bibfnamefont {E.}~\bibnamefont {{Di Valentino}}}, \bibinfo {author}
  {\bibfnamefont {J.~M.}\ \bibnamefont {{Diego}}}, \bibinfo {author}
  {\bibfnamefont {O.}~\bibnamefont {{Dor{\'e}}}}, \bibinfo {author}
  {\bibfnamefont {M.}~\bibnamefont {{Douspis}}}, \bibinfo {author}
  {\bibfnamefont {A.}~\bibnamefont {{Ducout}}}, \bibinfo {author}
  {\bibfnamefont {X.}~\bibnamefont {{Dupac}}}, \bibinfo {author} {\bibfnamefont
  {S.}~\bibnamefont {{Dusini}}}, \bibinfo {author} {\bibfnamefont
  {G.}~\bibnamefont {{Efstathiou}}}, \bibinfo {author} {\bibfnamefont
  {F.}~\bibnamefont {{Elsner}}}, \bibinfo {author} {\bibfnamefont {T.~A.}\
  \bibnamefont {{En{\ss}lin}}}, \bibinfo {author} {\bibfnamefont {H.~K.}\
  \bibnamefont {{Eriksen}}}, \bibinfo {author} {\bibfnamefont {Y.}~\bibnamefont
  {{Fantaye}}}, \bibinfo {author} {\bibfnamefont {M.}~\bibnamefont
  {{Farhang}}}, \bibinfo {author} {\bibfnamefont {J.}~\bibnamefont
  {{Fergusson}}}, \bibinfo {author} {\bibfnamefont {R.}~\bibnamefont
  {{Fernandez-Cobos}}}, \bibinfo {author} {\bibfnamefont {F.}~\bibnamefont
  {{Finelli}}}, \bibinfo {author} {\bibfnamefont {F.}~\bibnamefont
  {{Forastieri}}}, \bibinfo {author} {\bibfnamefont {M.}~\bibnamefont
  {{Frailis}}}, \bibinfo {author} {\bibfnamefont {A.~A.}\ \bibnamefont
  {{Fraisse}}}, \bibinfo {author} {\bibfnamefont {E.}~\bibnamefont
  {{Franceschi}}}, \bibinfo {author} {\bibfnamefont {A.}~\bibnamefont
  {{Frolov}}}, \bibinfo {author} {\bibfnamefont {S.}~\bibnamefont
  {{Galeotta}}}, \bibinfo {author} {\bibfnamefont {S.}~\bibnamefont {{Galli}}},
  \bibinfo {author} {\bibfnamefont {K.}~\bibnamefont {{Ganga}}}, \bibinfo
  {author} {\bibfnamefont {R.~T.}\ \bibnamefont {{G{\'e}nova-Santos}}},
  \bibinfo {author} {\bibfnamefont {M.}~\bibnamefont {{Gerbino}}}, \bibinfo
  {author} {\bibfnamefont {T.}~\bibnamefont {{Ghosh}}}, \bibinfo {author}
  {\bibfnamefont {J.}~\bibnamefont {{Gonz{\'a}lez-Nuevo}}}, \bibinfo {author}
  {\bibfnamefont {K.~M.}\ \bibnamefont {{G{\'o}rski}}}, \bibinfo {author}
  {\bibfnamefont {S.}~\bibnamefont {{Gratton}}}, \bibinfo {author}
  {\bibfnamefont {A.}~\bibnamefont {{Gruppuso}}}, \bibinfo {author}
  {\bibfnamefont {J.~E.}\ \bibnamefont {{Gudmundsson}}}, \bibinfo {author}
  {\bibfnamefont {J.}~\bibnamefont {{Hamann}}}, \bibinfo {author}
  {\bibfnamefont {W.}~\bibnamefont {{Handley}}}, \bibinfo {author}
  {\bibfnamefont {F.~K.}\ \bibnamefont {{Hansen}}}, \bibinfo {author}
  {\bibfnamefont {D.}~\bibnamefont {{Herranz}}}, \bibinfo {author}
  {\bibfnamefont {S.~R.}\ \bibnamefont {{Hildebrandt}}}, \bibinfo {author}
  {\bibfnamefont {E.}~\bibnamefont {{Hivon}}}, \bibinfo {author} {\bibfnamefont
  {Z.}~\bibnamefont {{Huang}}}, \bibinfo {author} {\bibfnamefont {A.~H.}\
  \bibnamefont {{Jaffe}}}, \bibinfo {author} {\bibfnamefont {W.~C.}\
  \bibnamefont {{Jones}}}, \bibinfo {author} {\bibfnamefont {A.}~\bibnamefont
  {{Karakci}}}, \bibinfo {author} {\bibfnamefont {E.}~\bibnamefont
  {{Keih{\"a}nen}}}, \bibinfo {author} {\bibfnamefont {R.}~\bibnamefont
  {{Keskitalo}}}, \bibinfo {author} {\bibfnamefont {K.}~\bibnamefont
  {{Kiiveri}}}, \bibinfo {author} {\bibfnamefont {J.}~\bibnamefont {{Kim}}},
  \bibinfo {author} {\bibfnamefont {T.~S.}\ \bibnamefont {{Kisner}}}, \bibinfo
  {author} {\bibfnamefont {L.}~\bibnamefont {{Knox}}}, \bibinfo {author}
  {\bibfnamefont {N.}~\bibnamefont {{Krachmalnicoff}}}, \bibinfo {author}
  {\bibfnamefont {M.}~\bibnamefont {{Kunz}}}, \bibinfo {author} {\bibfnamefont
  {H.}~\bibnamefont {{Kurki-Suonio}}}, \bibinfo {author} {\bibfnamefont
  {G.}~\bibnamefont {{Lagache}}}, \bibinfo {author} {\bibfnamefont {J.~M.}\
  \bibnamefont {{Lamarre}}}, \bibinfo {author} {\bibfnamefont {A.}~\bibnamefont
  {{Lasenby}}}, \bibinfo {author} {\bibfnamefont {M.}~\bibnamefont
  {{Lattanzi}}}, \bibinfo {author} {\bibfnamefont {C.~R.}\ \bibnamefont
  {{Lawrence}}}, \bibinfo {author} {\bibfnamefont {M.}~\bibnamefont {{Le
  Jeune}}}, \bibinfo {author} {\bibfnamefont {P.}~\bibnamefont {{Lemos}}},
  \bibinfo {author} {\bibfnamefont {J.}~\bibnamefont {{Lesgourgues}}}, \bibinfo
  {author} {\bibfnamefont {F.}~\bibnamefont {{Levrier}}}, \bibinfo {author}
  {\bibfnamefont {A.}~\bibnamefont {{Lewis}}}, \bibinfo {author} {\bibfnamefont
  {M.}~\bibnamefont {{Liguori}}}, \bibinfo {author} {\bibfnamefont {P.~B.}\
  \bibnamefont {{Lilje}}}, \bibinfo {author} {\bibfnamefont {M.}~\bibnamefont
  {{Lilley}}}, \bibinfo {author} {\bibfnamefont {V.}~\bibnamefont
  {{Lindholm}}}, \bibinfo {author} {\bibfnamefont {M.}~\bibnamefont
  {{L{\'o}pez-Caniego}}}, \bibinfo {author} {\bibfnamefont {P.~M.}\
  \bibnamefont {{Lubin}}}, \bibinfo {author} {\bibfnamefont {Y.~Z.}\
  \bibnamefont {{Ma}}}, \bibinfo {author} {\bibfnamefont {J.~F.}\ \bibnamefont
  {{Mac{\'\i}as-P{\'e}rez}}}, \bibinfo {author} {\bibfnamefont
  {G.}~\bibnamefont {{Maggio}}}, \bibinfo {author} {\bibfnamefont
  {D.}~\bibnamefont {{Maino}}}, \bibinfo {author} {\bibfnamefont
  {N.}~\bibnamefont {{Mandolesi}}}, \bibinfo {author} {\bibfnamefont
  {A.}~\bibnamefont {{Mangilli}}}, \bibinfo {author} {\bibfnamefont
  {A.}~\bibnamefont {{Marcos-Caballero}}}, \bibinfo {author} {\bibfnamefont
  {M.}~\bibnamefont {{Maris}}}, \bibinfo {author} {\bibfnamefont {P.~G.}\
  \bibnamefont {{Martin}}}, \bibinfo {author} {\bibfnamefont {M.}~\bibnamefont
  {{Martinelli}}}, \bibinfo {author} {\bibfnamefont {E.}~\bibnamefont
  {{Mart{\'\i}nez-Gonz{\'a}lez}}}, \bibinfo {author} {\bibfnamefont
  {S.}~\bibnamefont {{Matarrese}}}, \bibinfo {author} {\bibfnamefont
  {N.}~\bibnamefont {{Mauri}}}, \bibinfo {author} {\bibfnamefont {J.~D.}\
  \bibnamefont {{McEwen}}}, \bibinfo {author} {\bibfnamefont {P.~R.}\
  \bibnamefont {{Meinhold}}}, \bibinfo {author} {\bibfnamefont
  {A.}~\bibnamefont {{Melchiorri}}}, \bibinfo {author} {\bibfnamefont
  {A.}~\bibnamefont {{Mennella}}}, \bibinfo {author} {\bibfnamefont
  {M.}~\bibnamefont {{Migliaccio}}}, \bibinfo {author} {\bibfnamefont
  {M.}~\bibnamefont {{Millea}}}, \bibinfo {author} {\bibfnamefont
  {S.}~\bibnamefont {{Mitra}}}, \bibinfo {author} {\bibfnamefont {M.~A.}\
  \bibnamefont {{Miville-Desch{\^e}nes}}}, \bibinfo {author} {\bibfnamefont
  {D.}~\bibnamefont {{Molinari}}}, \bibinfo {author} {\bibfnamefont
  {L.}~\bibnamefont {{Montier}}}, \bibinfo {author} {\bibfnamefont
  {G.}~\bibnamefont {{Morgante}}}, \bibinfo {author} {\bibfnamefont
  {A.}~\bibnamefont {{Moss}}}, \bibinfo {author} {\bibfnamefont
  {P.}~\bibnamefont {{Natoli}}}, \bibinfo {author} {\bibfnamefont {H.~U.}\
  \bibnamefont {{N{\o}rgaard-Nielsen}}}, \bibinfo {author} {\bibfnamefont
  {L.}~\bibnamefont {{Pagano}}}, \bibinfo {author} {\bibfnamefont
  {D.}~\bibnamefont {{Paoletti}}}, \bibinfo {author} {\bibfnamefont
  {B.}~\bibnamefont {{Partridge}}}, \bibinfo {author} {\bibfnamefont
  {G.}~\bibnamefont {{Patanchon}}}, \bibinfo {author} {\bibfnamefont {H.~V.}\
  \bibnamefont {{Peiris}}}, \bibinfo {author} {\bibfnamefont {F.}~\bibnamefont
  {{Perrotta}}}, \bibinfo {author} {\bibfnamefont {V.}~\bibnamefont
  {{Pettorino}}}, \bibinfo {author} {\bibfnamefont {F.}~\bibnamefont
  {{Piacentini}}}, \bibinfo {author} {\bibfnamefont {L.}~\bibnamefont
  {{Polastri}}}, \bibinfo {author} {\bibfnamefont {G.}~\bibnamefont
  {{Polenta}}}, \bibinfo {author} {\bibfnamefont {J.~L.}\ \bibnamefont
  {{Puget}}}, \bibinfo {author} {\bibfnamefont {J.~P.}\ \bibnamefont
  {{Rachen}}}, \bibinfo {author} {\bibfnamefont {M.}~\bibnamefont
  {{Reinecke}}}, \bibinfo {author} {\bibfnamefont {M.}~\bibnamefont
  {{Remazeilles}}}, \bibinfo {author} {\bibfnamefont {A.}~\bibnamefont
  {{Renzi}}}, \bibinfo {author} {\bibfnamefont {G.}~\bibnamefont {{Rocha}}},
  \bibinfo {author} {\bibfnamefont {C.}~\bibnamefont {{Rosset}}}, \bibinfo
  {author} {\bibfnamefont {G.}~\bibnamefont {{Roudier}}}, \bibinfo {author}
  {\bibfnamefont {J.~A.}\ \bibnamefont {{Rubi{\~n}o-Mart{\'\i}n}}}, \bibinfo
  {author} {\bibfnamefont {B.}~\bibnamefont {{Ruiz-Granados}}}, \bibinfo
  {author} {\bibfnamefont {L.}~\bibnamefont {{Salvati}}}, \bibinfo {author}
  {\bibfnamefont {M.}~\bibnamefont {{Sandri}}}, \bibinfo {author}
  {\bibfnamefont {M.}~\bibnamefont {{Savelainen}}}, \bibinfo {author}
  {\bibfnamefont {D.}~\bibnamefont {{Scott}}}, \bibinfo {author} {\bibfnamefont
  {E.~P.~S.}\ \bibnamefont {{Shellard}}}, \bibinfo {author} {\bibfnamefont
  {C.}~\bibnamefont {{Sirignano}}}, \bibinfo {author} {\bibfnamefont
  {G.}~\bibnamefont {{Sirri}}}, \bibinfo {author} {\bibfnamefont {L.~D.}\
  \bibnamefont {{Spencer}}}, \bibinfo {author} {\bibfnamefont {R.}~\bibnamefont
  {{Sunyaev}}}, \bibinfo {author} {\bibfnamefont {A.~S.}\ \bibnamefont
  {{Suur-Uski}}}, \bibinfo {author} {\bibfnamefont {J.~A.}\ \bibnamefont
  {{Tauber}}}, \bibinfo {author} {\bibfnamefont {D.}~\bibnamefont
  {{Tavagnacco}}}, \bibinfo {author} {\bibfnamefont {M.}~\bibnamefont
  {{Tenti}}}, \bibinfo {author} {\bibfnamefont {L.}~\bibnamefont
  {{Toffolatti}}}, \bibinfo {author} {\bibfnamefont {M.}~\bibnamefont
  {{Tomasi}}}, \bibinfo {author} {\bibfnamefont {T.}~\bibnamefont
  {{Trombetti}}}, \bibinfo {author} {\bibfnamefont {L.}~\bibnamefont
  {{Valenziano}}}, \bibinfo {author} {\bibfnamefont {J.}~\bibnamefont
  {{Valiviita}}}, \bibinfo {author} {\bibfnamefont {B.}~\bibnamefont {{Van
  Tent}}}, \bibinfo {author} {\bibfnamefont {L.}~\bibnamefont {{Vibert}}},
  \bibinfo {author} {\bibfnamefont {P.}~\bibnamefont {{Vielva}}}, \bibinfo
  {author} {\bibfnamefont {F.}~\bibnamefont {{Villa}}}, \bibinfo {author}
  {\bibfnamefont {N.}~\bibnamefont {{Vittorio}}}, \bibinfo {author}
  {\bibfnamefont {B.~D.}\ \bibnamefont {{Wandelt}}}, \bibinfo {author}
  {\bibfnamefont {I.~K.}\ \bibnamefont {{Wehus}}}, \bibinfo {author}
  {\bibfnamefont {M.}~\bibnamefont {{White}}}, \bibinfo {author} {\bibfnamefont
  {S.~D.~M.}\ \bibnamefont {{White}}}, \bibinfo {author} {\bibfnamefont
  {A.}~\bibnamefont {{Zacchei}}},\ and\ \bibinfo {author} {\bibfnamefont
  {A.}~\bibnamefont {{Zonca}}},\ }\bibfield  {title} {\bibinfo {title} {{Planck
  2018 results. VI. Cosmological parameters}},\ }\href
  {https://doi.org/10.1051/0004-6361/201833910} {\bibfield  {journal} {\bibinfo
   {journal} {\aap}\ }\textbf {\bibinfo {volume} {641}},\ \bibinfo {eid} {A6}
  (\bibinfo {year} {2020}{\natexlab{a}})},\ \Eprint
  {https://arxiv.org/abs/1807.06209} {arXiv:1807.06209 [astro-ph.CO]}
  \BibitemShut {NoStop}%
\bibitem [{\citenamefont {{McDonald}}\ \emph {et~al.}(2006)\citenamefont
  {{McDonald}}, \citenamefont {{Seljak}}, \citenamefont {{Burles}},
  \citenamefont {{Schlegel}}, \citenamefont {{Weinberg}}, \citenamefont
  {{Cen}}, \citenamefont {{Shih}}, \citenamefont {{Schaye}}, \citenamefont
  {{Schneider}}, \citenamefont {{Bahcall}}, \citenamefont {{Briggs}},
  \citenamefont {{Brinkmann}}, \citenamefont {{Brunner}}, \citenamefont
  {{Fukugita}}, \citenamefont {{Gunn}}, \citenamefont {{Ivezi{\'c}}},
  \citenamefont {{Kent}}, \citenamefont {{Lupton}},\ and\ \citenamefont
  {{Vanden Berk}}}]{2006ApJS..163...80M}%
  \BibitemOpen
  \bibfield  {author} {\bibinfo {author} {\bibfnamefont {P.}~\bibnamefont
  {{McDonald}}}, \bibinfo {author} {\bibfnamefont {U.}~\bibnamefont
  {{Seljak}}}, \bibinfo {author} {\bibfnamefont {S.}~\bibnamefont {{Burles}}},
  \bibinfo {author} {\bibfnamefont {D.~J.}\ \bibnamefont {{Schlegel}}},
  \bibinfo {author} {\bibfnamefont {D.~H.}\ \bibnamefont {{Weinberg}}},
  \bibinfo {author} {\bibfnamefont {R.}~\bibnamefont {{Cen}}}, \bibinfo
  {author} {\bibfnamefont {D.}~\bibnamefont {{Shih}}}, \bibinfo {author}
  {\bibfnamefont {J.}~\bibnamefont {{Schaye}}}, \bibinfo {author}
  {\bibfnamefont {D.~P.}\ \bibnamefont {{Schneider}}}, \bibinfo {author}
  {\bibfnamefont {N.~A.}\ \bibnamefont {{Bahcall}}}, \bibinfo {author}
  {\bibfnamefont {J.~W.}\ \bibnamefont {{Briggs}}}, \bibinfo {author}
  {\bibfnamefont {J.}~\bibnamefont {{Brinkmann}}}, \bibinfo {author}
  {\bibfnamefont {R.~J.}\ \bibnamefont {{Brunner}}}, \bibinfo {author}
  {\bibfnamefont {M.}~\bibnamefont {{Fukugita}}}, \bibinfo {author}
  {\bibfnamefont {J.~E.}\ \bibnamefont {{Gunn}}}, \bibinfo {author}
  {\bibfnamefont {{\v{Z}}.}~\bibnamefont {{Ivezi{\'c}}}}, \bibinfo {author}
  {\bibfnamefont {S.}~\bibnamefont {{Kent}}}, \bibinfo {author} {\bibfnamefont
  {R.~H.}\ \bibnamefont {{Lupton}}},\ and\ \bibinfo {author} {\bibfnamefont
  {D.~E.}\ \bibnamefont {{Vanden Berk}}},\ }\bibfield  {title} {\bibinfo
  {title} {{The Ly{\ensuremath{\alpha}} Forest Power Spectrum from the Sloan
  Digital Sky Survey}},\ }\href {https://doi.org/10.1086/444361} {\bibfield
  {journal} {\bibinfo  {journal} {\apjs}\ }\textbf {\bibinfo {volume} {163}},\
  \bibinfo {pages} {80} (\bibinfo {year} {2006})},\ \Eprint
  {https://arxiv.org/abs/astro-ph/0405013} {arXiv:astro-ph/0405013 [astro-ph]}
  \BibitemShut {NoStop}%
\bibitem [{\citenamefont {{Chabanier}}\ \emph {et~al.}(2019)\citenamefont
  {{Chabanier}}, \citenamefont {{Palanque-Delabrouille}}, \citenamefont
  {{Y{\`e}che}}, \citenamefont {{Le Goff}}, \citenamefont {{Armengaud}},
  \citenamefont {{Bautista}}, \citenamefont {{Blomqvist}}, \citenamefont
  {{Busca}}, \citenamefont {{Dawson}}, \citenamefont {{Etourneau}},
  \citenamefont {{Font-Ribera}}, \citenamefont {{Lee}}, \citenamefont {{du Mas
  des Bourboux}}, \citenamefont {{Pieri}}, \citenamefont {{Rich}},
  \citenamefont {{Rossi}}, \citenamefont {{Schneider}},\ and\ \citenamefont
  {{Slosar}}}]{2019JCAP...07..017C}%
  \BibitemOpen
  \bibfield  {author} {\bibinfo {author} {\bibfnamefont {S.}~\bibnamefont
  {{Chabanier}}}, \bibinfo {author} {\bibfnamefont {N.}~\bibnamefont
  {{Palanque-Delabrouille}}}, \bibinfo {author} {\bibfnamefont
  {C.}~\bibnamefont {{Y{\`e}che}}}, \bibinfo {author} {\bibfnamefont {J.-M.}\
  \bibnamefont {{Le Goff}}}, \bibinfo {author} {\bibfnamefont {E.}~\bibnamefont
  {{Armengaud}}}, \bibinfo {author} {\bibfnamefont {J.}~\bibnamefont
  {{Bautista}}}, \bibinfo {author} {\bibfnamefont {M.}~\bibnamefont
  {{Blomqvist}}}, \bibinfo {author} {\bibfnamefont {N.}~\bibnamefont
  {{Busca}}}, \bibinfo {author} {\bibfnamefont {K.}~\bibnamefont {{Dawson}}},
  \bibinfo {author} {\bibfnamefont {T.}~\bibnamefont {{Etourneau}}}, \bibinfo
  {author} {\bibfnamefont {A.}~\bibnamefont {{Font-Ribera}}}, \bibinfo {author}
  {\bibfnamefont {Y.}~\bibnamefont {{Lee}}}, \bibinfo {author} {\bibfnamefont
  {H.}~\bibnamefont {{du Mas des Bourboux}}}, \bibinfo {author} {\bibfnamefont
  {M.}~\bibnamefont {{Pieri}}}, \bibinfo {author} {\bibfnamefont
  {J.}~\bibnamefont {{Rich}}}, \bibinfo {author} {\bibfnamefont
  {G.}~\bibnamefont {{Rossi}}}, \bibinfo {author} {\bibfnamefont
  {D.}~\bibnamefont {{Schneider}}},\ and\ \bibinfo {author} {\bibfnamefont
  {A.}~\bibnamefont {{Slosar}}},\ }\bibfield  {title} {\bibinfo {title} {{The
  one-dimensional power spectrum from the SDSS DR14 Ly{\ensuremath{\alpha}}
  forests}},\ }\href {https://doi.org/10.1088/1475-7516/2019/07/017} {\bibfield
   {journal} {\bibinfo  {journal} {\jcap}\ }\textbf {\bibinfo {volume}
  {2019}},\ \bibinfo {eid} {017} (\bibinfo {year} {2019})},\ \Eprint
  {https://arxiv.org/abs/1812.03554} {arXiv:1812.03554 [astro-ph.CO]}
  \BibitemShut {NoStop}%
\bibitem [{\citenamefont {{Joachimi}}\ \emph {et~al.}(2021)\citenamefont
  {{Joachimi}}, \citenamefont {{Lin}}, \citenamefont {{Asgari}}, \citenamefont
  {{Tr{\"o}ster}}, \citenamefont {{Heymans}}, \citenamefont {{Hildebrandt}},
  \citenamefont {{K{\"o}hlinger}}, \citenamefont {{S{\'a}nchez}}, \citenamefont
  {{Wright}}, \citenamefont {{Bilicki}}, \citenamefont {{Blake}}, \citenamefont
  {{van den Busch}}, \citenamefont {{Crocce}}, \citenamefont {{Dvornik}},
  \citenamefont {{Erben}}, \citenamefont {{Getman}}, \citenamefont {{Giblin}},
  \citenamefont {{Hoekstra}}, \citenamefont {{Kannawadi}}, \citenamefont
  {{Kuijken}}, \citenamefont {{Napolitano}}, \citenamefont {{Schneider}},
  \citenamefont {{Scoccimarro}}, \citenamefont {{Sellentin}}, \citenamefont
  {{Shan}}, \citenamefont {{von Wietersheim-Kramsta}},\ and\ \citenamefont
  {{Zuntz}}}]{2021A&A...646A.129J}%
  \BibitemOpen
  \bibfield  {author} {\bibinfo {author} {\bibfnamefont {B.}~\bibnamefont
  {{Joachimi}}}, \bibinfo {author} {\bibfnamefont {C.~A.}\ \bibnamefont
  {{Lin}}}, \bibinfo {author} {\bibfnamefont {M.}~\bibnamefont {{Asgari}}},
  \bibinfo {author} {\bibfnamefont {T.}~\bibnamefont {{Tr{\"o}ster}}}, \bibinfo
  {author} {\bibfnamefont {C.}~\bibnamefont {{Heymans}}}, \bibinfo {author}
  {\bibfnamefont {H.}~\bibnamefont {{Hildebrandt}}}, \bibinfo {author}
  {\bibfnamefont {F.}~\bibnamefont {{K{\"o}hlinger}}}, \bibinfo {author}
  {\bibfnamefont {A.~G.}\ \bibnamefont {{S{\'a}nchez}}}, \bibinfo {author}
  {\bibfnamefont {A.~H.}\ \bibnamefont {{Wright}}}, \bibinfo {author}
  {\bibfnamefont {M.}~\bibnamefont {{Bilicki}}}, \bibinfo {author}
  {\bibfnamefont {C.}~\bibnamefont {{Blake}}}, \bibinfo {author} {\bibfnamefont
  {J.~L.}\ \bibnamefont {{van den Busch}}}, \bibinfo {author} {\bibfnamefont
  {M.}~\bibnamefont {{Crocce}}}, \bibinfo {author} {\bibfnamefont
  {A.}~\bibnamefont {{Dvornik}}}, \bibinfo {author} {\bibfnamefont
  {T.}~\bibnamefont {{Erben}}}, \bibinfo {author} {\bibfnamefont
  {F.}~\bibnamefont {{Getman}}}, \bibinfo {author} {\bibfnamefont
  {B.}~\bibnamefont {{Giblin}}}, \bibinfo {author} {\bibfnamefont
  {H.}~\bibnamefont {{Hoekstra}}}, \bibinfo {author} {\bibfnamefont
  {A.}~\bibnamefont {{Kannawadi}}}, \bibinfo {author} {\bibfnamefont
  {K.}~\bibnamefont {{Kuijken}}}, \bibinfo {author} {\bibfnamefont {N.~R.}\
  \bibnamefont {{Napolitano}}}, \bibinfo {author} {\bibfnamefont
  {P.}~\bibnamefont {{Schneider}}}, \bibinfo {author} {\bibfnamefont
  {R.}~\bibnamefont {{Scoccimarro}}}, \bibinfo {author} {\bibfnamefont
  {E.}~\bibnamefont {{Sellentin}}}, \bibinfo {author} {\bibfnamefont {H.~Y.}\
  \bibnamefont {{Shan}}}, \bibinfo {author} {\bibfnamefont {M.}~\bibnamefont
  {{von Wietersheim-Kramsta}}},\ and\ \bibinfo {author} {\bibfnamefont
  {J.}~\bibnamefont {{Zuntz}}},\ }\bibfield  {title} {\bibinfo {title}
  {{KiDS-1000 methodology: Modelling and inference for joint weak gravitational
  lensing and spectroscopic galaxy clustering analysis}},\ }\href
  {https://doi.org/10.1051/0004-6361/202038831} {\bibfield  {journal} {\bibinfo
   {journal} {\aap}\ }\textbf {\bibinfo {volume} {646}},\ \bibinfo {eid} {A129}
  (\bibinfo {year} {2021})},\ \Eprint {https://arxiv.org/abs/2007.01844}
  {arXiv:2007.01844 [astro-ph.CO]} \BibitemShut {NoStop}%
\bibitem [{\citenamefont {{Costanzi}}\ \emph {et~al.}(2021)\citenamefont
  {{Costanzi}}, \citenamefont {{Saro}}, \citenamefont {{Bocquet}},
  \citenamefont {{Abbott}}, \citenamefont {{Aguena}}, \citenamefont {{Allam}},
  \citenamefont {{Amara}}, \citenamefont {{Annis}}, \citenamefont {{Avila}},
  \citenamefont {{Bacon}}, \citenamefont {{Benson}}, \citenamefont
  {{Bhargava}}, \citenamefont {{Brooks}}, \citenamefont {{Buckley-Geer}},
  \citenamefont {{Burke}}, \citenamefont {{Carnero Rosell}}, \citenamefont
  {{Carrasco Kind}}, \citenamefont {{Carretero}}, \citenamefont {{Choi}},
  \citenamefont {{da Costa}}, \citenamefont {{Pereira}}, \citenamefont {{De
  Vicente}}, \citenamefont {{Desai}}, \citenamefont {{Diehl}}, \citenamefont
  {{Dietrich}}, \citenamefont {{Doel}}, \citenamefont {{Eifler}}, \citenamefont
  {{Everett}}, \citenamefont {{Ferrero}}, \citenamefont {{Fert{\'e}}},
  \citenamefont {{Flaugher}}, \citenamefont {{Fosalba}}, \citenamefont
  {{Frieman}}, \citenamefont {{Garc{\'\i}a-Bellido}}, \citenamefont
  {{Gaztanaga}}, \citenamefont {{Gerdes}}, \citenamefont {{Giannantonio}},
  \citenamefont {{Giles}}, \citenamefont {{Grandis}}, \citenamefont {{Gruen}},
  \citenamefont {{Gruendl}}, \citenamefont {{Gupta}}, \citenamefont
  {{Gutierrez}}, \citenamefont {{Hartley}}, \citenamefont {{Hinton}},
  \citenamefont {{Hollowood}}, \citenamefont {{Honscheid}}, \citenamefont
  {{James}}, \citenamefont {{Jeltema}}, \citenamefont {{Krause}}, \citenamefont
  {{Kuehn}}, \citenamefont {{Kuropatkin}}, \citenamefont {{Lahav}},
  \citenamefont {{Lima}}, \citenamefont {{MacCrann}}, \citenamefont {{Maia}},
  \citenamefont {{Marshall}}, \citenamefont {{Menanteau}}, \citenamefont
  {{Miquel}}, \citenamefont {{Mohr}}, \citenamefont {{Morgan}}, \citenamefont
  {{Myles}}, \citenamefont {{Ogando}}, \citenamefont {{Palmese}}, \citenamefont
  {{Paz-Chinch{\'o}n}}, \citenamefont {{Plazas}}, \citenamefont {{Rapetti}},
  \citenamefont {{Reichardt}}, \citenamefont {{Romer}}, \citenamefont
  {{Roodman}}, \citenamefont {{Ruppin}}, \citenamefont {{Salvati}},
  \citenamefont {{Samuroff}}, \citenamefont {{Sanchez}}, \citenamefont
  {{Scarpine}}, \citenamefont {{Serrano}}, \citenamefont {{Sevilla-Noarbe}},
  \citenamefont {{Singh}}, \citenamefont {{Smith}}, \citenamefont
  {{Soares-Santos}}, \citenamefont {{Stark}}, \citenamefont {{Suchyta}},
  \citenamefont {{Swanson}}, \citenamefont {{Tarle}}, \citenamefont {{Thomas}},
  \citenamefont {{To}}, \citenamefont {{Tucker}}, \citenamefont {{Varga}},
  \citenamefont {{Wechsler}}, \citenamefont {{Zhang}}, \citenamefont {{DES}},\
  and\ \citenamefont {{SPT Collaborations}}}]{2021PhRvD.103d3522C}%
  \BibitemOpen
  \bibfield  {author} {\bibinfo {author} {\bibfnamefont {M.}~\bibnamefont
  {{Costanzi}}}, \bibinfo {author} {\bibfnamefont {A.}~\bibnamefont {{Saro}}},
  \bibinfo {author} {\bibfnamefont {S.}~\bibnamefont {{Bocquet}}}, \bibinfo
  {author} {\bibfnamefont {T.~M.~C.}\ \bibnamefont {{Abbott}}}, \bibinfo
  {author} {\bibfnamefont {M.}~\bibnamefont {{Aguena}}}, \bibinfo {author}
  {\bibfnamefont {S.}~\bibnamefont {{Allam}}}, \bibinfo {author} {\bibfnamefont
  {A.}~\bibnamefont {{Amara}}}, \bibinfo {author} {\bibfnamefont
  {J.}~\bibnamefont {{Annis}}}, \bibinfo {author} {\bibfnamefont
  {S.}~\bibnamefont {{Avila}}}, \bibinfo {author} {\bibfnamefont
  {D.}~\bibnamefont {{Bacon}}}, \bibinfo {author} {\bibfnamefont {B.~A.}\
  \bibnamefont {{Benson}}}, \bibinfo {author} {\bibfnamefont {S.}~\bibnamefont
  {{Bhargava}}}, \bibinfo {author} {\bibfnamefont {D.}~\bibnamefont
  {{Brooks}}}, \bibinfo {author} {\bibfnamefont {E.}~\bibnamefont
  {{Buckley-Geer}}}, \bibinfo {author} {\bibfnamefont {D.~L.}\ \bibnamefont
  {{Burke}}}, \bibinfo {author} {\bibfnamefont {A.}~\bibnamefont {{Carnero
  Rosell}}}, \bibinfo {author} {\bibfnamefont {M.}~\bibnamefont {{Carrasco
  Kind}}}, \bibinfo {author} {\bibfnamefont {J.}~\bibnamefont {{Carretero}}},
  \bibinfo {author} {\bibfnamefont {A.}~\bibnamefont {{Choi}}}, \bibinfo
  {author} {\bibfnamefont {L.~N.}\ \bibnamefont {{da Costa}}}, \bibinfo
  {author} {\bibfnamefont {M.~E.~S.}\ \bibnamefont {{Pereira}}}, \bibinfo
  {author} {\bibfnamefont {J.}~\bibnamefont {{De Vicente}}}, \bibinfo {author}
  {\bibfnamefont {S.}~\bibnamefont {{Desai}}}, \bibinfo {author} {\bibfnamefont
  {H.~T.}\ \bibnamefont {{Diehl}}}, \bibinfo {author} {\bibfnamefont {J.~P.}\
  \bibnamefont {{Dietrich}}}, \bibinfo {author} {\bibfnamefont
  {P.}~\bibnamefont {{Doel}}}, \bibinfo {author} {\bibfnamefont {T.~F.}\
  \bibnamefont {{Eifler}}}, \bibinfo {author} {\bibfnamefont {S.}~\bibnamefont
  {{Everett}}}, \bibinfo {author} {\bibfnamefont {I.}~\bibnamefont
  {{Ferrero}}}, \bibinfo {author} {\bibfnamefont {A.}~\bibnamefont
  {{Fert{\'e}}}}, \bibinfo {author} {\bibfnamefont {B.}~\bibnamefont
  {{Flaugher}}}, \bibinfo {author} {\bibfnamefont {P.}~\bibnamefont
  {{Fosalba}}}, \bibinfo {author} {\bibfnamefont {J.}~\bibnamefont
  {{Frieman}}}, \bibinfo {author} {\bibfnamefont {J.}~\bibnamefont
  {{Garc{\'\i}a-Bellido}}}, \bibinfo {author} {\bibfnamefont {E.}~\bibnamefont
  {{Gaztanaga}}}, \bibinfo {author} {\bibfnamefont {D.~W.}\ \bibnamefont
  {{Gerdes}}}, \bibinfo {author} {\bibfnamefont {T.}~\bibnamefont
  {{Giannantonio}}}, \bibinfo {author} {\bibfnamefont {P.}~\bibnamefont
  {{Giles}}}, \bibinfo {author} {\bibfnamefont {S.}~\bibnamefont {{Grandis}}},
  \bibinfo {author} {\bibfnamefont {D.}~\bibnamefont {{Gruen}}}, \bibinfo
  {author} {\bibfnamefont {R.~A.}\ \bibnamefont {{Gruendl}}}, \bibinfo {author}
  {\bibfnamefont {N.}~\bibnamefont {{Gupta}}}, \bibinfo {author} {\bibfnamefont
  {G.}~\bibnamefont {{Gutierrez}}}, \bibinfo {author} {\bibfnamefont {W.~G.}\
  \bibnamefont {{Hartley}}}, \bibinfo {author} {\bibfnamefont {S.~R.}\
  \bibnamefont {{Hinton}}}, \bibinfo {author} {\bibfnamefont {D.~L.}\
  \bibnamefont {{Hollowood}}}, \bibinfo {author} {\bibfnamefont
  {K.}~\bibnamefont {{Honscheid}}}, \bibinfo {author} {\bibfnamefont {D.~J.}\
  \bibnamefont {{James}}}, \bibinfo {author} {\bibfnamefont {T.}~\bibnamefont
  {{Jeltema}}}, \bibinfo {author} {\bibfnamefont {E.}~\bibnamefont {{Krause}}},
  \bibinfo {author} {\bibfnamefont {K.}~\bibnamefont {{Kuehn}}}, \bibinfo
  {author} {\bibfnamefont {N.}~\bibnamefont {{Kuropatkin}}}, \bibinfo {author}
  {\bibfnamefont {O.}~\bibnamefont {{Lahav}}}, \bibinfo {author} {\bibfnamefont
  {M.}~\bibnamefont {{Lima}}}, \bibinfo {author} {\bibfnamefont
  {N.}~\bibnamefont {{MacCrann}}}, \bibinfo {author} {\bibfnamefont {M.~A.~G.}\
  \bibnamefont {{Maia}}}, \bibinfo {author} {\bibfnamefont {J.~L.}\
  \bibnamefont {{Marshall}}}, \bibinfo {author} {\bibfnamefont
  {F.}~\bibnamefont {{Menanteau}}}, \bibinfo {author} {\bibfnamefont
  {R.}~\bibnamefont {{Miquel}}}, \bibinfo {author} {\bibfnamefont {J.~J.}\
  \bibnamefont {{Mohr}}}, \bibinfo {author} {\bibfnamefont {R.}~\bibnamefont
  {{Morgan}}}, \bibinfo {author} {\bibfnamefont {J.}~\bibnamefont {{Myles}}},
  \bibinfo {author} {\bibfnamefont {R.~L.~C.}\ \bibnamefont {{Ogando}}},
  \bibinfo {author} {\bibfnamefont {A.}~\bibnamefont {{Palmese}}}, \bibinfo
  {author} {\bibfnamefont {F.}~\bibnamefont {{Paz-Chinch{\'o}n}}}, \bibinfo
  {author} {\bibfnamefont {A.~A.}\ \bibnamefont {{Plazas}}}, \bibinfo {author}
  {\bibfnamefont {D.}~\bibnamefont {{Rapetti}}}, \bibinfo {author}
  {\bibfnamefont {C.~L.}\ \bibnamefont {{Reichardt}}}, \bibinfo {author}
  {\bibfnamefont {A.~K.}\ \bibnamefont {{Romer}}}, \bibinfo {author}
  {\bibfnamefont {A.}~\bibnamefont {{Roodman}}}, \bibinfo {author}
  {\bibfnamefont {F.}~\bibnamefont {{Ruppin}}}, \bibinfo {author}
  {\bibfnamefont {L.}~\bibnamefont {{Salvati}}}, \bibinfo {author}
  {\bibfnamefont {S.}~\bibnamefont {{Samuroff}}}, \bibinfo {author}
  {\bibfnamefont {E.}~\bibnamefont {{Sanchez}}}, \bibinfo {author}
  {\bibfnamefont {V.}~\bibnamefont {{Scarpine}}}, \bibinfo {author}
  {\bibfnamefont {S.}~\bibnamefont {{Serrano}}}, \bibinfo {author}
  {\bibfnamefont {I.}~\bibnamefont {{Sevilla-Noarbe}}}, \bibinfo {author}
  {\bibfnamefont {P.}~\bibnamefont {{Singh}}}, \bibinfo {author} {\bibfnamefont
  {M.}~\bibnamefont {{Smith}}}, \bibinfo {author} {\bibfnamefont
  {M.}~\bibnamefont {{Soares-Santos}}}, \bibinfo {author} {\bibfnamefont
  {A.~A.}\ \bibnamefont {{Stark}}}, \bibinfo {author} {\bibfnamefont
  {E.}~\bibnamefont {{Suchyta}}}, \bibinfo {author} {\bibfnamefont {M.~E.~C.}\
  \bibnamefont {{Swanson}}}, \bibinfo {author} {\bibfnamefont {G.}~\bibnamefont
  {{Tarle}}}, \bibinfo {author} {\bibfnamefont {D.}~\bibnamefont {{Thomas}}},
  \bibinfo {author} {\bibfnamefont {C.}~\bibnamefont {{To}}}, \bibinfo {author}
  {\bibfnamefont {D.~L.}\ \bibnamefont {{Tucker}}}, \bibinfo {author}
  {\bibfnamefont {T.~N.}\ \bibnamefont {{Varga}}}, \bibinfo {author}
  {\bibfnamefont {R.~H.}\ \bibnamefont {{Wechsler}}}, \bibinfo {author}
  {\bibfnamefont {Z.}~\bibnamefont {{Zhang}}}, \bibinfo {author} {\bibnamefont
  {{DES}}},\ and\ \bibinfo {author} {\bibnamefont {{SPT Collaborations}}},\
  }\bibfield  {title} {\bibinfo {title} {{Cosmological constraints from DES Y1
  cluster abundances and SPT multiwavelength data}},\ }\href
  {https://doi.org/10.1103/PhysRevD.103.043522} {\bibfield  {journal} {\bibinfo
   {journal} {\prd}\ }\textbf {\bibinfo {volume} {103}},\ \bibinfo {eid}
  {043522} (\bibinfo {year} {2021})},\ \Eprint
  {https://arxiv.org/abs/2010.13800} {arXiv:2010.13800 [astro-ph.CO]}
  \BibitemShut {NoStop}%
\bibitem [{\citenamefont {{Hilton}}\ \emph {et~al.}(2021)\citenamefont
  {{Hilton}}, \citenamefont {{Sif{\'o}n}}, \citenamefont {{Naess}},
  \citenamefont {{Madhavacheril}}, \citenamefont {{Oguri}}, \citenamefont
  {{Rozo}}, \citenamefont {{Rykoff}}, \citenamefont {{Abbott}}, \citenamefont
  {{Adhikari}}, \citenamefont {{Aguena}}, \citenamefont {{Aiola}},
  \citenamefont {{Allam}}, \citenamefont {{Amodeo}}, \citenamefont {{Amon}},
  \citenamefont {{Annis}}, \citenamefont {{Ansarinejad}}, \citenamefont
  {{Aros-Bunster}}, \citenamefont {{Austermann}}, \citenamefont {{Avila}},
  \citenamefont {{Bacon}}, \citenamefont {{Battaglia}}, \citenamefont
  {{Beall}}, \citenamefont {{Becker}}, \citenamefont {{Bernstein}},
  \citenamefont {{Bertin}}, \citenamefont {{Bhandarkar}}, \citenamefont
  {{Bhargava}}, \citenamefont {{Bond}}, \citenamefont {{Brooks}}, \citenamefont
  {{Burke}}, \citenamefont {{Calabrese}}, \citenamefont {{Carrasco Kind}},
  \citenamefont {{Carretero}}, \citenamefont {{Choi}}, \citenamefont {{Choi}},
  \citenamefont {{Conselice}}, \citenamefont {{da Costa}}, \citenamefont
  {{Costanzi}}, \citenamefont {{Crichton}}, \citenamefont {{Crowley}},
  \citenamefont {{D{\"u}nner}}, \citenamefont {{Denison}}, \citenamefont
  {{Devlin}}, \citenamefont {{Dicker}}, \citenamefont {{Diehl}}, \citenamefont
  {{Dietrich}}, \citenamefont {{Doel}}, \citenamefont {{Duff}}, \citenamefont
  {{Duivenvoorden}}, \citenamefont {{Dunkley}}, \citenamefont {{Everett}},
  \citenamefont {{Ferraro}}, \citenamefont {{Ferrero}}, \citenamefont
  {{Fert{\'e}}}, \citenamefont {{Flaugher}}, \citenamefont {{Frieman}},
  \citenamefont {{Gallardo}}, \citenamefont {{Garc{\'\i}a-Bellido}},
  \citenamefont {{Gaztanaga}}, \citenamefont {{Gerdes}}, \citenamefont
  {{Giles}}, \citenamefont {{Golec}}, \citenamefont {{Gralla}}, \citenamefont
  {{Grandis}}, \citenamefont {{Gruen}}, \citenamefont {{Gruendl}},
  \citenamefont {{Gschwend}}, \citenamefont {{Gutierrez}}, \citenamefont
  {{Han}}, \citenamefont {{Hartley}}, \citenamefont {{Hasselfield}},
  \citenamefont {{Hill}}, \citenamefont {{Hilton}}, \citenamefont {{Hincks}},
  \citenamefont {{Hinton}}, \citenamefont {{Ho}}, \citenamefont {{Honscheid}},
  \citenamefont {{Hoyle}}, \citenamefont {{Hubmayr}}, \citenamefont
  {{Huffenberger}}, \citenamefont {{Hughes}}, \citenamefont {{Jaelani}},
  \citenamefont {{Jain}}, \citenamefont {{James}}, \citenamefont {{Jeltema}},
  \citenamefont {{Kent}}, \citenamefont {{Knowles}}, \citenamefont {{Koopman}},
  \citenamefont {{Kuehn}}, \citenamefont {{Lahav}}, \citenamefont {{Lima}},
  \citenamefont {{Lin}}, \citenamefont {{Lokken}}, \citenamefont {{Loubser}},
  \citenamefont {{MacCrann}}, \citenamefont {{Maia}}, \citenamefont
  {{Marriage}}, \citenamefont {{Martin}}, \citenamefont {{McMahon}},
  \citenamefont {{Melchior}}, \citenamefont {{Menanteau}}, \citenamefont
  {{Miquel}}, \citenamefont {{Miyatake}}, \citenamefont {{Moodley}},
  \citenamefont {{Morgan}}, \citenamefont {{Mroczkowski}}, \citenamefont
  {{Nati}}, \citenamefont {{Newburgh}}, \citenamefont {{Niemack}},
  \citenamefont {{Nishizawa}}, \citenamefont {{Ogando}}, \citenamefont
  {{Orlowski-Scherer}}, \citenamefont {{Page}}, \citenamefont {{Palmese}},
  \citenamefont {{Partridge}}, \citenamefont {{Paz-Chinch{\'o}n}},
  \citenamefont {{Phakathi}}, \citenamefont {{Plazas}}, \citenamefont
  {{Robertson}}, \citenamefont {{Romer}}, \citenamefont {{Carnero Rosell}},
  \citenamefont {{Salatino}}, \citenamefont {{Sanchez}}, \citenamefont
  {{Schaan}}, \citenamefont {{Schillaci}}, \citenamefont {{Sehgal}},
  \citenamefont {{Serrano}}, \citenamefont {{Shin}}, \citenamefont {{Simon}},
  \citenamefont {{Smith}}, \citenamefont {{Soares-Santos}}, \citenamefont
  {{Spergel}}, \citenamefont {{Staggs}}, \citenamefont {{Storer}},
  \citenamefont {{Suchyta}}, \citenamefont {{Swanson}}, \citenamefont
  {{Tarle}}, \citenamefont {{Thomas}}, \citenamefont {{To}}, \citenamefont
  {{Trac}}, \citenamefont {{Ullom}}, \citenamefont {{Vale}}, \citenamefont
  {{Van Lanen}}, \citenamefont {{Vavagiakis}}, \citenamefont {{De Vicente}},
  \citenamefont {{Wilkinson}}, \citenamefont {{Wollack}}, \citenamefont
  {{Xu}},\ and\ \citenamefont {{Zhang}}}]{2021ApJS..253....3H}%
  \BibitemOpen
  \bibfield  {author} {\bibinfo {author} {\bibfnamefont {M.}~\bibnamefont
  {{Hilton}}}, \bibinfo {author} {\bibfnamefont {C.}~\bibnamefont
  {{Sif{\'o}n}}}, \bibinfo {author} {\bibfnamefont {S.}~\bibnamefont
  {{Naess}}}, \bibinfo {author} {\bibfnamefont {M.}~\bibnamefont
  {{Madhavacheril}}}, \bibinfo {author} {\bibfnamefont {M.}~\bibnamefont
  {{Oguri}}}, \bibinfo {author} {\bibfnamefont {E.}~\bibnamefont {{Rozo}}},
  \bibinfo {author} {\bibfnamefont {E.}~\bibnamefont {{Rykoff}}}, \bibinfo
  {author} {\bibfnamefont {T.~M.~C.}\ \bibnamefont {{Abbott}}}, \bibinfo
  {author} {\bibfnamefont {S.}~\bibnamefont {{Adhikari}}}, \bibinfo {author}
  {\bibfnamefont {M.}~\bibnamefont {{Aguena}}}, \bibinfo {author}
  {\bibfnamefont {S.}~\bibnamefont {{Aiola}}}, \bibinfo {author} {\bibfnamefont
  {S.}~\bibnamefont {{Allam}}}, \bibinfo {author} {\bibfnamefont
  {S.}~\bibnamefont {{Amodeo}}}, \bibinfo {author} {\bibfnamefont
  {A.}~\bibnamefont {{Amon}}}, \bibinfo {author} {\bibfnamefont
  {J.}~\bibnamefont {{Annis}}}, \bibinfo {author} {\bibfnamefont
  {B.}~\bibnamefont {{Ansarinejad}}}, \bibinfo {author} {\bibfnamefont
  {C.}~\bibnamefont {{Aros-Bunster}}}, \bibinfo {author} {\bibfnamefont
  {J.~E.}\ \bibnamefont {{Austermann}}}, \bibinfo {author} {\bibfnamefont
  {S.}~\bibnamefont {{Avila}}}, \bibinfo {author} {\bibfnamefont
  {D.}~\bibnamefont {{Bacon}}}, \bibinfo {author} {\bibfnamefont
  {N.}~\bibnamefont {{Battaglia}}}, \bibinfo {author} {\bibfnamefont {J.~A.}\
  \bibnamefont {{Beall}}}, \bibinfo {author} {\bibfnamefont {D.~T.}\
  \bibnamefont {{Becker}}}, \bibinfo {author} {\bibfnamefont {G.~M.}\
  \bibnamefont {{Bernstein}}}, \bibinfo {author} {\bibfnamefont
  {E.}~\bibnamefont {{Bertin}}}, \bibinfo {author} {\bibfnamefont
  {T.}~\bibnamefont {{Bhandarkar}}}, \bibinfo {author} {\bibfnamefont
  {S.}~\bibnamefont {{Bhargava}}}, \bibinfo {author} {\bibfnamefont {J.~R.}\
  \bibnamefont {{Bond}}}, \bibinfo {author} {\bibfnamefont {D.}~\bibnamefont
  {{Brooks}}}, \bibinfo {author} {\bibfnamefont {D.~L.}\ \bibnamefont
  {{Burke}}}, \bibinfo {author} {\bibfnamefont {E.}~\bibnamefont
  {{Calabrese}}}, \bibinfo {author} {\bibfnamefont {M.}~\bibnamefont {{Carrasco
  Kind}}}, \bibinfo {author} {\bibfnamefont {J.}~\bibnamefont {{Carretero}}},
  \bibinfo {author} {\bibfnamefont {S.~K.}\ \bibnamefont {{Choi}}}, \bibinfo
  {author} {\bibfnamefont {A.}~\bibnamefont {{Choi}}}, \bibinfo {author}
  {\bibfnamefont {C.}~\bibnamefont {{Conselice}}}, \bibinfo {author}
  {\bibfnamefont {L.~N.}\ \bibnamefont {{da Costa}}}, \bibinfo {author}
  {\bibfnamefont {M.}~\bibnamefont {{Costanzi}}}, \bibinfo {author}
  {\bibfnamefont {D.}~\bibnamefont {{Crichton}}}, \bibinfo {author}
  {\bibfnamefont {K.~T.}\ \bibnamefont {{Crowley}}}, \bibinfo {author}
  {\bibfnamefont {R.}~\bibnamefont {{D{\"u}nner}}}, \bibinfo {author}
  {\bibfnamefont {E.~V.}\ \bibnamefont {{Denison}}}, \bibinfo {author}
  {\bibfnamefont {M.~J.}\ \bibnamefont {{Devlin}}}, \bibinfo {author}
  {\bibfnamefont {S.~R.}\ \bibnamefont {{Dicker}}}, \bibinfo {author}
  {\bibfnamefont {H.~T.}\ \bibnamefont {{Diehl}}}, \bibinfo {author}
  {\bibfnamefont {J.~P.}\ \bibnamefont {{Dietrich}}}, \bibinfo {author}
  {\bibfnamefont {P.}~\bibnamefont {{Doel}}}, \bibinfo {author} {\bibfnamefont
  {S.~M.}\ \bibnamefont {{Duff}}}, \bibinfo {author} {\bibfnamefont {A.~J.}\
  \bibnamefont {{Duivenvoorden}}}, \bibinfo {author} {\bibfnamefont
  {J.}~\bibnamefont {{Dunkley}}}, \bibinfo {author} {\bibfnamefont
  {S.}~\bibnamefont {{Everett}}}, \bibinfo {author} {\bibfnamefont
  {S.}~\bibnamefont {{Ferraro}}}, \bibinfo {author} {\bibfnamefont
  {I.}~\bibnamefont {{Ferrero}}}, \bibinfo {author} {\bibfnamefont
  {A.}~\bibnamefont {{Fert{\'e}}}}, \bibinfo {author} {\bibfnamefont
  {B.}~\bibnamefont {{Flaugher}}}, \bibinfo {author} {\bibfnamefont
  {J.}~\bibnamefont {{Frieman}}}, \bibinfo {author} {\bibfnamefont {P.~A.}\
  \bibnamefont {{Gallardo}}}, \bibinfo {author} {\bibfnamefont
  {J.}~\bibnamefont {{Garc{\'\i}a-Bellido}}}, \bibinfo {author} {\bibfnamefont
  {E.}~\bibnamefont {{Gaztanaga}}}, \bibinfo {author} {\bibfnamefont {D.~W.}\
  \bibnamefont {{Gerdes}}}, \bibinfo {author} {\bibfnamefont {P.}~\bibnamefont
  {{Giles}}}, \bibinfo {author} {\bibfnamefont {J.~E.}\ \bibnamefont
  {{Golec}}}, \bibinfo {author} {\bibfnamefont {M.~B.}\ \bibnamefont
  {{Gralla}}}, \bibinfo {author} {\bibfnamefont {S.}~\bibnamefont {{Grandis}}},
  \bibinfo {author} {\bibfnamefont {D.}~\bibnamefont {{Gruen}}}, \bibinfo
  {author} {\bibfnamefont {R.~A.}\ \bibnamefont {{Gruendl}}}, \bibinfo {author}
  {\bibfnamefont {J.}~\bibnamefont {{Gschwend}}}, \bibinfo {author}
  {\bibfnamefont {G.}~\bibnamefont {{Gutierrez}}}, \bibinfo {author}
  {\bibfnamefont {D.}~\bibnamefont {{Han}}}, \bibinfo {author} {\bibfnamefont
  {W.~G.}\ \bibnamefont {{Hartley}}}, \bibinfo {author} {\bibfnamefont
  {M.}~\bibnamefont {{Hasselfield}}}, \bibinfo {author} {\bibfnamefont {J.~C.}\
  \bibnamefont {{Hill}}}, \bibinfo {author} {\bibfnamefont {G.~C.}\
  \bibnamefont {{Hilton}}}, \bibinfo {author} {\bibfnamefont {A.~D.}\
  \bibnamefont {{Hincks}}}, \bibinfo {author} {\bibfnamefont {S.~R.}\
  \bibnamefont {{Hinton}}}, \bibinfo {author} {\bibfnamefont {S.~P.~P.}\
  \bibnamefont {{Ho}}}, \bibinfo {author} {\bibfnamefont {K.}~\bibnamefont
  {{Honscheid}}}, \bibinfo {author} {\bibfnamefont {B.}~\bibnamefont
  {{Hoyle}}}, \bibinfo {author} {\bibfnamefont {J.}~\bibnamefont {{Hubmayr}}},
  \bibinfo {author} {\bibfnamefont {K.~M.}\ \bibnamefont {{Huffenberger}}},
  \bibinfo {author} {\bibfnamefont {J.~P.}\ \bibnamefont {{Hughes}}}, \bibinfo
  {author} {\bibfnamefont {A.~T.}\ \bibnamefont {{Jaelani}}}, \bibinfo {author}
  {\bibfnamefont {B.}~\bibnamefont {{Jain}}}, \bibinfo {author} {\bibfnamefont
  {D.~J.}\ \bibnamefont {{James}}}, \bibinfo {author} {\bibfnamefont
  {T.}~\bibnamefont {{Jeltema}}}, \bibinfo {author} {\bibfnamefont
  {S.}~\bibnamefont {{Kent}}}, \bibinfo {author} {\bibfnamefont
  {K.}~\bibnamefont {{Knowles}}}, \bibinfo {author} {\bibfnamefont {B.~J.}\
  \bibnamefont {{Koopman}}}, \bibinfo {author} {\bibfnamefont {K.}~\bibnamefont
  {{Kuehn}}}, \bibinfo {author} {\bibfnamefont {O.}~\bibnamefont {{Lahav}}},
  \bibinfo {author} {\bibfnamefont {M.}~\bibnamefont {{Lima}}}, \bibinfo
  {author} {\bibfnamefont {Y.~T.}\ \bibnamefont {{Lin}}}, \bibinfo {author}
  {\bibfnamefont {M.}~\bibnamefont {{Lokken}}}, \bibinfo {author}
  {\bibfnamefont {S.~I.}\ \bibnamefont {{Loubser}}}, \bibinfo {author}
  {\bibfnamefont {N.}~\bibnamefont {{MacCrann}}}, \bibinfo {author}
  {\bibfnamefont {M.~A.~G.}\ \bibnamefont {{Maia}}}, \bibinfo {author}
  {\bibfnamefont {T.~A.}\ \bibnamefont {{Marriage}}}, \bibinfo {author}
  {\bibfnamefont {J.}~\bibnamefont {{Martin}}}, \bibinfo {author}
  {\bibfnamefont {J.}~\bibnamefont {{McMahon}}}, \bibinfo {author}
  {\bibfnamefont {P.}~\bibnamefont {{Melchior}}}, \bibinfo {author}
  {\bibfnamefont {F.}~\bibnamefont {{Menanteau}}}, \bibinfo {author}
  {\bibfnamefont {R.}~\bibnamefont {{Miquel}}}, \bibinfo {author}
  {\bibfnamefont {H.}~\bibnamefont {{Miyatake}}}, \bibinfo {author}
  {\bibfnamefont {K.}~\bibnamefont {{Moodley}}}, \bibinfo {author}
  {\bibfnamefont {R.}~\bibnamefont {{Morgan}}}, \bibinfo {author}
  {\bibfnamefont {T.}~\bibnamefont {{Mroczkowski}}}, \bibinfo {author}
  {\bibfnamefont {F.}~\bibnamefont {{Nati}}}, \bibinfo {author} {\bibfnamefont
  {L.~B.}\ \bibnamefont {{Newburgh}}}, \bibinfo {author} {\bibfnamefont
  {M.~D.}\ \bibnamefont {{Niemack}}}, \bibinfo {author} {\bibfnamefont {A.~J.}\
  \bibnamefont {{Nishizawa}}}, \bibinfo {author} {\bibfnamefont {R.~L.~C.}\
  \bibnamefont {{Ogando}}}, \bibinfo {author} {\bibfnamefont {J.}~\bibnamefont
  {{Orlowski-Scherer}}}, \bibinfo {author} {\bibfnamefont {L.~A.}\ \bibnamefont
  {{Page}}}, \bibinfo {author} {\bibfnamefont {A.}~\bibnamefont {{Palmese}}},
  \bibinfo {author} {\bibfnamefont {B.}~\bibnamefont {{Partridge}}}, \bibinfo
  {author} {\bibfnamefont {F.}~\bibnamefont {{Paz-Chinch{\'o}n}}}, \bibinfo
  {author} {\bibfnamefont {P.}~\bibnamefont {{Phakathi}}}, \bibinfo {author}
  {\bibfnamefont {A.~A.}\ \bibnamefont {{Plazas}}}, \bibinfo {author}
  {\bibfnamefont {N.~C.}\ \bibnamefont {{Robertson}}}, \bibinfo {author}
  {\bibfnamefont {A.~K.}\ \bibnamefont {{Romer}}}, \bibinfo {author}
  {\bibfnamefont {A.}~\bibnamefont {{Carnero Rosell}}}, \bibinfo {author}
  {\bibfnamefont {M.}~\bibnamefont {{Salatino}}}, \bibinfo {author}
  {\bibfnamefont {E.}~\bibnamefont {{Sanchez}}}, \bibinfo {author}
  {\bibfnamefont {E.}~\bibnamefont {{Schaan}}}, \bibinfo {author}
  {\bibfnamefont {A.}~\bibnamefont {{Schillaci}}}, \bibinfo {author}
  {\bibfnamefont {N.}~\bibnamefont {{Sehgal}}}, \bibinfo {author}
  {\bibfnamefont {S.}~\bibnamefont {{Serrano}}}, \bibinfo {author}
  {\bibfnamefont {T.}~\bibnamefont {{Shin}}}, \bibinfo {author} {\bibfnamefont
  {S.~M.}\ \bibnamefont {{Simon}}}, \bibinfo {author} {\bibfnamefont
  {M.}~\bibnamefont {{Smith}}}, \bibinfo {author} {\bibfnamefont
  {M.}~\bibnamefont {{Soares-Santos}}}, \bibinfo {author} {\bibfnamefont
  {D.~N.}\ \bibnamefont {{Spergel}}}, \bibinfo {author} {\bibfnamefont {S.~T.}\
  \bibnamefont {{Staggs}}}, \bibinfo {author} {\bibfnamefont {E.~R.}\
  \bibnamefont {{Storer}}}, \bibinfo {author} {\bibfnamefont {E.}~\bibnamefont
  {{Suchyta}}}, \bibinfo {author} {\bibfnamefont {M.~E.~C.}\ \bibnamefont
  {{Swanson}}}, \bibinfo {author} {\bibfnamefont {G.}~\bibnamefont {{Tarle}}},
  \bibinfo {author} {\bibfnamefont {D.}~\bibnamefont {{Thomas}}}, \bibinfo
  {author} {\bibfnamefont {C.}~\bibnamefont {{To}}}, \bibinfo {author}
  {\bibfnamefont {H.}~\bibnamefont {{Trac}}}, \bibinfo {author} {\bibfnamefont
  {J.~N.}\ \bibnamefont {{Ullom}}}, \bibinfo {author} {\bibfnamefont {L.~R.}\
  \bibnamefont {{Vale}}}, \bibinfo {author} {\bibfnamefont {J.}~\bibnamefont
  {{Van Lanen}}}, \bibinfo {author} {\bibfnamefont {E.~M.}\ \bibnamefont
  {{Vavagiakis}}}, \bibinfo {author} {\bibfnamefont {J.}~\bibnamefont {{De
  Vicente}}}, \bibinfo {author} {\bibfnamefont {R.~D.}\ \bibnamefont
  {{Wilkinson}}}, \bibinfo {author} {\bibfnamefont {E.~J.}\ \bibnamefont
  {{Wollack}}}, \bibinfo {author} {\bibfnamefont {Z.}~\bibnamefont {{Xu}}},\
  and\ \bibinfo {author} {\bibfnamefont {Y.}~\bibnamefont {{Zhang}}},\
  }\bibfield  {title} {\bibinfo {title} {{The Atacama Cosmology Telescope: A
  Catalog of >4000 Sunyaev{\textendash}Zel{\textquoteright}dovich Galaxy
  Clusters}},\ }\href {https://doi.org/10.3847/1538-4365/abd023} {\bibfield
  {journal} {\bibinfo  {journal} {\apjs}\ }\textbf {\bibinfo {volume} {253}},\
  \bibinfo {eid} {3} (\bibinfo {year} {2021})},\ \Eprint
  {https://arxiv.org/abs/2009.11043} {arXiv:2009.11043 [astro-ph.CO]}
  \BibitemShut {NoStop}%
\bibitem [{\citenamefont {{DES Collaboration}}\ \emph
  {et~al.}(2021)\citenamefont {{DES Collaboration}}, \citenamefont {{Abbott}},
  \citenamefont {{Aguena}}, \citenamefont {{Alarcon}}, \citenamefont {{Allam}},
  \citenamefont {{Alves}}, \citenamefont {{Amon}}, \citenamefont
  {{Andrade-Oliveira}}, \citenamefont {{Annis}}, \citenamefont {{Avila}},
  \citenamefont {{Bacon}}, \citenamefont {{Baxter}}, \citenamefont {{Bechtol}},
  \citenamefont {{Becker}}, \citenamefont {{Bernstein}}, \citenamefont
  {{Bhargava}}, \citenamefont {{Birrer}}, \citenamefont {{Blazek}},
  \citenamefont {{Brandao-Souza}}, \citenamefont {{Bridle}}, \citenamefont
  {{Brooks}}, \citenamefont {{Buckley-Geer}}, \citenamefont {{Burke}},
  \citenamefont {{Camacho}}, \citenamefont {{Campos}}, \citenamefont {{Carnero
  Rosell}}, \citenamefont {{Carrasco Kind}}, \citenamefont {{Carretero}},
  \citenamefont {{Castander}}, \citenamefont {{Cawthon}}, \citenamefont
  {{Chang}}, \citenamefont {{Chen}}, \citenamefont {{Chen}}, \citenamefont
  {{Choi}}, \citenamefont {{Conselice}}, \citenamefont {{Cordero}},
  \citenamefont {{Costanzi}}, \citenamefont {{Crocce}}, \citenamefont {{da
  Costa}}, \citenamefont {{da Silva Pereira}}, \citenamefont {{Davis}},
  \citenamefont {{Davis}}, \citenamefont {{De Vicente}}, \citenamefont
  {{DeRose}}, \citenamefont {{Desai}}, \citenamefont {{Di Valentino}},
  \citenamefont {{Diehl}}, \citenamefont {{Dietrich}}, \citenamefont
  {{Dodelson}}, \citenamefont {{Doel}}, \citenamefont {{Doux}}, \citenamefont
  {{Drlica-Wagner}}, \citenamefont {{Eckert}}, \citenamefont {{Eifler}},
  \citenamefont {{Elsner}}, \citenamefont {{Elvin-Poole}}, \citenamefont
  {{Everett}}, \citenamefont {{Evrard}}, \citenamefont {{Fang}}, \citenamefont
  {{Farahi}}, \citenamefont {{Fernandez}}, \citenamefont {{Ferrero}},
  \citenamefont {{Fert{\'e}}}, \citenamefont {{Fosalba}}, \citenamefont
  {{Friedrich}}, \citenamefont {{Frieman}}, \citenamefont
  {{Garc{\'\i}a-Bellido}}, \citenamefont {{Gatti}}, \citenamefont
  {{Gaztanaga}}, \citenamefont {{Gerdes}}, \citenamefont {{Giannantonio}},
  \citenamefont {{Giannini}}, \citenamefont {{Gruen}}, \citenamefont
  {{Gruendl}}, \citenamefont {{Gschwend}}, \citenamefont {{Gutierrez}},
  \citenamefont {{Harrison}}, \citenamefont {{Hartley}}, \citenamefont
  {{Herner}}, \citenamefont {{Hinton}}, \citenamefont {{Hollowood}},
  \citenamefont {{Honscheid}}, \citenamefont {{Hoyle}}, \citenamefont {{Huff}},
  \citenamefont {{Huterer}}, \citenamefont {{Jain}}, \citenamefont {{James}},
  \citenamefont {{Jarvis}}, \citenamefont {{Jeffrey}}, \citenamefont
  {{Jeltema}}, \citenamefont {{Kovacs}}, \citenamefont {{Krause}},
  \citenamefont {{Kron}}, \citenamefont {{Kuehn}}, \citenamefont
  {{Kuropatkin}}, \citenamefont {{Lahav}}, \citenamefont {{Leget}},
  \citenamefont {{Lemos}}, \citenamefont {{Liddle}}, \citenamefont {{Lidman}},
  \citenamefont {{Lima}}, \citenamefont {{Lin}}, \citenamefont {{MacCrann}},
  \citenamefont {{Maia}}, \citenamefont {{Marshall}}, \citenamefont
  {{Martini}}, \citenamefont {{McCullough}}, \citenamefont {{Melchior}},
  \citenamefont {{Mena-Fern{\'a}ndez}}, \citenamefont {{Menanteau}},
  \citenamefont {{Miquel}}, \citenamefont {{Mohr}}, \citenamefont {{Morgan}},
  \citenamefont {{Muir}}, \citenamefont {{Myles}}, \citenamefont {{Nadathur}},
  \citenamefont {{Navarro-Alsina}}, \citenamefont {{Nichol}}, \citenamefont
  {{Ogando}}, \citenamefont {{Omori}}, \citenamefont {{Palmese}}, \citenamefont
  {{Pandey}}, \citenamefont {{Park}}, \citenamefont {{Paz-Chinch{\'o}n}},
  \citenamefont {{Petravick}}, \citenamefont {{Pieres}}, \citenamefont {{Plazas
  Malag{\'o}n}}, \citenamefont {{Porredon}}, \citenamefont {{Prat}},
  \citenamefont {{Raveri}}, \citenamefont {{Rodriguez-Monroy}}, \citenamefont
  {{Rollins}}, \citenamefont {{Romer}}, \citenamefont {{Roodman}},
  \citenamefont {{Rosenfeld}}, \citenamefont {{Ross}}, \citenamefont
  {{Rykoff}}, \citenamefont {{Samuroff}}, \citenamefont {{S{\'a}nchez}},
  \citenamefont {{Sanchez}}, \citenamefont {{Sanchez}}, \citenamefont {{Sanchez
  Cid}}, \citenamefont {{Scarpine}}, \citenamefont {{Schubnell}}, \citenamefont
  {{Scolnic}}, \citenamefont {{Secco}}, \citenamefont {{Serrano}},
  \citenamefont {{Sevilla-Noarbe}}, \citenamefont {{Sheldon}}, \citenamefont
  {{Shin}}, \citenamefont {{Smith}}, \citenamefont {{Soares-Santos}},
  \citenamefont {{Suchyta}}, \citenamefont {{Swanson}}, \citenamefont
  {{Tabbutt}}, \citenamefont {{Tarle}}, \citenamefont {{Thomas}}, \citenamefont
  {{To}}, \citenamefont {{Troja}}, \citenamefont {{Troxel}}, \citenamefont
  {{Tucker}}, \citenamefont {{Tutusaus}}, \citenamefont {{Varga}},
  \citenamefont {{Walker}}, \citenamefont {{Weaverdyck}}, \citenamefont
  {{Weller}}, \citenamefont {{Yanny}}, \citenamefont {{Yin}}, \citenamefont
  {{Zhang}},\ and\ \citenamefont {{Zuntz}}}]{2021arXiv210513549D}%
  \BibitemOpen
  \bibfield  {author} {\bibinfo {author} {\bibnamefont {{DES Collaboration}}},
  \bibinfo {author} {\bibfnamefont {T.~M.~C.}\ \bibnamefont {{Abbott}}},
  \bibinfo {author} {\bibfnamefont {M.}~\bibnamefont {{Aguena}}}, \bibinfo
  {author} {\bibfnamefont {A.}~\bibnamefont {{Alarcon}}}, \bibinfo {author}
  {\bibfnamefont {S.}~\bibnamefont {{Allam}}}, \bibinfo {author} {\bibfnamefont
  {O.}~\bibnamefont {{Alves}}}, \bibinfo {author} {\bibfnamefont
  {A.}~\bibnamefont {{Amon}}}, \bibinfo {author} {\bibfnamefont
  {F.}~\bibnamefont {{Andrade-Oliveira}}}, \bibinfo {author} {\bibfnamefont
  {J.}~\bibnamefont {{Annis}}}, \bibinfo {author} {\bibfnamefont
  {S.}~\bibnamefont {{Avila}}}, \bibinfo {author} {\bibfnamefont
  {D.}~\bibnamefont {{Bacon}}}, \bibinfo {author} {\bibfnamefont
  {E.}~\bibnamefont {{Baxter}}}, \bibinfo {author} {\bibfnamefont
  {K.}~\bibnamefont {{Bechtol}}}, \bibinfo {author} {\bibfnamefont {M.~R.}\
  \bibnamefont {{Becker}}}, \bibinfo {author} {\bibfnamefont {G.~M.}\
  \bibnamefont {{Bernstein}}}, \bibinfo {author} {\bibfnamefont
  {S.}~\bibnamefont {{Bhargava}}}, \bibinfo {author} {\bibfnamefont
  {S.}~\bibnamefont {{Birrer}}}, \bibinfo {author} {\bibfnamefont
  {J.}~\bibnamefont {{Blazek}}}, \bibinfo {author} {\bibfnamefont
  {A.}~\bibnamefont {{Brandao-Souza}}}, \bibinfo {author} {\bibfnamefont
  {S.~L.}\ \bibnamefont {{Bridle}}}, \bibinfo {author} {\bibfnamefont
  {D.}~\bibnamefont {{Brooks}}}, \bibinfo {author} {\bibfnamefont
  {E.}~\bibnamefont {{Buckley-Geer}}}, \bibinfo {author} {\bibfnamefont
  {D.~L.}\ \bibnamefont {{Burke}}}, \bibinfo {author} {\bibfnamefont
  {H.}~\bibnamefont {{Camacho}}}, \bibinfo {author} {\bibfnamefont
  {A.}~\bibnamefont {{Campos}}}, \bibinfo {author} {\bibfnamefont
  {A.}~\bibnamefont {{Carnero Rosell}}}, \bibinfo {author} {\bibfnamefont
  {M.}~\bibnamefont {{Carrasco Kind}}}, \bibinfo {author} {\bibfnamefont
  {J.}~\bibnamefont {{Carretero}}}, \bibinfo {author} {\bibfnamefont {F.~J.}\
  \bibnamefont {{Castander}}}, \bibinfo {author} {\bibfnamefont
  {R.}~\bibnamefont {{Cawthon}}}, \bibinfo {author} {\bibfnamefont
  {C.}~\bibnamefont {{Chang}}}, \bibinfo {author} {\bibfnamefont
  {A.}~\bibnamefont {{Chen}}}, \bibinfo {author} {\bibfnamefont
  {R.}~\bibnamefont {{Chen}}}, \bibinfo {author} {\bibfnamefont
  {A.}~\bibnamefont {{Choi}}}, \bibinfo {author} {\bibfnamefont
  {C.}~\bibnamefont {{Conselice}}}, \bibinfo {author} {\bibfnamefont
  {J.}~\bibnamefont {{Cordero}}}, \bibinfo {author} {\bibfnamefont
  {M.}~\bibnamefont {{Costanzi}}}, \bibinfo {author} {\bibfnamefont
  {M.}~\bibnamefont {{Crocce}}}, \bibinfo {author} {\bibfnamefont {L.~N.}\
  \bibnamefont {{da Costa}}}, \bibinfo {author} {\bibfnamefont {M.~E.}\
  \bibnamefont {{da Silva Pereira}}}, \bibinfo {author} {\bibfnamefont
  {C.}~\bibnamefont {{Davis}}}, \bibinfo {author} {\bibfnamefont {T.~M.}\
  \bibnamefont {{Davis}}}, \bibinfo {author} {\bibfnamefont {J.}~\bibnamefont
  {{De Vicente}}}, \bibinfo {author} {\bibfnamefont {J.}~\bibnamefont
  {{DeRose}}}, \bibinfo {author} {\bibfnamefont {S.}~\bibnamefont {{Desai}}},
  \bibinfo {author} {\bibfnamefont {E.}~\bibnamefont {{Di Valentino}}},
  \bibinfo {author} {\bibfnamefont {H.~T.}\ \bibnamefont {{Diehl}}}, \bibinfo
  {author} {\bibfnamefont {J.~P.}\ \bibnamefont {{Dietrich}}}, \bibinfo
  {author} {\bibfnamefont {S.}~\bibnamefont {{Dodelson}}}, \bibinfo {author}
  {\bibfnamefont {P.}~\bibnamefont {{Doel}}}, \bibinfo {author} {\bibfnamefont
  {C.}~\bibnamefont {{Doux}}}, \bibinfo {author} {\bibfnamefont
  {A.}~\bibnamefont {{Drlica-Wagner}}}, \bibinfo {author} {\bibfnamefont
  {K.}~\bibnamefont {{Eckert}}}, \bibinfo {author} {\bibfnamefont {T.~F.}\
  \bibnamefont {{Eifler}}}, \bibinfo {author} {\bibfnamefont {F.}~\bibnamefont
  {{Elsner}}}, \bibinfo {author} {\bibfnamefont {J.}~\bibnamefont
  {{Elvin-Poole}}}, \bibinfo {author} {\bibfnamefont {S.}~\bibnamefont
  {{Everett}}}, \bibinfo {author} {\bibfnamefont {A.~E.}\ \bibnamefont
  {{Evrard}}}, \bibinfo {author} {\bibfnamefont {X.}~\bibnamefont {{Fang}}},
  \bibinfo {author} {\bibfnamefont {A.}~\bibnamefont {{Farahi}}}, \bibinfo
  {author} {\bibfnamefont {E.}~\bibnamefont {{Fernandez}}}, \bibinfo {author}
  {\bibfnamefont {I.}~\bibnamefont {{Ferrero}}}, \bibinfo {author}
  {\bibfnamefont {A.}~\bibnamefont {{Fert{\'e}}}}, \bibinfo {author}
  {\bibfnamefont {P.}~\bibnamefont {{Fosalba}}}, \bibinfo {author}
  {\bibfnamefont {O.}~\bibnamefont {{Friedrich}}}, \bibinfo {author}
  {\bibfnamefont {J.}~\bibnamefont {{Frieman}}}, \bibinfo {author}
  {\bibfnamefont {J.}~\bibnamefont {{Garc{\'\i}a-Bellido}}}, \bibinfo {author}
  {\bibfnamefont {M.}~\bibnamefont {{Gatti}}}, \bibinfo {author} {\bibfnamefont
  {E.}~\bibnamefont {{Gaztanaga}}}, \bibinfo {author} {\bibfnamefont {D.~W.}\
  \bibnamefont {{Gerdes}}}, \bibinfo {author} {\bibfnamefont {T.}~\bibnamefont
  {{Giannantonio}}}, \bibinfo {author} {\bibfnamefont {G.}~\bibnamefont
  {{Giannini}}}, \bibinfo {author} {\bibfnamefont {D.}~\bibnamefont {{Gruen}}},
  \bibinfo {author} {\bibfnamefont {R.~A.}\ \bibnamefont {{Gruendl}}}, \bibinfo
  {author} {\bibfnamefont {J.}~\bibnamefont {{Gschwend}}}, \bibinfo {author}
  {\bibfnamefont {G.}~\bibnamefont {{Gutierrez}}}, \bibinfo {author}
  {\bibfnamefont {I.}~\bibnamefont {{Harrison}}}, \bibinfo {author}
  {\bibfnamefont {W.~G.}\ \bibnamefont {{Hartley}}}, \bibinfo {author}
  {\bibfnamefont {K.}~\bibnamefont {{Herner}}}, \bibinfo {author}
  {\bibfnamefont {S.~R.}\ \bibnamefont {{Hinton}}}, \bibinfo {author}
  {\bibfnamefont {D.~L.}\ \bibnamefont {{Hollowood}}}, \bibinfo {author}
  {\bibfnamefont {K.}~\bibnamefont {{Honscheid}}}, \bibinfo {author}
  {\bibfnamefont {B.}~\bibnamefont {{Hoyle}}}, \bibinfo {author} {\bibfnamefont
  {E.~M.}\ \bibnamefont {{Huff}}}, \bibinfo {author} {\bibfnamefont
  {D.}~\bibnamefont {{Huterer}}}, \bibinfo {author} {\bibfnamefont
  {B.}~\bibnamefont {{Jain}}}, \bibinfo {author} {\bibfnamefont {D.~J.}\
  \bibnamefont {{James}}}, \bibinfo {author} {\bibfnamefont {M.}~\bibnamefont
  {{Jarvis}}}, \bibinfo {author} {\bibfnamefont {N.}~\bibnamefont {{Jeffrey}}},
  \bibinfo {author} {\bibfnamefont {T.}~\bibnamefont {{Jeltema}}}, \bibinfo
  {author} {\bibfnamefont {A.}~\bibnamefont {{Kovacs}}}, \bibinfo {author}
  {\bibfnamefont {E.}~\bibnamefont {{Krause}}}, \bibinfo {author}
  {\bibfnamefont {R.}~\bibnamefont {{Kron}}}, \bibinfo {author} {\bibfnamefont
  {K.}~\bibnamefont {{Kuehn}}}, \bibinfo {author} {\bibfnamefont
  {N.}~\bibnamefont {{Kuropatkin}}}, \bibinfo {author} {\bibfnamefont
  {O.}~\bibnamefont {{Lahav}}}, \bibinfo {author} {\bibfnamefont {P.~F.}\
  \bibnamefont {{Leget}}}, \bibinfo {author} {\bibfnamefont {P.}~\bibnamefont
  {{Lemos}}}, \bibinfo {author} {\bibfnamefont {A.~R.}\ \bibnamefont
  {{Liddle}}}, \bibinfo {author} {\bibfnamefont {C.}~\bibnamefont {{Lidman}}},
  \bibinfo {author} {\bibfnamefont {M.}~\bibnamefont {{Lima}}}, \bibinfo
  {author} {\bibfnamefont {H.}~\bibnamefont {{Lin}}}, \bibinfo {author}
  {\bibfnamefont {N.}~\bibnamefont {{MacCrann}}}, \bibinfo {author}
  {\bibfnamefont {M.~A.~G.}\ \bibnamefont {{Maia}}}, \bibinfo {author}
  {\bibfnamefont {J.~L.}\ \bibnamefont {{Marshall}}}, \bibinfo {author}
  {\bibfnamefont {P.}~\bibnamefont {{Martini}}}, \bibinfo {author}
  {\bibfnamefont {J.}~\bibnamefont {{McCullough}}}, \bibinfo {author}
  {\bibfnamefont {P.}~\bibnamefont {{Melchior}}}, \bibinfo {author}
  {\bibfnamefont {J.}~\bibnamefont {{Mena-Fern{\'a}ndez}}}, \bibinfo {author}
  {\bibfnamefont {F.}~\bibnamefont {{Menanteau}}}, \bibinfo {author}
  {\bibfnamefont {R.}~\bibnamefont {{Miquel}}}, \bibinfo {author}
  {\bibfnamefont {J.~J.}\ \bibnamefont {{Mohr}}}, \bibinfo {author}
  {\bibfnamefont {R.}~\bibnamefont {{Morgan}}}, \bibinfo {author}
  {\bibfnamefont {J.}~\bibnamefont {{Muir}}}, \bibinfo {author} {\bibfnamefont
  {J.}~\bibnamefont {{Myles}}}, \bibinfo {author} {\bibfnamefont
  {S.}~\bibnamefont {{Nadathur}}}, \bibinfo {author} {\bibfnamefont
  {A.}~\bibnamefont {{Navarro-Alsina}}}, \bibinfo {author} {\bibfnamefont
  {R.~C.}\ \bibnamefont {{Nichol}}}, \bibinfo {author} {\bibfnamefont
  {R.~L.~C.}\ \bibnamefont {{Ogando}}}, \bibinfo {author} {\bibfnamefont
  {Y.}~\bibnamefont {{Omori}}}, \bibinfo {author} {\bibfnamefont
  {A.}~\bibnamefont {{Palmese}}}, \bibinfo {author} {\bibfnamefont
  {S.}~\bibnamefont {{Pandey}}}, \bibinfo {author} {\bibfnamefont
  {Y.}~\bibnamefont {{Park}}}, \bibinfo {author} {\bibfnamefont
  {F.}~\bibnamefont {{Paz-Chinch{\'o}n}}}, \bibinfo {author} {\bibfnamefont
  {D.}~\bibnamefont {{Petravick}}}, \bibinfo {author} {\bibfnamefont
  {A.}~\bibnamefont {{Pieres}}}, \bibinfo {author} {\bibfnamefont {A.~A.}\
  \bibnamefont {{Plazas Malag{\'o}n}}}, \bibinfo {author} {\bibfnamefont
  {A.}~\bibnamefont {{Porredon}}}, \bibinfo {author} {\bibfnamefont
  {J.}~\bibnamefont {{Prat}}}, \bibinfo {author} {\bibfnamefont
  {M.}~\bibnamefont {{Raveri}}}, \bibinfo {author} {\bibfnamefont
  {M.}~\bibnamefont {{Rodriguez-Monroy}}}, \bibinfo {author} {\bibfnamefont
  {R.~P.}\ \bibnamefont {{Rollins}}}, \bibinfo {author} {\bibfnamefont {A.~K.}\
  \bibnamefont {{Romer}}}, \bibinfo {author} {\bibfnamefont {A.}~\bibnamefont
  {{Roodman}}}, \bibinfo {author} {\bibfnamefont {R.}~\bibnamefont
  {{Rosenfeld}}}, \bibinfo {author} {\bibfnamefont {A.~J.}\ \bibnamefont
  {{Ross}}}, \bibinfo {author} {\bibfnamefont {E.~S.}\ \bibnamefont
  {{Rykoff}}}, \bibinfo {author} {\bibfnamefont {S.}~\bibnamefont
  {{Samuroff}}}, \bibinfo {author} {\bibfnamefont {C.}~\bibnamefont
  {{S{\'a}nchez}}}, \bibinfo {author} {\bibfnamefont {E.}~\bibnamefont
  {{Sanchez}}}, \bibinfo {author} {\bibfnamefont {J.}~\bibnamefont
  {{Sanchez}}}, \bibinfo {author} {\bibfnamefont {D.}~\bibnamefont {{Sanchez
  Cid}}}, \bibinfo {author} {\bibfnamefont {V.}~\bibnamefont {{Scarpine}}},
  \bibinfo {author} {\bibfnamefont {M.}~\bibnamefont {{Schubnell}}}, \bibinfo
  {author} {\bibfnamefont {D.}~\bibnamefont {{Scolnic}}}, \bibinfo {author}
  {\bibfnamefont {L.~F.}\ \bibnamefont {{Secco}}}, \bibinfo {author}
  {\bibfnamefont {S.}~\bibnamefont {{Serrano}}}, \bibinfo {author}
  {\bibfnamefont {I.}~\bibnamefont {{Sevilla-Noarbe}}}, \bibinfo {author}
  {\bibfnamefont {E.}~\bibnamefont {{Sheldon}}}, \bibinfo {author}
  {\bibfnamefont {T.}~\bibnamefont {{Shin}}}, \bibinfo {author} {\bibfnamefont
  {M.}~\bibnamefont {{Smith}}}, \bibinfo {author} {\bibfnamefont
  {M.}~\bibnamefont {{Soares-Santos}}}, \bibinfo {author} {\bibfnamefont
  {E.}~\bibnamefont {{Suchyta}}}, \bibinfo {author} {\bibfnamefont {M.~E.~C.}\
  \bibnamefont {{Swanson}}}, \bibinfo {author} {\bibfnamefont {M.}~\bibnamefont
  {{Tabbutt}}}, \bibinfo {author} {\bibfnamefont {G.}~\bibnamefont {{Tarle}}},
  \bibinfo {author} {\bibfnamefont {D.}~\bibnamefont {{Thomas}}}, \bibinfo
  {author} {\bibfnamefont {C.}~\bibnamefont {{To}}}, \bibinfo {author}
  {\bibfnamefont {A.}~\bibnamefont {{Troja}}}, \bibinfo {author} {\bibfnamefont
  {M.~A.}\ \bibnamefont {{Troxel}}}, \bibinfo {author} {\bibfnamefont {D.~L.}\
  \bibnamefont {{Tucker}}}, \bibinfo {author} {\bibfnamefont {I.}~\bibnamefont
  {{Tutusaus}}}, \bibinfo {author} {\bibfnamefont {T.~N.}\ \bibnamefont
  {{Varga}}}, \bibinfo {author} {\bibfnamefont {A.~R.}\ \bibnamefont
  {{Walker}}}, \bibinfo {author} {\bibfnamefont {N.}~\bibnamefont
  {{Weaverdyck}}}, \bibinfo {author} {\bibfnamefont {J.}~\bibnamefont
  {{Weller}}}, \bibinfo {author} {\bibfnamefont {B.}~\bibnamefont {{Yanny}}},
  \bibinfo {author} {\bibfnamefont {B.}~\bibnamefont {{Yin}}}, \bibinfo
  {author} {\bibfnamefont {Y.}~\bibnamefont {{Zhang}}},\ and\ \bibinfo {author}
  {\bibfnamefont {J.}~\bibnamefont {{Zuntz}}},\ }\bibfield  {title} {\bibinfo
  {title} {{Dark Energy Survey Year 3 Results: Cosmological Constraints from
  Galaxy Clustering and Weak Lensing}},\ }\href@noop {} {\bibfield  {journal}
  {\bibinfo  {journal} {arXiv e-prints}\ ,\ \bibinfo {eid} {arXiv:2105.13549}}
  (\bibinfo {year} {2021})},\ \Eprint {https://arxiv.org/abs/2105.13549}
  {arXiv:2105.13549 [astro-ph.CO]} \BibitemShut {NoStop}%
\bibitem [{\citenamefont {{Starobinsky}}(1982)}]{1982PhLB..117..175S}%
  \BibitemOpen
  \bibfield  {author} {\bibinfo {author} {\bibfnamefont {A.~A.}\ \bibnamefont
  {{Starobinsky}}},\ }\bibfield  {title} {\bibinfo {title} {{Dynamics of phase
  transition in the new inflationary universe scenario and generation of
  perturbations}},\ }\href {https://doi.org/10.1016/0370-2693(82)90541-X}
  {\bibfield  {journal} {\bibinfo  {journal} {Physics Letters B}\ }\textbf
  {\bibinfo {volume} {117}},\ \bibinfo {pages} {175} (\bibinfo {year}
  {1982})}\BibitemShut {NoStop}%
\bibitem [{\citenamefont {{Guth}}\ and\ \citenamefont
  {{Pi}}(1982)}]{1982PhRvL..49.1110G}%
  \BibitemOpen
  \bibfield  {author} {\bibinfo {author} {\bibfnamefont {A.~H.}\ \bibnamefont
  {{Guth}}}\ and\ \bibinfo {author} {\bibfnamefont {S.~Y.}\ \bibnamefont
  {{Pi}}},\ }\bibfield  {title} {\bibinfo {title} {{Fluctuations in the New
  Inflationary Universe}},\ }\href
  {https://doi.org/10.1103/PhysRevLett.49.1110} {\bibfield  {journal} {\bibinfo
   {journal} {\prl}\ }\textbf {\bibinfo {volume} {49}},\ \bibinfo {pages}
  {1110} (\bibinfo {year} {1982})}\BibitemShut {NoStop}%
\bibitem [{\citenamefont {{Ashoorioon}}\ \emph {et~al.}(2009)\citenamefont
  {{Ashoorioon}}, \citenamefont {{Krause}},\ and\ \citenamefont
  {{Turzynski}}}]{2009JCAP...02..014A}%
  \BibitemOpen
  \bibfield  {author} {\bibinfo {author} {\bibfnamefont {A.}~\bibnamefont
  {{Ashoorioon}}}, \bibinfo {author} {\bibfnamefont {A.}~\bibnamefont
  {{Krause}}},\ and\ \bibinfo {author} {\bibfnamefont {K.}~\bibnamefont
  {{Turzynski}}},\ }\bibfield  {title} {\bibinfo {title} {{Energy transfer in
  multi field inflation and cosmological perturbations}},\ }\href
  {https://doi.org/10.1088/1475-7516/2009/02/014} {\bibfield  {journal}
  {\bibinfo  {journal} {\jcap}\ }\textbf {\bibinfo {volume} {2009}},\ \bibinfo
  {eid} {014} (\bibinfo {year} {2009})},\ \Eprint
  {https://arxiv.org/abs/0810.4660} {arXiv:0810.4660 [hep-th]} \BibitemShut
  {NoStop}%
\bibitem [{\citenamefont {{Schmid}}\ \emph {et~al.}(1997)\citenamefont
  {{Schmid}}, \citenamefont {{Schwarz}},\ and\ \citenamefont
  {{Widerin}}}]{1997PhRvL..78..791S}%
  \BibitemOpen
  \bibfield  {author} {\bibinfo {author} {\bibfnamefont {C.}~\bibnamefont
  {{Schmid}}}, \bibinfo {author} {\bibfnamefont {D.~J.}\ \bibnamefont
  {{Schwarz}}},\ and\ \bibinfo {author} {\bibfnamefont {P.}~\bibnamefont
  {{Widerin}}},\ }\bibfield  {title} {\bibinfo {title} {{Peaks above the
  Harrison-Zel'dovich Spectrum due to the Quark-Gluon to Hadron Transition}},\
  }\href {https://doi.org/10.1103/PhysRevLett.78.791} {\bibfield  {journal}
  {\bibinfo  {journal} {\prl}\ }\textbf {\bibinfo {volume} {78}},\ \bibinfo
  {pages} {791} (\bibinfo {year} {1997})},\ \Eprint
  {https://arxiv.org/abs/astro-ph/9606125} {arXiv:astro-ph/9606125 [astro-ph]}
  \BibitemShut {NoStop}%
\bibitem [{\citenamefont {{Barriga}}\ \emph {et~al.}(2001)\citenamefont
  {{Barriga}}, \citenamefont {{Gazta{\~n}aga}}, \citenamefont {{Santos}},\ and\
  \citenamefont {{Sarkar}}}]{2001MNRAS.324..977B}%
  \BibitemOpen
  \bibfield  {author} {\bibinfo {author} {\bibfnamefont {J.}~\bibnamefont
  {{Barriga}}}, \bibinfo {author} {\bibfnamefont {E.}~\bibnamefont
  {{Gazta{\~n}aga}}}, \bibinfo {author} {\bibfnamefont {M.~G.}\ \bibnamefont
  {{Santos}}},\ and\ \bibinfo {author} {\bibfnamefont {S.}~\bibnamefont
  {{Sarkar}}},\ }\bibfield  {title} {\bibinfo {title} {{On the APM power
  spectrum and the CMB anisotropy: evidence for a phase transition during
  inflation?}},\ }\href {https://doi.org/10.1046/j.1365-8711.2001.04373.x}
  {\bibfield  {journal} {\bibinfo  {journal} {\mnras}\ }\textbf {\bibinfo
  {volume} {324}},\ \bibinfo {pages} {977} (\bibinfo {year} {2001})},\ \Eprint
  {https://arxiv.org/abs/astro-ph/0011398} {arXiv:astro-ph/0011398 [astro-ph]}
  \BibitemShut {NoStop}%
\bibitem [{\citenamefont {{Barenboim}}\ and\ \citenamefont
  {{Rasero}}(2014)}]{2014JHEP...04..138B}%
  \BibitemOpen
  \bibfield  {author} {\bibinfo {author} {\bibfnamefont {G.}~\bibnamefont
  {{Barenboim}}}\ and\ \bibinfo {author} {\bibfnamefont {J.}~\bibnamefont
  {{Rasero}}},\ }\bibfield  {title} {\bibinfo {title} {{Structure formation
  during an early period of matter domination}},\ }\href
  {https://doi.org/10.1007/JHEP04(2014)138} {\bibfield  {journal} {\bibinfo
  {journal} {Journal of High Energy Physics}\ }\textbf {\bibinfo {volume}
  {2014}},\ \bibinfo {eid} {138} (\bibinfo {year} {2014})},\ \Eprint
  {https://arxiv.org/abs/1311.4034} {arXiv:1311.4034 [hep-ph]} \BibitemShut
  {NoStop}%
\bibitem [{\citenamefont {{Redmond}}\ \emph {et~al.}(2018)\citenamefont
  {{Redmond}}, \citenamefont {{Trezza}},\ and\ \citenamefont
  {{Erickcek}}}]{2018PhRvD..98f3504R}%
  \BibitemOpen
  \bibfield  {author} {\bibinfo {author} {\bibfnamefont {K.}~\bibnamefont
  {{Redmond}}}, \bibinfo {author} {\bibfnamefont {A.}~\bibnamefont
  {{Trezza}}},\ and\ \bibinfo {author} {\bibfnamefont {A.~L.}\ \bibnamefont
  {{Erickcek}}},\ }\bibfield  {title} {\bibinfo {title} {{Growth of dark matter
  perturbations during kination}},\ }\href
  {https://doi.org/10.1103/PhysRevD.98.063504} {\bibfield  {journal} {\bibinfo
  {journal} {\prd}\ }\textbf {\bibinfo {volume} {98}},\ \bibinfo {eid} {063504}
  (\bibinfo {year} {2018})},\ \Eprint {https://arxiv.org/abs/1807.01327}
  {arXiv:1807.01327 [astro-ph.CO]} \BibitemShut {NoStop}%
\bibitem [{\citenamefont {{Chluba}}\ and\ \citenamefont
  {{Sunyaev}}(2012)}]{2012MNRAS.419.1294C}%
  \BibitemOpen
  \bibfield  {author} {\bibinfo {author} {\bibfnamefont {J.}~\bibnamefont
  {{Chluba}}}\ and\ \bibinfo {author} {\bibfnamefont {R.~A.}\ \bibnamefont
  {{Sunyaev}}},\ }\bibfield  {title} {\bibinfo {title} {{The evolution of CMB
  spectral distortions in the early Universe}},\ }\href
  {https://doi.org/10.1111/j.1365-2966.2011.19786.x} {\bibfield  {journal}
  {\bibinfo  {journal} {\mnras}\ }\textbf {\bibinfo {volume} {419}},\ \bibinfo
  {pages} {1294} (\bibinfo {year} {2012})},\ \Eprint
  {https://arxiv.org/abs/1109.6552} {arXiv:1109.6552 [astro-ph.CO]}
  \BibitemShut {NoStop}%
\bibitem [{\citenamefont {{Tashiro}}(2014)}]{2014PTEP.2014fB107T}%
  \BibitemOpen
  \bibfield  {author} {\bibinfo {author} {\bibfnamefont {H.}~\bibnamefont
  {{Tashiro}}},\ }\bibfield  {title} {\bibinfo {title} {{CMB spectral
  distortions and energy release in the early universe}},\ }\href
  {https://doi.org/10.1093/ptep/ptu066} {\bibfield  {journal} {\bibinfo
  {journal} {Progress of Theoretical and Experimental Physics}\ }\textbf
  {\bibinfo {volume} {2014}},\ \bibinfo {eid} {06B107} (\bibinfo {year}
  {2014})}\BibitemShut {NoStop}%
\bibitem [{\citenamefont {{Chluba}}\ and\ \citenamefont
  {{Jeong}}(2014)}]{2014MNRAS.438.2065C}%
  \BibitemOpen
  \bibfield  {author} {\bibinfo {author} {\bibfnamefont {J.}~\bibnamefont
  {{Chluba}}}\ and\ \bibinfo {author} {\bibfnamefont {D.}~\bibnamefont
  {{Jeong}}},\ }\bibfield  {title} {\bibinfo {title} {{Teasing bits of
  information out of the CMB energy spectrum}},\ }\href
  {https://doi.org/10.1093/mnras/stt2327} {\bibfield  {journal} {\bibinfo
  {journal} {\mnras}\ }\textbf {\bibinfo {volume} {438}},\ \bibinfo {pages}
  {2065} (\bibinfo {year} {2014})},\ \Eprint {https://arxiv.org/abs/1306.5751}
  {arXiv:1306.5751 [astro-ph.CO]} \BibitemShut {NoStop}%
\bibitem [{\citenamefont {{Hu}}\ \emph {et~al.}(1994)\citenamefont {{Hu}},
  \citenamefont {{Scott}},\ and\ \citenamefont {{Silk}}}]{1994ApJ...430L...5H}%
  \BibitemOpen
  \bibfield  {author} {\bibinfo {author} {\bibfnamefont {W.}~\bibnamefont
  {{Hu}}}, \bibinfo {author} {\bibfnamefont {D.}~\bibnamefont {{Scott}}},\ and\
  \bibinfo {author} {\bibfnamefont {J.}~\bibnamefont {{Silk}}},\ }\bibfield
  {title} {\bibinfo {title} {{Power Spectrum Constraints from Spectral
  Distortions in the Cosmic Microwave Background}},\ }\href
  {https://doi.org/10.1086/187424} {\bibfield  {journal} {\bibinfo  {journal}
  {\apjl}\ }\textbf {\bibinfo {volume} {430}},\ \bibinfo {pages} {L5} (\bibinfo
  {year} {1994})},\ \Eprint {https://arxiv.org/abs/astro-ph/9402045}
  {arXiv:astro-ph/9402045 [astro-ph]} \BibitemShut {NoStop}%
\bibitem [{\citenamefont {{Chluba}}\ \emph {et~al.}(2012)\citenamefont
  {{Chluba}}, \citenamefont {{Erickcek}},\ and\ \citenamefont
  {{Ben-Dayan}}}]{2012ApJ...758...76C}%
  \BibitemOpen
  \bibfield  {author} {\bibinfo {author} {\bibfnamefont {J.}~\bibnamefont
  {{Chluba}}}, \bibinfo {author} {\bibfnamefont {A.~L.}\ \bibnamefont
  {{Erickcek}}},\ and\ \bibinfo {author} {\bibfnamefont {I.}~\bibnamefont
  {{Ben-Dayan}}},\ }\bibfield  {title} {\bibinfo {title} {{Probing the
  Inflaton: Small-scale Power Spectrum Constraints from Measurements of the
  Cosmic Microwave Background Energy Spectrum}},\ }\href
  {https://doi.org/10.1088/0004-637X/758/2/76} {\bibfield  {journal} {\bibinfo
  {journal} {\apj}\ }\textbf {\bibinfo {volume} {758}},\ \bibinfo {eid} {76}
  (\bibinfo {year} {2012})},\ \Eprint {https://arxiv.org/abs/1203.2681}
  {arXiv:1203.2681 [astro-ph.CO]} \BibitemShut {NoStop}%
\bibitem [{\citenamefont {{Pajer}}\ and\ \citenamefont
  {{Zaldarriaga}}(2012)}]{2012PhRvL.109b1302P}%
  \BibitemOpen
  \bibfield  {author} {\bibinfo {author} {\bibfnamefont {E.}~\bibnamefont
  {{Pajer}}}\ and\ \bibinfo {author} {\bibfnamefont {M.}~\bibnamefont
  {{Zaldarriaga}}},\ }\bibfield  {title} {\bibinfo {title} {{New Window on
  Primordial Non-Gaussianity}},\ }\href
  {https://doi.org/10.1103/PhysRevLett.109.021302} {\bibfield  {journal}
  {\bibinfo  {journal} {\prl}\ }\textbf {\bibinfo {volume} {109}},\ \bibinfo
  {eid} {021302} (\bibinfo {year} {2012})},\ \Eprint
  {https://arxiv.org/abs/1201.5375} {arXiv:1201.5375 [astro-ph.CO]}
  \BibitemShut {NoStop}%
\bibitem [{\citenamefont {{Dent}}\ \emph {et~al.}(2012)\citenamefont {{Dent}},
  \citenamefont {{Easson}},\ and\ \citenamefont
  {{Tashiro}}}]{2012PhRvD..86b3514D}%
  \BibitemOpen
  \bibfield  {author} {\bibinfo {author} {\bibfnamefont {J.~B.}\ \bibnamefont
  {{Dent}}}, \bibinfo {author} {\bibfnamefont {D.~A.}\ \bibnamefont
  {{Easson}}},\ and\ \bibinfo {author} {\bibfnamefont {H.}~\bibnamefont
  {{Tashiro}}},\ }\bibfield  {title} {\bibinfo {title} {{Cosmological
  constraints from CMB distortion}},\ }\href
  {https://doi.org/10.1103/PhysRevD.86.023514} {\bibfield  {journal} {\bibinfo
  {journal} {\prd}\ }\textbf {\bibinfo {volume} {86}},\ \bibinfo {eid} {023514}
  (\bibinfo {year} {2012})},\ \Eprint {https://arxiv.org/abs/1202.6066}
  {arXiv:1202.6066 [astro-ph.CO]} \BibitemShut {NoStop}%
\bibitem [{\citenamefont {{Chluba}}\ and\ \citenamefont
  {{Grin}}(2013)}]{2013MNRAS.434.1619C}%
  \BibitemOpen
  \bibfield  {author} {\bibinfo {author} {\bibfnamefont {J.}~\bibnamefont
  {{Chluba}}}\ and\ \bibinfo {author} {\bibfnamefont {D.}~\bibnamefont
  {{Grin}}},\ }\bibfield  {title} {\bibinfo {title} {{CMB spectral distortions
  from small-scale isocurvature fluctuations}},\ }\href
  {https://doi.org/10.1093/mnras/stt1129} {\bibfield  {journal} {\bibinfo
  {journal} {\mnras}\ }\textbf {\bibinfo {volume} {434}},\ \bibinfo {pages}
  {1619} (\bibinfo {year} {2013})},\ \Eprint {https://arxiv.org/abs/1304.4596}
  {arXiv:1304.4596 [astro-ph.CO]} \BibitemShut {NoStop}%
\bibitem [{\citenamefont {{Fixsen}}\ \emph {et~al.}(1996)\citenamefont
  {{Fixsen}}, \citenamefont {{Cheng}}, \citenamefont {{Gales}}, \citenamefont
  {{Mather}}, \citenamefont {{Shafer}},\ and\ \citenamefont
  {{Wright}}}]{1996ApJ...473..576F}%
  \BibitemOpen
  \bibfield  {author} {\bibinfo {author} {\bibfnamefont {D.~J.}\ \bibnamefont
  {{Fixsen}}}, \bibinfo {author} {\bibfnamefont {E.~S.}\ \bibnamefont
  {{Cheng}}}, \bibinfo {author} {\bibfnamefont {J.~M.}\ \bibnamefont
  {{Gales}}}, \bibinfo {author} {\bibfnamefont {J.~C.}\ \bibnamefont
  {{Mather}}}, \bibinfo {author} {\bibfnamefont {R.~A.}\ \bibnamefont
  {{Shafer}}},\ and\ \bibinfo {author} {\bibfnamefont {E.~L.}\ \bibnamefont
  {{Wright}}},\ }\bibfield  {title} {\bibinfo {title} {{The Cosmic Microwave
  Background Spectrum from the Full COBE FIRAS Data Set}},\ }\href
  {https://doi.org/10.1086/178173} {\bibfield  {journal} {\bibinfo  {journal}
  {\apj}\ }\textbf {\bibinfo {volume} {473}},\ \bibinfo {pages} {576} (\bibinfo
  {year} {1996})},\ \Eprint {https://arxiv.org/abs/astro-ph/9605054}
  {arXiv:astro-ph/9605054 [astro-ph]} \BibitemShut {NoStop}%
\bibitem [{\citenamefont {{Mather}}\ \emph {et~al.}(1994)\citenamefont
  {{Mather}}, \citenamefont {{Cheng}}, \citenamefont {{Cottingham}},
  \citenamefont {{Eplee}}, \citenamefont {{Fixsen}}, \citenamefont
  {{Hewagama}}, \citenamefont {{Isaacman}}, \citenamefont {{Jensen}},
  \citenamefont {{Meyer}}, \citenamefont {{Noerdlinger}}, \citenamefont
  {{Read}}, \citenamefont {{Rosen}}, \citenamefont {{Shafer}}, \citenamefont
  {{Wright}}, \citenamefont {{Bennett}}, \citenamefont {{Boggess}},
  \citenamefont {{Hauser}}, \citenamefont {{Kelsall}}, \citenamefont
  {{Moseley}}, \citenamefont {{Silverberg}}, \citenamefont {{Smoot}},
  \citenamefont {{Weiss}},\ and\ \citenamefont
  {{Wilkinson}}}]{1994ApJ...420..439M}%
  \BibitemOpen
  \bibfield  {author} {\bibinfo {author} {\bibfnamefont {J.~C.}\ \bibnamefont
  {{Mather}}}, \bibinfo {author} {\bibfnamefont {E.~S.}\ \bibnamefont
  {{Cheng}}}, \bibinfo {author} {\bibfnamefont {D.~A.}\ \bibnamefont
  {{Cottingham}}}, \bibinfo {author} {\bibfnamefont {J.}~\bibnamefont
  {{Eplee}}, \bibfnamefont {R.~E.}}, \bibinfo {author} {\bibfnamefont {D.~J.}\
  \bibnamefont {{Fixsen}}}, \bibinfo {author} {\bibfnamefont {T.}~\bibnamefont
  {{Hewagama}}}, \bibinfo {author} {\bibfnamefont {R.~B.}\ \bibnamefont
  {{Isaacman}}}, \bibinfo {author} {\bibfnamefont {K.~A.}\ \bibnamefont
  {{Jensen}}}, \bibinfo {author} {\bibfnamefont {S.~S.}\ \bibnamefont
  {{Meyer}}}, \bibinfo {author} {\bibfnamefont {P.~D.}\ \bibnamefont
  {{Noerdlinger}}}, \bibinfo {author} {\bibfnamefont {S.~M.}\ \bibnamefont
  {{Read}}}, \bibinfo {author} {\bibfnamefont {L.~P.}\ \bibnamefont {{Rosen}}},
  \bibinfo {author} {\bibfnamefont {R.~A.}\ \bibnamefont {{Shafer}}}, \bibinfo
  {author} {\bibfnamefont {E.~L.}\ \bibnamefont {{Wright}}}, \bibinfo {author}
  {\bibfnamefont {C.~L.}\ \bibnamefont {{Bennett}}}, \bibinfo {author}
  {\bibfnamefont {N.~W.}\ \bibnamefont {{Boggess}}}, \bibinfo {author}
  {\bibfnamefont {M.~G.}\ \bibnamefont {{Hauser}}}, \bibinfo {author}
  {\bibfnamefont {T.}~\bibnamefont {{Kelsall}}}, \bibinfo {author}
  {\bibfnamefont {J.}~\bibnamefont {{Moseley}}, \bibfnamefont {S.~H.}},
  \bibinfo {author} {\bibfnamefont {R.~F.}\ \bibnamefont {{Silverberg}}},
  \bibinfo {author} {\bibfnamefont {G.~F.}\ \bibnamefont {{Smoot}}}, \bibinfo
  {author} {\bibfnamefont {R.}~\bibnamefont {{Weiss}}},\ and\ \bibinfo {author}
  {\bibfnamefont {D.~T.}\ \bibnamefont {{Wilkinson}}},\ }\bibfield  {title}
  {\bibinfo {title} {{Measurement of the Cosmic Microwave Background Spectrum
  by the COBE FIRAS Instrument}},\ }\href {https://doi.org/10.1086/173574}
  {\bibfield  {journal} {\bibinfo  {journal} {\apj}\ }\textbf {\bibinfo
  {volume} {420}},\ \bibinfo {pages} {439} (\bibinfo {year}
  {1994})}\BibitemShut {NoStop}%
\bibitem [{\citenamefont {{Chluba}}\ \emph {et~al.}(2019)\citenamefont
  {{Chluba}}, \citenamefont {{Abitbol}}, \citenamefont {{Aghanim}},
  \citenamefont {{Ali-Haimoud}}, \citenamefont {{Alvarez}}, \citenamefont
  {{Basu}}, \citenamefont {{Bolliet}}, \citenamefont {{Burigana}},
  \citenamefont {{de Bernardis}}, \citenamefont {{Delabrouille}}, \citenamefont
  {{Dimastrogiovanni}}, \citenamefont {{Finelli}}, \citenamefont {{Fixsen}},
  \citenamefont {{Hart}}, \citenamefont {{Hernandez-Monteagudo}}, \citenamefont
  {{Hill}}, \citenamefont {{Kogut}}, \citenamefont {{Kohri}}, \citenamefont
  {{Lesgourgues}}, \citenamefont {{Maffei}}, \citenamefont {{Mather}},
  \citenamefont {{Mukherjee}}, \citenamefont {{Patil}}, \citenamefont
  {{Ravenni}}, \citenamefont {{Remazeilles}}, \citenamefont {{Rotti}},
  \citenamefont {{Rubino-Martin}}, \citenamefont {{Silk}}, \citenamefont
  {{Sunyaev}},\ and\ \citenamefont {{Switzer}}}]{2019arXiv190901593C}%
  \BibitemOpen
  \bibfield  {author} {\bibinfo {author} {\bibfnamefont {J.}~\bibnamefont
  {{Chluba}}}, \bibinfo {author} {\bibfnamefont {M.~H.}\ \bibnamefont
  {{Abitbol}}}, \bibinfo {author} {\bibfnamefont {N.}~\bibnamefont
  {{Aghanim}}}, \bibinfo {author} {\bibfnamefont {Y.}~\bibnamefont
  {{Ali-Haimoud}}}, \bibinfo {author} {\bibfnamefont {M.}~\bibnamefont
  {{Alvarez}}}, \bibinfo {author} {\bibfnamefont {K.}~\bibnamefont {{Basu}}},
  \bibinfo {author} {\bibfnamefont {B.}~\bibnamefont {{Bolliet}}}, \bibinfo
  {author} {\bibfnamefont {C.}~\bibnamefont {{Burigana}}}, \bibinfo {author}
  {\bibfnamefont {P.}~\bibnamefont {{de Bernardis}}}, \bibinfo {author}
  {\bibfnamefont {J.}~\bibnamefont {{Delabrouille}}}, \bibinfo {author}
  {\bibfnamefont {E.}~\bibnamefont {{Dimastrogiovanni}}}, \bibinfo {author}
  {\bibfnamefont {F.}~\bibnamefont {{Finelli}}}, \bibinfo {author}
  {\bibfnamefont {D.}~\bibnamefont {{Fixsen}}}, \bibinfo {author}
  {\bibfnamefont {L.}~\bibnamefont {{Hart}}}, \bibinfo {author} {\bibfnamefont
  {C.}~\bibnamefont {{Hernandez-Monteagudo}}}, \bibinfo {author} {\bibfnamefont
  {J.~C.}\ \bibnamefont {{Hill}}}, \bibinfo {author} {\bibfnamefont
  {A.}~\bibnamefont {{Kogut}}}, \bibinfo {author} {\bibfnamefont
  {K.}~\bibnamefont {{Kohri}}}, \bibinfo {author} {\bibfnamefont
  {J.}~\bibnamefont {{Lesgourgues}}}, \bibinfo {author} {\bibfnamefont
  {B.}~\bibnamefont {{Maffei}}}, \bibinfo {author} {\bibfnamefont
  {J.}~\bibnamefont {{Mather}}}, \bibinfo {author} {\bibfnamefont
  {S.}~\bibnamefont {{Mukherjee}}}, \bibinfo {author} {\bibfnamefont {S.~P.}\
  \bibnamefont {{Patil}}}, \bibinfo {author} {\bibfnamefont {A.}~\bibnamefont
  {{Ravenni}}}, \bibinfo {author} {\bibfnamefont {M.}~\bibnamefont
  {{Remazeilles}}}, \bibinfo {author} {\bibfnamefont {A.}~\bibnamefont
  {{Rotti}}}, \bibinfo {author} {\bibfnamefont {J.~A.}\ \bibnamefont
  {{Rubino-Martin}}}, \bibinfo {author} {\bibfnamefont {J.}~\bibnamefont
  {{Silk}}}, \bibinfo {author} {\bibfnamefont {R.~A.}\ \bibnamefont
  {{Sunyaev}}},\ and\ \bibinfo {author} {\bibfnamefont {E.~R.}\ \bibnamefont
  {{Switzer}}},\ }\bibfield  {title} {\bibinfo {title} {{New Horizons in
  Cosmology with Spectral Distortions of the Cosmic Microwave Background}},\
  }\href@noop {} {\bibfield  {journal} {\bibinfo  {journal} {arXiv e-prints}\
  ,\ \bibinfo {eid} {arXiv:1909.01593}} (\bibinfo {year} {2019})},\ \Eprint
  {https://arxiv.org/abs/1909.01593} {arXiv:1909.01593 [astro-ph.CO]}
  \BibitemShut {NoStop}%
\bibitem [{\citenamefont {{Loeb}}\ and\ \citenamefont
  {{Zaldarriaga}}(2004)}]{2004PhRvL..92u1301L}%
  \BibitemOpen
  \bibfield  {author} {\bibinfo {author} {\bibfnamefont {A.}~\bibnamefont
  {{Loeb}}}\ and\ \bibinfo {author} {\bibfnamefont {M.}~\bibnamefont
  {{Zaldarriaga}}},\ }\bibfield  {title} {\bibinfo {title} {{Measuring the
  Small-Scale Power Spectrum of Cosmic Density Fluctuations through 21cm
  Tomography Prior to the Epoch of Structure Formation}},\ }\href
  {https://doi.org/10.1103/PhysRevLett.92.211301} {\bibfield  {journal}
  {\bibinfo  {journal} {\prl}\ }\textbf {\bibinfo {volume} {92}},\ \bibinfo
  {eid} {211301} (\bibinfo {year} {2004})},\ \Eprint
  {https://arxiv.org/abs/astro-ph/0312134} {arXiv:astro-ph/0312134 [astro-ph]}
  \BibitemShut {NoStop}%
\bibitem [{\citenamefont {{Zel'dovich}}\ and\ \citenamefont
  {{Novikov}}(1967)}]{1967SvA....10..602Z}%
  \BibitemOpen
  \bibfield  {author} {\bibinfo {author} {\bibfnamefont {Y.~B.}\ \bibnamefont
  {{Zel'dovich}}}\ and\ \bibinfo {author} {\bibfnamefont {I.~D.}\ \bibnamefont
  {{Novikov}}},\ }\bibfield  {title} {\bibinfo {title} {{The Hypothesis of
  Cores Retarded during Expansion and the Hot Cosmological Model}},\
  }\href@noop {} {\bibfield  {journal} {\bibinfo  {journal} {\sovast}\ }\textbf
  {\bibinfo {volume} {10}},\ \bibinfo {pages} {602} (\bibinfo {year}
  {1967})}\BibitemShut {NoStop}%
\bibitem [{\citenamefont {{Hawking}}(1971)}]{1971MNRAS.152...75H}%
  \BibitemOpen
  \bibfield  {author} {\bibinfo {author} {\bibfnamefont {S.}~\bibnamefont
  {{Hawking}}},\ }\bibfield  {title} {\bibinfo {title} {{Gravitationally
  collapsed objects of very low mass}},\ }\href
  {https://doi.org/10.1093/mnras/152.1.75} {\bibfield  {journal} {\bibinfo
  {journal} {\mnras}\ }\textbf {\bibinfo {volume} {152}},\ \bibinfo {pages}
  {75} (\bibinfo {year} {1971})}\BibitemShut {NoStop}%
\bibitem [{\citenamefont {{Chapline}}(1975)}]{1975Natur.253..251C}%
  \BibitemOpen
  \bibfield  {author} {\bibinfo {author} {\bibfnamefont {G.~F.}\ \bibnamefont
  {{Chapline}}},\ }\bibfield  {title} {\bibinfo {title} {{Cosmological effects
  of primordial black holes}},\ }\href {https://doi.org/10.1038/253251a0}
  {\bibfield  {journal} {\bibinfo  {journal} {\nat}\ }\textbf {\bibinfo
  {volume} {253}},\ \bibinfo {pages} {251} (\bibinfo {year}
  {1975})}\BibitemShut {NoStop}%
\bibitem [{\citenamefont {{Abbott}}\ and\ \citenamefont {{(LIGO Scientific,
  Virgo)}}(2016)}]{2016PhRvL.116f1102A}%
  \BibitemOpen
  \bibfield  {author} {\bibinfo {author} {\bibfnamefont {B.~P.}\ \bibnamefont
  {{Abbott}}}\ and\ \bibinfo {author} {\bibnamefont {{(LIGO Scientific,
  Virgo)}}},\ }\bibfield  {title} {\bibinfo {title} {{Observation of
  Gravitational Waves from a Binary Black Hole Merger}},\ }\href
  {https://doi.org/10.1103/PhysRevLett.116.061102} {\bibfield  {journal}
  {\bibinfo  {journal} {\prl}\ }\textbf {\bibinfo {volume} {116}},\ \bibinfo
  {eid} {061102} (\bibinfo {year} {2016})},\ \Eprint
  {https://arxiv.org/abs/1602.03837} {arXiv:1602.03837 [gr-qc]} \BibitemShut
  {NoStop}%
\bibitem [{\citenamefont {{Carr}}\ \emph {et~al.}(2020)\citenamefont {{Carr}},
  \citenamefont {{Kohri}}, \citenamefont {{Sendouda}},\ and\ \citenamefont
  {{Yokoyama}}}]{2020arXiv200212778C}%
  \BibitemOpen
  \bibfield  {author} {\bibinfo {author} {\bibfnamefont {B.}~\bibnamefont
  {{Carr}}}, \bibinfo {author} {\bibfnamefont {K.}~\bibnamefont {{Kohri}}},
  \bibinfo {author} {\bibfnamefont {Y.}~\bibnamefont {{Sendouda}}},\ and\
  \bibinfo {author} {\bibfnamefont {J.}~\bibnamefont {{Yokoyama}}},\ }\bibfield
   {title} {\bibinfo {title} {{Constraints on Primordial Black Holes}},\
  }\href@noop {} {\bibfield  {journal} {\bibinfo  {journal} {arXiv e-prints}\
  ,\ \bibinfo {eid} {arXiv:2002.12778}} (\bibinfo {year} {2020})},\ \Eprint
  {https://arxiv.org/abs/2002.12778} {arXiv:2002.12778 [astro-ph.CO]}
  \BibitemShut {NoStop}%
\bibitem [{\citenamefont {{Emami}}\ and\ \citenamefont
  {{Smoot}}(2018)}]{2018JCAP...01..007E}%
  \BibitemOpen
  \bibfield  {author} {\bibinfo {author} {\bibfnamefont {R.}~\bibnamefont
  {{Emami}}}\ and\ \bibinfo {author} {\bibfnamefont {G.~F.}\ \bibnamefont
  {{Smoot}}},\ }\bibfield  {title} {\bibinfo {title} {{Observational
  constraints on the primordial curvature power spectrum}},\ }\href
  {https://doi.org/10.1088/1475-7516/2018/01/007} {\bibfield  {journal}
  {\bibinfo  {journal} {\jcap}\ }\textbf {\bibinfo {volume} {2018}},\ \bibinfo
  {eid} {007} (\bibinfo {year} {2018})},\ \Eprint
  {https://arxiv.org/abs/1705.09924} {arXiv:1705.09924 [astro-ph.CO]}
  \BibitemShut {NoStop}%
\bibitem [{\citenamefont {{Ricotti}}\ and\ \citenamefont
  {{Gould}}(2009)}]{2009ApJ...707..979R}%
  \BibitemOpen
  \bibfield  {author} {\bibinfo {author} {\bibfnamefont {M.}~\bibnamefont
  {{Ricotti}}}\ and\ \bibinfo {author} {\bibfnamefont {A.}~\bibnamefont
  {{Gould}}},\ }\bibfield  {title} {\bibinfo {title} {{A New Probe of Dark
  Matter and High-Energy Universe Using Microlensing}},\ }\href
  {https://doi.org/10.1088/0004-637X/707/2/979} {\bibfield  {journal} {\bibinfo
   {journal} {\apj}\ }\textbf {\bibinfo {volume} {707}},\ \bibinfo {pages}
  {979} (\bibinfo {year} {2009})},\ \Eprint {https://arxiv.org/abs/0908.0735}
  {arXiv:0908.0735 [astro-ph.CO]} \BibitemShut {NoStop}%
\bibitem [{\citenamefont {{Scott}}\ and\ \citenamefont
  {{Sivertsson}}(2009)}]{2009PhRvL.103u1301S}%
  \BibitemOpen
  \bibfield  {author} {\bibinfo {author} {\bibfnamefont {P.}~\bibnamefont
  {{Scott}}}\ and\ \bibinfo {author} {\bibfnamefont {S.}~\bibnamefont
  {{Sivertsson}}},\ }\bibfield  {title} {\bibinfo {title} {{Gamma Rays from
  Ultracompact Primordial Dark Matter Minihalos}},\ }\href
  {https://doi.org/10.1103/PhysRevLett.103.211301} {\bibfield  {journal}
  {\bibinfo  {journal} {\prl}\ }\textbf {\bibinfo {volume} {103}},\ \bibinfo
  {eid} {211301} (\bibinfo {year} {2009})},\ \Eprint
  {https://arxiv.org/abs/0908.4082} {arXiv:0908.4082 [astro-ph.CO]}
  \BibitemShut {NoStop}%
\bibitem [{\citenamefont {{Josan}}\ and\ \citenamefont
  {{Green}}(2010)}]{2010PhRvD..82h3527J}%
  \BibitemOpen
  \bibfield  {author} {\bibinfo {author} {\bibfnamefont {A.~S.}\ \bibnamefont
  {{Josan}}}\ and\ \bibinfo {author} {\bibfnamefont {A.~M.}\ \bibnamefont
  {{Green}}},\ }\bibfield  {title} {\bibinfo {title} {{Gamma rays from
  ultracompact minihalos: Potential constraints on the primordial curvature
  perturbation}},\ }\href {https://doi.org/10.1103/PhysRevD.82.083527}
  {\bibfield  {journal} {\bibinfo  {journal} {\prd}\ }\textbf {\bibinfo
  {volume} {82}},\ \bibinfo {eid} {083527} (\bibinfo {year} {2010})},\ \Eprint
  {https://arxiv.org/abs/1006.4970} {arXiv:1006.4970 [astro-ph.CO]}
  \BibitemShut {NoStop}%
\bibitem [{\citenamefont {{Bringmann}}\ \emph {et~al.}(2012)\citenamefont
  {{Bringmann}}, \citenamefont {{Scott}},\ and\ \citenamefont
  {{Akrami}}}]{2012PhRvD..85l5027B}%
  \BibitemOpen
  \bibfield  {author} {\bibinfo {author} {\bibfnamefont {T.}~\bibnamefont
  {{Bringmann}}}, \bibinfo {author} {\bibfnamefont {P.}~\bibnamefont
  {{Scott}}},\ and\ \bibinfo {author} {\bibfnamefont {Y.}~\bibnamefont
  {{Akrami}}},\ }\bibfield  {title} {\bibinfo {title} {{Improved constraints on
  the primordial power spectrum at small scales from ultracompact minihalos}},\
  }\href {https://doi.org/10.1103/PhysRevD.85.125027} {\bibfield  {journal}
  {\bibinfo  {journal} {\prd}\ }\textbf {\bibinfo {volume} {85}},\ \bibinfo
  {eid} {125027} (\bibinfo {year} {2012})},\ \Eprint
  {https://arxiv.org/abs/1110.2484} {arXiv:1110.2484 [astro-ph.CO]}
  \BibitemShut {NoStop}%
\bibitem [{\citenamefont {{Iliev}}\ \emph {et~al.}(2002)\citenamefont
  {{Iliev}}, \citenamefont {{Shapiro}}, \citenamefont {{Ferrara}},\ and\
  \citenamefont {{Martel}}}]{2002ApJ...572L.123I}%
  \BibitemOpen
  \bibfield  {author} {\bibinfo {author} {\bibfnamefont {I.~T.}\ \bibnamefont
  {{Iliev}}}, \bibinfo {author} {\bibfnamefont {P.~R.}\ \bibnamefont
  {{Shapiro}}}, \bibinfo {author} {\bibfnamefont {A.}~\bibnamefont
  {{Ferrara}}},\ and\ \bibinfo {author} {\bibfnamefont {H.}~\bibnamefont
  {{Martel}}},\ }\bibfield  {title} {\bibinfo {title} {{On the Direct
  Detectability of the Cosmic Dark Ages: 21 Centimeter Emission from
  Minihalos}},\ }\href {https://doi.org/10.1086/341869} {\bibfield  {journal}
  {\bibinfo  {journal} {\apjl}\ }\textbf {\bibinfo {volume} {572}},\ \bibinfo
  {pages} {L123} (\bibinfo {year} {2002})},\ \Eprint
  {https://arxiv.org/abs/astro-ph/0202410} {arXiv:astro-ph/0202410 [astro-ph]}
  \BibitemShut {NoStop}%
\bibitem [{\citenamefont {{Sekiguchi}}\ \emph {et~al.}(2014)\citenamefont
  {{Sekiguchi}}, \citenamefont {{Tashiro}}, \citenamefont {{Silk}},\ and\
  \citenamefont {{Sugiyama}}}]{2014JCAP...03..001S}%
  \BibitemOpen
  \bibfield  {author} {\bibinfo {author} {\bibfnamefont {T.}~\bibnamefont
  {{Sekiguchi}}}, \bibinfo {author} {\bibfnamefont {H.}~\bibnamefont
  {{Tashiro}}}, \bibinfo {author} {\bibfnamefont {J.}~\bibnamefont {{Silk}}},\
  and\ \bibinfo {author} {\bibfnamefont {N.}~\bibnamefont {{Sugiyama}}},\
  }\bibfield  {title} {\bibinfo {title} {{Cosmological signatures of tilted
  isocurvature perturbations: reionization and 21cm fluctuations}},\ }\href
  {https://doi.org/10.1088/1475-7516/2014/03/001} {\bibfield  {journal}
  {\bibinfo  {journal} {\jcap}\ }\textbf {\bibinfo {volume} {2014}},\ \bibinfo
  {eid} {001} (\bibinfo {year} {2014})},\ \Eprint
  {https://arxiv.org/abs/1311.3294} {arXiv:1311.3294 [astro-ph.CO]}
  \BibitemShut {NoStop}%
\bibitem [{\citenamefont {{Shimabukuro}}\ \emph {et~al.}(2014)\citenamefont
  {{Shimabukuro}}, \citenamefont {{Ichiki}}, \citenamefont {{Inoue}},\ and\
  \citenamefont {{Yokoyama}}}]{2014PhRvD..90h3003S}%
  \BibitemOpen
  \bibfield  {author} {\bibinfo {author} {\bibfnamefont {H.}~\bibnamefont
  {{Shimabukuro}}}, \bibinfo {author} {\bibfnamefont {K.}~\bibnamefont
  {{Ichiki}}}, \bibinfo {author} {\bibfnamefont {S.}~\bibnamefont {{Inoue}}},\
  and\ \bibinfo {author} {\bibfnamefont {S.}~\bibnamefont {{Yokoyama}}},\
  }\bibfield  {title} {\bibinfo {title} {{Probing small-scale cosmological
  fluctuations with the 21 cm forest: Effects of neutrino mass, running
  spectral index, and warm dark matter}},\ }\href
  {https://doi.org/10.1103/PhysRevD.90.083003} {\bibfield  {journal} {\bibinfo
  {journal} {\prd}\ }\textbf {\bibinfo {volume} {90}},\ \bibinfo {eid} {083003}
  (\bibinfo {year} {2014})},\ \Eprint {https://arxiv.org/abs/1403.1605}
  {arXiv:1403.1605 [astro-ph.CO]} \BibitemShut {NoStop}%
\bibitem [{\citenamefont {{Sekiguchi}}\ \emph {et~al.}(2018)\citenamefont
  {{Sekiguchi}}, \citenamefont {{Takahashi}}, \citenamefont {{Tashiro}},\ and\
  \citenamefont {{Yokoyama}}}]{2018JCAP...02..053S}%
  \BibitemOpen
  \bibfield  {author} {\bibinfo {author} {\bibfnamefont {T.}~\bibnamefont
  {{Sekiguchi}}}, \bibinfo {author} {\bibfnamefont {T.}~\bibnamefont
  {{Takahashi}}}, \bibinfo {author} {\bibfnamefont {H.}~\bibnamefont
  {{Tashiro}}},\ and\ \bibinfo {author} {\bibfnamefont {S.}~\bibnamefont
  {{Yokoyama}}},\ }\bibfield  {title} {\bibinfo {title} {{21 cm angular power
  spectrum from minihalos as a probe of primordial spectral runnings}},\ }\href
  {https://doi.org/10.1088/1475-7516/2018/02/053} {\bibfield  {journal}
  {\bibinfo  {journal} {\jcap}\ }\textbf {\bibinfo {volume} {2018}},\ \bibinfo
  {eid} {053} (\bibinfo {year} {2018})},\ \Eprint
  {https://arxiv.org/abs/1705.00405} {arXiv:1705.00405 [astro-ph.CO]}
  \BibitemShut {NoStop}%
\bibitem [{\citenamefont {{Furugori}}\ \emph {et~al.}(2020)\citenamefont
  {{Furugori}}, \citenamefont {{Abe}}, \citenamefont {{Tanaka}}, \citenamefont
  {{Hashimoto}}, \citenamefont {{Tashiro}},\ and\ \citenamefont
  {{Hasegawa}}}]{2020MNRAS.494.4334F}%
  \BibitemOpen
  \bibfield  {author} {\bibinfo {author} {\bibfnamefont {K.}~\bibnamefont
  {{Furugori}}}, \bibinfo {author} {\bibfnamefont {K.~T.}\ \bibnamefont
  {{Abe}}}, \bibinfo {author} {\bibfnamefont {T.}~\bibnamefont {{Tanaka}}},
  \bibinfo {author} {\bibfnamefont {D.}~\bibnamefont {{Hashimoto}}}, \bibinfo
  {author} {\bibfnamefont {H.}~\bibnamefont {{Tashiro}}},\ and\ \bibinfo
  {author} {\bibfnamefont {K.}~\bibnamefont {{Hasegawa}}},\ }\bibfield  {title}
  {\bibinfo {title} {{The 21-cm signals from ultracompact minihaloes as a probe
  of primordial small-scale fluctuations}},\ }\href
  {https://doi.org/10.1093/mnras/staa1033} {\bibfield  {journal} {\bibinfo
  {journal} {\mnras}\ }\textbf {\bibinfo {volume} {494}},\ \bibinfo {pages}
  {4334} (\bibinfo {year} {2020})},\ \Eprint {https://arxiv.org/abs/2002.04817}
  {arXiv:2002.04817 [astro-ph.CO]} \BibitemShut {NoStop}%
\bibitem [{\citenamefont {{Planck Collaboration}}\ \emph
  {et~al.}(2016)\citenamefont {{Planck Collaboration}}, \citenamefont {{Adam}},
  \citenamefont {{Ade}}, \citenamefont {{Aghanim}}, \citenamefont {{Alves}},
  \citenamefont {{Arnaud}}, \citenamefont {{Ashdown}}, \citenamefont
  {{Aumont}}, \citenamefont {{Baccigalupi}}, \citenamefont {{Banday}},
  \citenamefont {{Barreiro}}, \citenamefont {{Bartlett}}, \citenamefont
  {{Bartolo}}, \citenamefont {{Battaner}}, \citenamefont {{Benabed}},
  \citenamefont {{Beno{\^\i}t}}, \citenamefont {{Benoit-L{\'e}vy}},
  \citenamefont {{Bernard}}, \citenamefont {{Bersanelli}}, \citenamefont
  {{Bielewicz}}, \citenamefont {{Bock}}, \citenamefont {{Bonaldi}},
  \citenamefont {{Bonavera}}, \citenamefont {{Bond}}, \citenamefont
  {{Borrill}}, \citenamefont {{Bouchet}}, \citenamefont {{Boulanger}},
  \citenamefont {{Bucher}}, \citenamefont {{Burigana}}, \citenamefont
  {{Butler}}, \citenamefont {{Calabrese}}, \citenamefont {{Cardoso}},
  \citenamefont {{Catalano}}, \citenamefont {{Challinor}}, \citenamefont
  {{Chamballu}}, \citenamefont {{Chary}}, \citenamefont {{Chiang}},
  \citenamefont {{Christensen}}, \citenamefont {{Clements}}, \citenamefont
  {{Colombi}}, \citenamefont {{Colombo}}, \citenamefont {{Combet}},
  \citenamefont {{Couchot}}, \citenamefont {{Coulais}}, \citenamefont
  {{Crill}}, \citenamefont {{Curto}}, \citenamefont {{Cuttaia}}, \citenamefont
  {{Danese}}, \citenamefont {{Davies}}, \citenamefont {{Davis}}, \citenamefont
  {{de Bernardis}}, \citenamefont {{de Rosa}}, \citenamefont {{de Zotti}},
  \citenamefont {{Delabrouille}}, \citenamefont {{D{\'e}sert}}, \citenamefont
  {{Dickinson}}, \citenamefont {{Diego}}, \citenamefont {{Dole}}, \citenamefont
  {{Donzelli}}, \citenamefont {{Dor{\'e}}}, \citenamefont {{Douspis}},
  \citenamefont {{Ducout}}, \citenamefont {{Dupac}}, \citenamefont
  {{Efstathiou}}, \citenamefont {{Elsner}}, \citenamefont {{En{\ss}lin}},
  \citenamefont {{Eriksen}}, \citenamefont {{Falgarone}}, \citenamefont
  {{Fergusson}}, \citenamefont {{Finelli}}, \citenamefont {{Forni}},
  \citenamefont {{Frailis}}, \citenamefont {{Fraisse}}, \citenamefont
  {{Franceschi}}, \citenamefont {{Frejsel}}, \citenamefont {{Galeotta}},
  \citenamefont {{Galli}}, \citenamefont {{Ganga}}, \citenamefont {{Ghosh}},
  \citenamefont {{Giard}}, \citenamefont {{Giraud-H{\'e}raud}}, \citenamefont
  {{Gjerl{\o}w}}, \citenamefont {{Gonz{\'a}lez-Nuevo}}, \citenamefont
  {{G{\'o}rski}}, \citenamefont {{Gratton}}, \citenamefont {{Gregorio}},
  \citenamefont {{Gruppuso}}, \citenamefont {{Gudmundsson}}, \citenamefont
  {{Hansen}}, \citenamefont {{Hanson}}, \citenamefont {{Harrison}},
  \citenamefont {{Helou}}, \citenamefont {{Henrot-Versill{\'e}}}, \citenamefont
  {{Hern{\'a}ndez-Monteagudo}}, \citenamefont {{Herranz}}, \citenamefont
  {{Hildebrandt}}, \citenamefont {{Hivon}}, \citenamefont {{Hobson}},
  \citenamefont {{Holmes}}, \citenamefont {{Hornstrup}}, \citenamefont
  {{Hovest}}, \citenamefont {{Huffenberger}}, \citenamefont {{Hurier}},
  \citenamefont {{Jaffe}}, \citenamefont {{Jaffe}}, \citenamefont {{Jones}},
  \citenamefont {{Juvela}}, \citenamefont {{Keih{\"a}nen}}, \citenamefont
  {{Keskitalo}}, \citenamefont {{Kisner}}, \citenamefont {{Kneissl}},
  \citenamefont {{Knoche}}, \citenamefont {{Kunz}}, \citenamefont
  {{Kurki-Suonio}}, \citenamefont {{Lagache}}, \citenamefont
  {{L{\"a}hteenm{\"a}ki}}, \citenamefont {{Lamarre}}, \citenamefont
  {{Lasenby}}, \citenamefont {{Lattanzi}}, \citenamefont {{Lawrence}},
  \citenamefont {{Le Jeune}}, \citenamefont {{Leahy}}, \citenamefont
  {{Leonardi}}, \citenamefont {{Lesgourgues}}, \citenamefont {{Levrier}},
  \citenamefont {{Liguori}}, \citenamefont {{Lilje}}, \citenamefont
  {{Linden-V{\o}rnle}}, \citenamefont {{L{\'o}pez-Caniego}}, \citenamefont
  {{Lubin}}, \citenamefont {{Mac{\'\i}as-P{\'e}rez}}, \citenamefont {{Maggio}},
  \citenamefont {{Maino}}, \citenamefont {{Mandolesi}}, \citenamefont
  {{Mangilli}}, \citenamefont {{Maris}}, \citenamefont {{Marshall}},
  \citenamefont {{Martin}}, \citenamefont {{Mart{\'\i}nez-Gonz{\'a}lez}},
  \citenamefont {{Masi}}, \citenamefont {{Matarrese}}, \citenamefont
  {{McGehee}}, \citenamefont {{Meinhold}}, \citenamefont {{Melchiorri}},
  \citenamefont {{Mendes}}, \citenamefont {{Mennella}}, \citenamefont
  {{Migliaccio}}, \citenamefont {{Mitra}}, \citenamefont
  {{Miville-Desch{\^e}nes}}, \citenamefont {{Moneti}}, \citenamefont
  {{Montier}}, \citenamefont {{Morgante}}, \citenamefont {{Mortlock}},
  \citenamefont {{Moss}}, \citenamefont {{Munshi}}, \citenamefont {{Murphy}},
  \citenamefont {{Naselsky}}, \citenamefont {{Nati}}, \citenamefont {{Natoli}},
  \citenamefont {{Netterfield}}, \citenamefont {{N{\o}rgaard-Nielsen}},
  \citenamefont {{Noviello}}, \citenamefont {{Novikov}}, \citenamefont
  {{Novikov}}, \citenamefont {{Orlando}}, \citenamefont {{Oxborrow}},
  \citenamefont {{Paci}}, \citenamefont {{Pagano}}, \citenamefont {{Pajot}},
  \citenamefont {{Paladini}}, \citenamefont {{Paoletti}}, \citenamefont
  {{Partridge}}, \citenamefont {{Pasian}}, \citenamefont {{Patanchon}},
  \citenamefont {{Pearson}}, \citenamefont {{Perdereau}}, \citenamefont
  {{Perotto}}, \citenamefont {{Perrotta}}, \citenamefont {{Pettorino}},
  \citenamefont {{Piacentini}}, \citenamefont {{Piat}}, \citenamefont
  {{Pierpaoli}}, \citenamefont {{Pietrobon}}, \citenamefont {{Plaszczynski}},
  \citenamefont {{Pointecouteau}}, \citenamefont {{Polenta}}, \citenamefont
  {{Pratt}}, \citenamefont {{Pr{\'e}zeau}}, \citenamefont {{Prunet}},
  \citenamefont {{Puget}}, \citenamefont {{Rachen}}, \citenamefont {{Reach}},
  \citenamefont {{Rebolo}}, \citenamefont {{Reinecke}}, \citenamefont
  {{Remazeilles}}, \citenamefont {{Renault}}, \citenamefont {{Renzi}},
  \citenamefont {{Ristorcelli}}, \citenamefont {{Rocha}}, \citenamefont
  {{Rosset}}, \citenamefont {{Rossetti}}, \citenamefont {{Roudier}},
  \citenamefont {{Rubi{\~n}o-Mart{\'\i}n}}, \citenamefont {{Rusholme}},
  \citenamefont {{Sandri}}, \citenamefont {{Santos}}, \citenamefont
  {{Savelainen}}, \citenamefont {{Savini}}, \citenamefont {{Scott}},
  \citenamefont {{Seiffert}}, \citenamefont {{Shellard}}, \citenamefont
  {{Spencer}}, \citenamefont {{Stolyarov}}, \citenamefont {{Stompor}},
  \citenamefont {{Strong}}, \citenamefont {{Sudiwala}}, \citenamefont
  {{Sunyaev}}, \citenamefont {{Sutton}}, \citenamefont {{Suur-Uski}},
  \citenamefont {{Sygnet}}, \citenamefont {{Tauber}}, \citenamefont
  {{Terenzi}}, \citenamefont {{Toffolatti}}, \citenamefont {{Tomasi}},
  \citenamefont {{Tristram}}, \citenamefont {{Tucci}}, \citenamefont
  {{Tuovinen}}, \citenamefont {{Umana}}, \citenamefont {{Valenziano}},
  \citenamefont {{Valiviita}}, \citenamefont {{Van Tent}}, \citenamefont
  {{Vielva}}, \citenamefont {{Villa}}, \citenamefont {{Wade}}, \citenamefont
  {{Wandelt}}, \citenamefont {{Wehus}}, \citenamefont {{Wilkinson}},
  \citenamefont {{Yvon}}, \citenamefont {{Zacchei}},\ and\ \citenamefont
  {{Zonca}}}]{2016A&A...594A..10P}%
  \BibitemOpen
  \bibfield  {author} {\bibinfo {author} {\bibnamefont {{Planck
  Collaboration}}}, \bibinfo {author} {\bibfnamefont {R.}~\bibnamefont
  {{Adam}}}, \bibinfo {author} {\bibfnamefont {P.~A.~R.}\ \bibnamefont
  {{Ade}}}, \bibinfo {author} {\bibfnamefont {N.}~\bibnamefont {{Aghanim}}},
  \bibinfo {author} {\bibfnamefont {M.~I.~R.}\ \bibnamefont {{Alves}}},
  \bibinfo {author} {\bibfnamefont {M.}~\bibnamefont {{Arnaud}}}, \bibinfo
  {author} {\bibfnamefont {M.}~\bibnamefont {{Ashdown}}}, \bibinfo {author}
  {\bibfnamefont {J.}~\bibnamefont {{Aumont}}}, \bibinfo {author}
  {\bibfnamefont {C.}~\bibnamefont {{Baccigalupi}}}, \bibinfo {author}
  {\bibfnamefont {A.~J.}\ \bibnamefont {{Banday}}}, \bibinfo {author}
  {\bibfnamefont {R.~B.}\ \bibnamefont {{Barreiro}}}, \bibinfo {author}
  {\bibfnamefont {J.~G.}\ \bibnamefont {{Bartlett}}}, \bibinfo {author}
  {\bibfnamefont {N.}~\bibnamefont {{Bartolo}}}, \bibinfo {author}
  {\bibfnamefont {E.}~\bibnamefont {{Battaner}}}, \bibinfo {author}
  {\bibfnamefont {K.}~\bibnamefont {{Benabed}}}, \bibinfo {author}
  {\bibfnamefont {A.}~\bibnamefont {{Beno{\^\i}t}}}, \bibinfo {author}
  {\bibfnamefont {A.}~\bibnamefont {{Benoit-L{\'e}vy}}}, \bibinfo {author}
  {\bibfnamefont {J.~P.}\ \bibnamefont {{Bernard}}}, \bibinfo {author}
  {\bibfnamefont {M.}~\bibnamefont {{Bersanelli}}}, \bibinfo {author}
  {\bibfnamefont {P.}~\bibnamefont {{Bielewicz}}}, \bibinfo {author}
  {\bibfnamefont {J.~J.}\ \bibnamefont {{Bock}}}, \bibinfo {author}
  {\bibfnamefont {A.}~\bibnamefont {{Bonaldi}}}, \bibinfo {author}
  {\bibfnamefont {L.}~\bibnamefont {{Bonavera}}}, \bibinfo {author}
  {\bibfnamefont {J.~R.}\ \bibnamefont {{Bond}}}, \bibinfo {author}
  {\bibfnamefont {J.}~\bibnamefont {{Borrill}}}, \bibinfo {author}
  {\bibfnamefont {F.~R.}\ \bibnamefont {{Bouchet}}}, \bibinfo {author}
  {\bibfnamefont {F.}~\bibnamefont {{Boulanger}}}, \bibinfo {author}
  {\bibfnamefont {M.}~\bibnamefont {{Bucher}}}, \bibinfo {author}
  {\bibfnamefont {C.}~\bibnamefont {{Burigana}}}, \bibinfo {author}
  {\bibfnamefont {R.~C.}\ \bibnamefont {{Butler}}}, \bibinfo {author}
  {\bibfnamefont {E.}~\bibnamefont {{Calabrese}}}, \bibinfo {author}
  {\bibfnamefont {J.~F.}\ \bibnamefont {{Cardoso}}}, \bibinfo {author}
  {\bibfnamefont {A.}~\bibnamefont {{Catalano}}}, \bibinfo {author}
  {\bibfnamefont {A.}~\bibnamefont {{Challinor}}}, \bibinfo {author}
  {\bibfnamefont {A.}~\bibnamefont {{Chamballu}}}, \bibinfo {author}
  {\bibfnamefont {R.~R.}\ \bibnamefont {{Chary}}}, \bibinfo {author}
  {\bibfnamefont {H.~C.}\ \bibnamefont {{Chiang}}}, \bibinfo {author}
  {\bibfnamefont {P.~R.}\ \bibnamefont {{Christensen}}}, \bibinfo {author}
  {\bibfnamefont {D.~L.}\ \bibnamefont {{Clements}}}, \bibinfo {author}
  {\bibfnamefont {S.}~\bibnamefont {{Colombi}}}, \bibinfo {author}
  {\bibfnamefont {L.~P.~L.}\ \bibnamefont {{Colombo}}}, \bibinfo {author}
  {\bibfnamefont {C.}~\bibnamefont {{Combet}}}, \bibinfo {author}
  {\bibfnamefont {F.}~\bibnamefont {{Couchot}}}, \bibinfo {author}
  {\bibfnamefont {A.}~\bibnamefont {{Coulais}}}, \bibinfo {author}
  {\bibfnamefont {B.~P.}\ \bibnamefont {{Crill}}}, \bibinfo {author}
  {\bibfnamefont {A.}~\bibnamefont {{Curto}}}, \bibinfo {author} {\bibfnamefont
  {F.}~\bibnamefont {{Cuttaia}}}, \bibinfo {author} {\bibfnamefont
  {L.}~\bibnamefont {{Danese}}}, \bibinfo {author} {\bibfnamefont {R.~D.}\
  \bibnamefont {{Davies}}}, \bibinfo {author} {\bibfnamefont {R.~J.}\
  \bibnamefont {{Davis}}}, \bibinfo {author} {\bibfnamefont {P.}~\bibnamefont
  {{de Bernardis}}}, \bibinfo {author} {\bibfnamefont {A.}~\bibnamefont {{de
  Rosa}}}, \bibinfo {author} {\bibfnamefont {G.}~\bibnamefont {{de Zotti}}},
  \bibinfo {author} {\bibfnamefont {J.}~\bibnamefont {{Delabrouille}}},
  \bibinfo {author} {\bibfnamefont {F.~X.}\ \bibnamefont {{D{\'e}sert}}},
  \bibinfo {author} {\bibfnamefont {C.}~\bibnamefont {{Dickinson}}}, \bibinfo
  {author} {\bibfnamefont {J.~M.}\ \bibnamefont {{Diego}}}, \bibinfo {author}
  {\bibfnamefont {H.}~\bibnamefont {{Dole}}}, \bibinfo {author} {\bibfnamefont
  {S.}~\bibnamefont {{Donzelli}}}, \bibinfo {author} {\bibfnamefont
  {O.}~\bibnamefont {{Dor{\'e}}}}, \bibinfo {author} {\bibfnamefont
  {M.}~\bibnamefont {{Douspis}}}, \bibinfo {author} {\bibfnamefont
  {A.}~\bibnamefont {{Ducout}}}, \bibinfo {author} {\bibfnamefont
  {X.}~\bibnamefont {{Dupac}}}, \bibinfo {author} {\bibfnamefont
  {G.}~\bibnamefont {{Efstathiou}}}, \bibinfo {author} {\bibfnamefont
  {F.}~\bibnamefont {{Elsner}}}, \bibinfo {author} {\bibfnamefont {T.~A.}\
  \bibnamefont {{En{\ss}lin}}}, \bibinfo {author} {\bibfnamefont {H.~K.}\
  \bibnamefont {{Eriksen}}}, \bibinfo {author} {\bibfnamefont {E.}~\bibnamefont
  {{Falgarone}}}, \bibinfo {author} {\bibfnamefont {J.}~\bibnamefont
  {{Fergusson}}}, \bibinfo {author} {\bibfnamefont {F.}~\bibnamefont
  {{Finelli}}}, \bibinfo {author} {\bibfnamefont {O.}~\bibnamefont {{Forni}}},
  \bibinfo {author} {\bibfnamefont {M.}~\bibnamefont {{Frailis}}}, \bibinfo
  {author} {\bibfnamefont {A.~A.}\ \bibnamefont {{Fraisse}}}, \bibinfo {author}
  {\bibfnamefont {E.}~\bibnamefont {{Franceschi}}}, \bibinfo {author}
  {\bibfnamefont {A.}~\bibnamefont {{Frejsel}}}, \bibinfo {author}
  {\bibfnamefont {S.}~\bibnamefont {{Galeotta}}}, \bibinfo {author}
  {\bibfnamefont {S.}~\bibnamefont {{Galli}}}, \bibinfo {author} {\bibfnamefont
  {K.}~\bibnamefont {{Ganga}}}, \bibinfo {author} {\bibfnamefont
  {T.}~\bibnamefont {{Ghosh}}}, \bibinfo {author} {\bibfnamefont
  {M.}~\bibnamefont {{Giard}}}, \bibinfo {author} {\bibfnamefont
  {Y.}~\bibnamefont {{Giraud-H{\'e}raud}}}, \bibinfo {author} {\bibfnamefont
  {E.}~\bibnamefont {{Gjerl{\o}w}}}, \bibinfo {author} {\bibfnamefont
  {J.}~\bibnamefont {{Gonz{\'a}lez-Nuevo}}}, \bibinfo {author} {\bibfnamefont
  {K.~M.}\ \bibnamefont {{G{\'o}rski}}}, \bibinfo {author} {\bibfnamefont
  {S.}~\bibnamefont {{Gratton}}}, \bibinfo {author} {\bibfnamefont
  {A.}~\bibnamefont {{Gregorio}}}, \bibinfo {author} {\bibfnamefont
  {A.}~\bibnamefont {{Gruppuso}}}, \bibinfo {author} {\bibfnamefont {J.~E.}\
  \bibnamefont {{Gudmundsson}}}, \bibinfo {author} {\bibfnamefont {F.~K.}\
  \bibnamefont {{Hansen}}}, \bibinfo {author} {\bibfnamefont {D.}~\bibnamefont
  {{Hanson}}}, \bibinfo {author} {\bibfnamefont {D.~L.}\ \bibnamefont
  {{Harrison}}}, \bibinfo {author} {\bibfnamefont {G.}~\bibnamefont {{Helou}}},
  \bibinfo {author} {\bibfnamefont {S.}~\bibnamefont {{Henrot-Versill{\'e}}}},
  \bibinfo {author} {\bibfnamefont {C.}~\bibnamefont
  {{Hern{\'a}ndez-Monteagudo}}}, \bibinfo {author} {\bibfnamefont
  {D.}~\bibnamefont {{Herranz}}}, \bibinfo {author} {\bibfnamefont {S.~R.}\
  \bibnamefont {{Hildebrandt}}}, \bibinfo {author} {\bibfnamefont
  {E.}~\bibnamefont {{Hivon}}}, \bibinfo {author} {\bibfnamefont
  {M.}~\bibnamefont {{Hobson}}}, \bibinfo {author} {\bibfnamefont {W.~A.}\
  \bibnamefont {{Holmes}}}, \bibinfo {author} {\bibfnamefont {A.}~\bibnamefont
  {{Hornstrup}}}, \bibinfo {author} {\bibfnamefont {W.}~\bibnamefont
  {{Hovest}}}, \bibinfo {author} {\bibfnamefont {K.~M.}\ \bibnamefont
  {{Huffenberger}}}, \bibinfo {author} {\bibfnamefont {G.}~\bibnamefont
  {{Hurier}}}, \bibinfo {author} {\bibfnamefont {A.~H.}\ \bibnamefont
  {{Jaffe}}}, \bibinfo {author} {\bibfnamefont {T.~R.}\ \bibnamefont
  {{Jaffe}}}, \bibinfo {author} {\bibfnamefont {W.~C.}\ \bibnamefont
  {{Jones}}}, \bibinfo {author} {\bibfnamefont {M.}~\bibnamefont {{Juvela}}},
  \bibinfo {author} {\bibfnamefont {E.}~\bibnamefont {{Keih{\"a}nen}}},
  \bibinfo {author} {\bibfnamefont {R.}~\bibnamefont {{Keskitalo}}}, \bibinfo
  {author} {\bibfnamefont {T.~S.}\ \bibnamefont {{Kisner}}}, \bibinfo {author}
  {\bibfnamefont {R.}~\bibnamefont {{Kneissl}}}, \bibinfo {author}
  {\bibfnamefont {J.}~\bibnamefont {{Knoche}}}, \bibinfo {author}
  {\bibfnamefont {M.}~\bibnamefont {{Kunz}}}, \bibinfo {author} {\bibfnamefont
  {H.}~\bibnamefont {{Kurki-Suonio}}}, \bibinfo {author} {\bibfnamefont
  {G.}~\bibnamefont {{Lagache}}}, \bibinfo {author} {\bibfnamefont
  {A.}~\bibnamefont {{L{\"a}hteenm{\"a}ki}}}, \bibinfo {author} {\bibfnamefont
  {J.~M.}\ \bibnamefont {{Lamarre}}}, \bibinfo {author} {\bibfnamefont
  {A.}~\bibnamefont {{Lasenby}}}, \bibinfo {author} {\bibfnamefont
  {M.}~\bibnamefont {{Lattanzi}}}, \bibinfo {author} {\bibfnamefont {C.~R.}\
  \bibnamefont {{Lawrence}}}, \bibinfo {author} {\bibfnamefont
  {M.}~\bibnamefont {{Le Jeune}}}, \bibinfo {author} {\bibfnamefont {J.~P.}\
  \bibnamefont {{Leahy}}}, \bibinfo {author} {\bibfnamefont {R.}~\bibnamefont
  {{Leonardi}}}, \bibinfo {author} {\bibfnamefont {J.}~\bibnamefont
  {{Lesgourgues}}}, \bibinfo {author} {\bibfnamefont {F.}~\bibnamefont
  {{Levrier}}}, \bibinfo {author} {\bibfnamefont {M.}~\bibnamefont
  {{Liguori}}}, \bibinfo {author} {\bibfnamefont {P.~B.}\ \bibnamefont
  {{Lilje}}}, \bibinfo {author} {\bibfnamefont {M.}~\bibnamefont
  {{Linden-V{\o}rnle}}}, \bibinfo {author} {\bibfnamefont {M.}~\bibnamefont
  {{L{\'o}pez-Caniego}}}, \bibinfo {author} {\bibfnamefont {P.~M.}\
  \bibnamefont {{Lubin}}}, \bibinfo {author} {\bibfnamefont {J.~F.}\
  \bibnamefont {{Mac{\'\i}as-P{\'e}rez}}}, \bibinfo {author} {\bibfnamefont
  {G.}~\bibnamefont {{Maggio}}}, \bibinfo {author} {\bibfnamefont
  {D.}~\bibnamefont {{Maino}}}, \bibinfo {author} {\bibfnamefont
  {N.}~\bibnamefont {{Mandolesi}}}, \bibinfo {author} {\bibfnamefont
  {A.}~\bibnamefont {{Mangilli}}}, \bibinfo {author} {\bibfnamefont
  {M.}~\bibnamefont {{Maris}}}, \bibinfo {author} {\bibfnamefont {D.~J.}\
  \bibnamefont {{Marshall}}}, \bibinfo {author} {\bibfnamefont {P.~G.}\
  \bibnamefont {{Martin}}}, \bibinfo {author} {\bibfnamefont {E.}~\bibnamefont
  {{Mart{\'\i}nez-Gonz{\'a}lez}}}, \bibinfo {author} {\bibfnamefont
  {S.}~\bibnamefont {{Masi}}}, \bibinfo {author} {\bibfnamefont
  {S.}~\bibnamefont {{Matarrese}}}, \bibinfo {author} {\bibfnamefont
  {P.}~\bibnamefont {{McGehee}}}, \bibinfo {author} {\bibfnamefont {P.~R.}\
  \bibnamefont {{Meinhold}}}, \bibinfo {author} {\bibfnamefont
  {A.}~\bibnamefont {{Melchiorri}}}, \bibinfo {author} {\bibfnamefont
  {L.}~\bibnamefont {{Mendes}}}, \bibinfo {author} {\bibfnamefont
  {A.}~\bibnamefont {{Mennella}}}, \bibinfo {author} {\bibfnamefont
  {M.}~\bibnamefont {{Migliaccio}}}, \bibinfo {author} {\bibfnamefont
  {S.}~\bibnamefont {{Mitra}}}, \bibinfo {author} {\bibfnamefont {M.~A.}\
  \bibnamefont {{Miville-Desch{\^e}nes}}}, \bibinfo {author} {\bibfnamefont
  {A.}~\bibnamefont {{Moneti}}}, \bibinfo {author} {\bibfnamefont
  {L.}~\bibnamefont {{Montier}}}, \bibinfo {author} {\bibfnamefont
  {G.}~\bibnamefont {{Morgante}}}, \bibinfo {author} {\bibfnamefont
  {D.}~\bibnamefont {{Mortlock}}}, \bibinfo {author} {\bibfnamefont
  {A.}~\bibnamefont {{Moss}}}, \bibinfo {author} {\bibfnamefont
  {D.}~\bibnamefont {{Munshi}}}, \bibinfo {author} {\bibfnamefont {J.~A.}\
  \bibnamefont {{Murphy}}}, \bibinfo {author} {\bibfnamefont {P.}~\bibnamefont
  {{Naselsky}}}, \bibinfo {author} {\bibfnamefont {F.}~\bibnamefont {{Nati}}},
  \bibinfo {author} {\bibfnamefont {P.}~\bibnamefont {{Natoli}}}, \bibinfo
  {author} {\bibfnamefont {C.~B.}\ \bibnamefont {{Netterfield}}}, \bibinfo
  {author} {\bibfnamefont {H.~U.}\ \bibnamefont {{N{\o}rgaard-Nielsen}}},
  \bibinfo {author} {\bibfnamefont {F.}~\bibnamefont {{Noviello}}}, \bibinfo
  {author} {\bibfnamefont {D.}~\bibnamefont {{Novikov}}}, \bibinfo {author}
  {\bibfnamefont {I.}~\bibnamefont {{Novikov}}}, \bibinfo {author}
  {\bibfnamefont {E.}~\bibnamefont {{Orlando}}}, \bibinfo {author}
  {\bibfnamefont {C.~A.}\ \bibnamefont {{Oxborrow}}}, \bibinfo {author}
  {\bibfnamefont {F.}~\bibnamefont {{Paci}}}, \bibinfo {author} {\bibfnamefont
  {L.}~\bibnamefont {{Pagano}}}, \bibinfo {author} {\bibfnamefont
  {F.}~\bibnamefont {{Pajot}}}, \bibinfo {author} {\bibfnamefont
  {R.}~\bibnamefont {{Paladini}}}, \bibinfo {author} {\bibfnamefont
  {D.}~\bibnamefont {{Paoletti}}}, \bibinfo {author} {\bibfnamefont
  {B.}~\bibnamefont {{Partridge}}}, \bibinfo {author} {\bibfnamefont
  {F.}~\bibnamefont {{Pasian}}}, \bibinfo {author} {\bibfnamefont
  {G.}~\bibnamefont {{Patanchon}}}, \bibinfo {author} {\bibfnamefont {T.~J.}\
  \bibnamefont {{Pearson}}}, \bibinfo {author} {\bibfnamefont {O.}~\bibnamefont
  {{Perdereau}}}, \bibinfo {author} {\bibfnamefont {L.}~\bibnamefont
  {{Perotto}}}, \bibinfo {author} {\bibfnamefont {F.}~\bibnamefont
  {{Perrotta}}}, \bibinfo {author} {\bibfnamefont {V.}~\bibnamefont
  {{Pettorino}}}, \bibinfo {author} {\bibfnamefont {F.}~\bibnamefont
  {{Piacentini}}}, \bibinfo {author} {\bibfnamefont {M.}~\bibnamefont
  {{Piat}}}, \bibinfo {author} {\bibfnamefont {E.}~\bibnamefont {{Pierpaoli}}},
  \bibinfo {author} {\bibfnamefont {D.}~\bibnamefont {{Pietrobon}}}, \bibinfo
  {author} {\bibfnamefont {S.}~\bibnamefont {{Plaszczynski}}}, \bibinfo
  {author} {\bibfnamefont {E.}~\bibnamefont {{Pointecouteau}}}, \bibinfo
  {author} {\bibfnamefont {G.}~\bibnamefont {{Polenta}}}, \bibinfo {author}
  {\bibfnamefont {G.~W.}\ \bibnamefont {{Pratt}}}, \bibinfo {author}
  {\bibfnamefont {G.}~\bibnamefont {{Pr{\'e}zeau}}}, \bibinfo {author}
  {\bibfnamefont {S.}~\bibnamefont {{Prunet}}}, \bibinfo {author}
  {\bibfnamefont {J.~L.}\ \bibnamefont {{Puget}}}, \bibinfo {author}
  {\bibfnamefont {J.~P.}\ \bibnamefont {{Rachen}}}, \bibinfo {author}
  {\bibfnamefont {W.~T.}\ \bibnamefont {{Reach}}}, \bibinfo {author}
  {\bibfnamefont {R.}~\bibnamefont {{Rebolo}}}, \bibinfo {author}
  {\bibfnamefont {M.}~\bibnamefont {{Reinecke}}}, \bibinfo {author}
  {\bibfnamefont {M.}~\bibnamefont {{Remazeilles}}}, \bibinfo {author}
  {\bibfnamefont {C.}~\bibnamefont {{Renault}}}, \bibinfo {author}
  {\bibfnamefont {A.}~\bibnamefont {{Renzi}}}, \bibinfo {author} {\bibfnamefont
  {I.}~\bibnamefont {{Ristorcelli}}}, \bibinfo {author} {\bibfnamefont
  {G.}~\bibnamefont {{Rocha}}}, \bibinfo {author} {\bibfnamefont
  {C.}~\bibnamefont {{Rosset}}}, \bibinfo {author} {\bibfnamefont
  {M.}~\bibnamefont {{Rossetti}}}, \bibinfo {author} {\bibfnamefont
  {G.}~\bibnamefont {{Roudier}}}, \bibinfo {author} {\bibfnamefont {J.~A.}\
  \bibnamefont {{Rubi{\~n}o-Mart{\'\i}n}}}, \bibinfo {author} {\bibfnamefont
  {B.}~\bibnamefont {{Rusholme}}}, \bibinfo {author} {\bibfnamefont
  {M.}~\bibnamefont {{Sandri}}}, \bibinfo {author} {\bibfnamefont
  {D.}~\bibnamefont {{Santos}}}, \bibinfo {author} {\bibfnamefont
  {M.}~\bibnamefont {{Savelainen}}}, \bibinfo {author} {\bibfnamefont
  {G.}~\bibnamefont {{Savini}}}, \bibinfo {author} {\bibfnamefont
  {D.}~\bibnamefont {{Scott}}}, \bibinfo {author} {\bibfnamefont {M.~D.}\
  \bibnamefont {{Seiffert}}}, \bibinfo {author} {\bibfnamefont {E.~P.~S.}\
  \bibnamefont {{Shellard}}}, \bibinfo {author} {\bibfnamefont {L.~D.}\
  \bibnamefont {{Spencer}}}, \bibinfo {author} {\bibfnamefont {V.}~\bibnamefont
  {{Stolyarov}}}, \bibinfo {author} {\bibfnamefont {R.}~\bibnamefont
  {{Stompor}}}, \bibinfo {author} {\bibfnamefont {A.~W.}\ \bibnamefont
  {{Strong}}}, \bibinfo {author} {\bibfnamefont {R.}~\bibnamefont
  {{Sudiwala}}}, \bibinfo {author} {\bibfnamefont {R.}~\bibnamefont
  {{Sunyaev}}}, \bibinfo {author} {\bibfnamefont {D.}~\bibnamefont {{Sutton}}},
  \bibinfo {author} {\bibfnamefont {A.~S.}\ \bibnamefont {{Suur-Uski}}},
  \bibinfo {author} {\bibfnamefont {J.~F.}\ \bibnamefont {{Sygnet}}}, \bibinfo
  {author} {\bibfnamefont {J.~A.}\ \bibnamefont {{Tauber}}}, \bibinfo {author}
  {\bibfnamefont {L.}~\bibnamefont {{Terenzi}}}, \bibinfo {author}
  {\bibfnamefont {L.}~\bibnamefont {{Toffolatti}}}, \bibinfo {author}
  {\bibfnamefont {M.}~\bibnamefont {{Tomasi}}}, \bibinfo {author}
  {\bibfnamefont {M.}~\bibnamefont {{Tristram}}}, \bibinfo {author}
  {\bibfnamefont {M.}~\bibnamefont {{Tucci}}}, \bibinfo {author} {\bibfnamefont
  {J.}~\bibnamefont {{Tuovinen}}}, \bibinfo {author} {\bibfnamefont
  {G.}~\bibnamefont {{Umana}}}, \bibinfo {author} {\bibfnamefont
  {L.}~\bibnamefont {{Valenziano}}}, \bibinfo {author} {\bibfnamefont
  {J.}~\bibnamefont {{Valiviita}}}, \bibinfo {author} {\bibfnamefont
  {F.}~\bibnamefont {{Van Tent}}}, \bibinfo {author} {\bibfnamefont
  {P.}~\bibnamefont {{Vielva}}}, \bibinfo {author} {\bibfnamefont
  {F.}~\bibnamefont {{Villa}}}, \bibinfo {author} {\bibfnamefont {L.~A.}\
  \bibnamefont {{Wade}}}, \bibinfo {author} {\bibfnamefont {B.~D.}\
  \bibnamefont {{Wandelt}}}, \bibinfo {author} {\bibfnamefont {I.~K.}\
  \bibnamefont {{Wehus}}}, \bibinfo {author} {\bibfnamefont {A.}~\bibnamefont
  {{Wilkinson}}}, \bibinfo {author} {\bibfnamefont {D.}~\bibnamefont {{Yvon}}},
  \bibinfo {author} {\bibfnamefont {A.}~\bibnamefont {{Zacchei}}},\ and\
  \bibinfo {author} {\bibfnamefont {A.}~\bibnamefont {{Zonca}}},\ }\bibfield
  {title} {\bibinfo {title} {{Planck 2015 results. X. Diffuse component
  separation: Foreground maps}},\ }\href
  {https://doi.org/10.1051/0004-6361/201525967} {\bibfield  {journal} {\bibinfo
   {journal} {\aap}\ }\textbf {\bibinfo {volume} {594}},\ \bibinfo {eid} {A10}
  (\bibinfo {year} {2016})},\ \Eprint {https://arxiv.org/abs/1502.01588}
  {arXiv:1502.01588 [astro-ph.CO]} \BibitemShut {NoStop}%
\bibitem [{\citenamefont {{Planck Collaboration}}\ \emph
  {et~al.}(2020{\natexlab{b}})\citenamefont {{Planck Collaboration}},
  \citenamefont {{Akrami}}, \citenamefont {{Ashdown}}, \citenamefont
  {{Aumont}}, \citenamefont {{Baccigalupi}}, \citenamefont {{Ballardini}},
  \citenamefont {{Banday}}, \citenamefont {{Barreiro}}, \citenamefont
  {{Bartolo}}, \citenamefont {{Basak}}, \citenamefont {{Benabed}},
  \citenamefont {{Bersanelli}}, \citenamefont {{Bielewicz}}, \citenamefont
  {{Bond}}, \citenamefont {{Borrill}}, \citenamefont {{Bouchet}}, \citenamefont
  {{Boulanger}}, \citenamefont {{Bucher}}, \citenamefont {{Burigana}},
  \citenamefont {{Calabrese}}, \citenamefont {{Cardoso}}, \citenamefont
  {{Carron}}, \citenamefont {{Casaponsa}}, \citenamefont {{Challinor}},
  \citenamefont {{Colombo}}, \citenamefont {{Combet}}, \citenamefont {{Crill}},
  \citenamefont {{Cuttaia}}, \citenamefont {{de Bernardis}}, \citenamefont {{de
  Rosa}}, \citenamefont {{de Zotti}}, \citenamefont {{Delabrouille}},
  \citenamefont {{Delouis}}, \citenamefont {{Di Valentino}}, \citenamefont
  {{Dickinson}}, \citenamefont {{Diego}}, \citenamefont {{Donzelli}},
  \citenamefont {{Dor{\'e}}}, \citenamefont {{Ducout}}, \citenamefont
  {{Dupac}}, \citenamefont {{Efstathiou}}, \citenamefont {{Elsner}},
  \citenamefont {{En{\ss}lin}}, \citenamefont {{Eriksen}}, \citenamefont
  {{Falgarone}}, \citenamefont {{Fernandez-Cobos}}, \citenamefont {{Finelli}},
  \citenamefont {{Forastieri}}, \citenamefont {{Frailis}}, \citenamefont
  {{Fraisse}}, \citenamefont {{Franceschi}}, \citenamefont {{Frolov}},
  \citenamefont {{Galeotta}}, \citenamefont {{Galli}}, \citenamefont {{Ganga}},
  \citenamefont {{G{\'e}nova-Santos}}, \citenamefont {{Gerbino}}, \citenamefont
  {{Ghosh}}, \citenamefont {{Gonz{\'a}lez-Nuevo}}, \citenamefont
  {{G{\'o}rski}}, \citenamefont {{Gratton}}, \citenamefont {{Gruppuso}},
  \citenamefont {{Gudmundsson}}, \citenamefont {{Handley}}, \citenamefont
  {{Hansen}}, \citenamefont {{Helou}}, \citenamefont {{Herranz}}, \citenamefont
  {{Hildebrandt}}, \citenamefont {{Huang}}, \citenamefont {{Jaffe}},
  \citenamefont {{Karakci}}, \citenamefont {{Keih{\"a}nen}}, \citenamefont
  {{Keskitalo}}, \citenamefont {{Kiiveri}}, \citenamefont {{Kim}},
  \citenamefont {{Kisner}}, \citenamefont {{Krachmalnicoff}}, \citenamefont
  {{Kunz}}, \citenamefont {{Kurki-Suonio}}, \citenamefont {{Lagache}},
  \citenamefont {{Lamarre}}, \citenamefont {{Lasenby}}, \citenamefont
  {{Lattanzi}}, \citenamefont {{Lawrence}}, \citenamefont {{Le Jeune}},
  \citenamefont {{Levrier}}, \citenamefont {{Liguori}}, \citenamefont
  {{Lilje}}, \citenamefont {{Lindholm}}, \citenamefont {{L{\'o}pez-Caniego}},
  \citenamefont {{Lubin}}, \citenamefont {{Ma}}, \citenamefont
  {{Mac{\'\i}as-P{\'e}rez}}, \citenamefont {{Maggio}}, \citenamefont {{Maino}},
  \citenamefont {{Mandolesi}}, \citenamefont {{Mangilli}}, \citenamefont
  {{Marcos-Caballero}}, \citenamefont {{Maris}}, \citenamefont {{Martin}},
  \citenamefont {{Mart{\'\i}nez-Gonz{\'a}lez}}, \citenamefont {{Matarrese}},
  \citenamefont {{Mauri}}, \citenamefont {{McEwen}}, \citenamefont
  {{Meinhold}}, \citenamefont {{Melchiorri}}, \citenamefont {{Mennella}},
  \citenamefont {{Migliaccio}}, \citenamefont {{Miville-Desch{\^e}nes}},
  \citenamefont {{Molinari}}, \citenamefont {{Moneti}}, \citenamefont
  {{Montier}}, \citenamefont {{Morgante}}, \citenamefont {{Natoli}},
  \citenamefont {{Oppizzi}}, \citenamefont {{Pagano}}, \citenamefont
  {{Paoletti}}, \citenamefont {{Partridge}}, \citenamefont {{Peel}},
  \citenamefont {{Pettorino}}, \citenamefont {{Piacentini}}, \citenamefont
  {{Polenta}}, \citenamefont {{Puget}}, \citenamefont {{Rachen}}, \citenamefont
  {{Reinecke}}, \citenamefont {{Remazeilles}}, \citenamefont {{Renzi}},
  \citenamefont {{Rocha}}, \citenamefont {{Roudier}}, \citenamefont
  {{Rubi{\~n}o-Mart{\'\i}n}}, \citenamefont {{Ruiz-Granados}}, \citenamefont
  {{Salvati}}, \citenamefont {{Sandri}}, \citenamefont {{Savelainen}},
  \citenamefont {{Scott}}, \citenamefont {{Seljebotn}}, \citenamefont
  {{Sirignano}}, \citenamefont {{Spencer}}, \citenamefont {{Suur-Uski}},
  \citenamefont {{Tauber}}, \citenamefont {{Tavagnacco}}, \citenamefont
  {{Tenti}}, \citenamefont {{Thommesen}}, \citenamefont {{Toffolatti}},
  \citenamefont {{Tomasi}}, \citenamefont {{Trombetti}}, \citenamefont
  {{Valiviita}}, \citenamefont {{Van Tent}}, \citenamefont {{Vielva}},
  \citenamefont {{Villa}}, \citenamefont {{Vittorio}}, \citenamefont
  {{Wandelt}}, \citenamefont {{Wehus}}, \citenamefont {{Zacchei}},\ and\
  \citenamefont {{Zonca}}}]{2020A&A...641A...4P}%
  \BibitemOpen
  \bibfield  {author} {\bibinfo {author} {\bibnamefont {{Planck
  Collaboration}}}, \bibinfo {author} {\bibfnamefont {Y.}~\bibnamefont
  {{Akrami}}}, \bibinfo {author} {\bibfnamefont {M.}~\bibnamefont {{Ashdown}}},
  \bibinfo {author} {\bibfnamefont {J.}~\bibnamefont {{Aumont}}}, \bibinfo
  {author} {\bibfnamefont {C.}~\bibnamefont {{Baccigalupi}}}, \bibinfo {author}
  {\bibfnamefont {M.}~\bibnamefont {{Ballardini}}}, \bibinfo {author}
  {\bibfnamefont {A.~J.}\ \bibnamefont {{Banday}}}, \bibinfo {author}
  {\bibfnamefont {R.~B.}\ \bibnamefont {{Barreiro}}}, \bibinfo {author}
  {\bibfnamefont {N.}~\bibnamefont {{Bartolo}}}, \bibinfo {author}
  {\bibfnamefont {S.}~\bibnamefont {{Basak}}}, \bibinfo {author} {\bibfnamefont
  {K.}~\bibnamefont {{Benabed}}}, \bibinfo {author} {\bibfnamefont
  {M.}~\bibnamefont {{Bersanelli}}}, \bibinfo {author} {\bibfnamefont
  {P.}~\bibnamefont {{Bielewicz}}}, \bibinfo {author} {\bibfnamefont {J.~R.}\
  \bibnamefont {{Bond}}}, \bibinfo {author} {\bibfnamefont {J.}~\bibnamefont
  {{Borrill}}}, \bibinfo {author} {\bibfnamefont {F.~R.}\ \bibnamefont
  {{Bouchet}}}, \bibinfo {author} {\bibfnamefont {F.}~\bibnamefont
  {{Boulanger}}}, \bibinfo {author} {\bibfnamefont {M.}~\bibnamefont
  {{Bucher}}}, \bibinfo {author} {\bibfnamefont {C.}~\bibnamefont
  {{Burigana}}}, \bibinfo {author} {\bibfnamefont {E.}~\bibnamefont
  {{Calabrese}}}, \bibinfo {author} {\bibfnamefont {J.~F.}\ \bibnamefont
  {{Cardoso}}}, \bibinfo {author} {\bibfnamefont {J.}~\bibnamefont {{Carron}}},
  \bibinfo {author} {\bibfnamefont {B.}~\bibnamefont {{Casaponsa}}}, \bibinfo
  {author} {\bibfnamefont {A.}~\bibnamefont {{Challinor}}}, \bibinfo {author}
  {\bibfnamefont {L.~P.~L.}\ \bibnamefont {{Colombo}}}, \bibinfo {author}
  {\bibfnamefont {C.}~\bibnamefont {{Combet}}}, \bibinfo {author}
  {\bibfnamefont {B.~P.}\ \bibnamefont {{Crill}}}, \bibinfo {author}
  {\bibfnamefont {F.}~\bibnamefont {{Cuttaia}}}, \bibinfo {author}
  {\bibfnamefont {P.}~\bibnamefont {{de Bernardis}}}, \bibinfo {author}
  {\bibfnamefont {A.}~\bibnamefont {{de Rosa}}}, \bibinfo {author}
  {\bibfnamefont {G.}~\bibnamefont {{de Zotti}}}, \bibinfo {author}
  {\bibfnamefont {J.}~\bibnamefont {{Delabrouille}}}, \bibinfo {author}
  {\bibfnamefont {J.~M.}\ \bibnamefont {{Delouis}}}, \bibinfo {author}
  {\bibfnamefont {E.}~\bibnamefont {{Di Valentino}}}, \bibinfo {author}
  {\bibfnamefont {C.}~\bibnamefont {{Dickinson}}}, \bibinfo {author}
  {\bibfnamefont {J.~M.}\ \bibnamefont {{Diego}}}, \bibinfo {author}
  {\bibfnamefont {S.}~\bibnamefont {{Donzelli}}}, \bibinfo {author}
  {\bibfnamefont {O.}~\bibnamefont {{Dor{\'e}}}}, \bibinfo {author}
  {\bibfnamefont {A.}~\bibnamefont {{Ducout}}}, \bibinfo {author}
  {\bibfnamefont {X.}~\bibnamefont {{Dupac}}}, \bibinfo {author} {\bibfnamefont
  {G.}~\bibnamefont {{Efstathiou}}}, \bibinfo {author} {\bibfnamefont
  {F.}~\bibnamefont {{Elsner}}}, \bibinfo {author} {\bibfnamefont {T.~A.}\
  \bibnamefont {{En{\ss}lin}}}, \bibinfo {author} {\bibfnamefont {H.~K.}\
  \bibnamefont {{Eriksen}}}, \bibinfo {author} {\bibfnamefont {E.}~\bibnamefont
  {{Falgarone}}}, \bibinfo {author} {\bibfnamefont {R.}~\bibnamefont
  {{Fernandez-Cobos}}}, \bibinfo {author} {\bibfnamefont {F.}~\bibnamefont
  {{Finelli}}}, \bibinfo {author} {\bibfnamefont {F.}~\bibnamefont
  {{Forastieri}}}, \bibinfo {author} {\bibfnamefont {M.}~\bibnamefont
  {{Frailis}}}, \bibinfo {author} {\bibfnamefont {A.~A.}\ \bibnamefont
  {{Fraisse}}}, \bibinfo {author} {\bibfnamefont {E.}~\bibnamefont
  {{Franceschi}}}, \bibinfo {author} {\bibfnamefont {A.}~\bibnamefont
  {{Frolov}}}, \bibinfo {author} {\bibfnamefont {S.}~\bibnamefont
  {{Galeotta}}}, \bibinfo {author} {\bibfnamefont {S.}~\bibnamefont {{Galli}}},
  \bibinfo {author} {\bibfnamefont {K.}~\bibnamefont {{Ganga}}}, \bibinfo
  {author} {\bibfnamefont {R.~T.}\ \bibnamefont {{G{\'e}nova-Santos}}},
  \bibinfo {author} {\bibfnamefont {M.}~\bibnamefont {{Gerbino}}}, \bibinfo
  {author} {\bibfnamefont {T.}~\bibnamefont {{Ghosh}}}, \bibinfo {author}
  {\bibfnamefont {J.}~\bibnamefont {{Gonz{\'a}lez-Nuevo}}}, \bibinfo {author}
  {\bibfnamefont {K.~M.}\ \bibnamefont {{G{\'o}rski}}}, \bibinfo {author}
  {\bibfnamefont {S.}~\bibnamefont {{Gratton}}}, \bibinfo {author}
  {\bibfnamefont {A.}~\bibnamefont {{Gruppuso}}}, \bibinfo {author}
  {\bibfnamefont {J.~E.}\ \bibnamefont {{Gudmundsson}}}, \bibinfo {author}
  {\bibfnamefont {W.}~\bibnamefont {{Handley}}}, \bibinfo {author}
  {\bibfnamefont {F.~K.}\ \bibnamefont {{Hansen}}}, \bibinfo {author}
  {\bibfnamefont {G.}~\bibnamefont {{Helou}}}, \bibinfo {author} {\bibfnamefont
  {D.}~\bibnamefont {{Herranz}}}, \bibinfo {author} {\bibfnamefont {S.~R.}\
  \bibnamefont {{Hildebrandt}}}, \bibinfo {author} {\bibfnamefont
  {Z.}~\bibnamefont {{Huang}}}, \bibinfo {author} {\bibfnamefont {A.~H.}\
  \bibnamefont {{Jaffe}}}, \bibinfo {author} {\bibfnamefont {A.}~\bibnamefont
  {{Karakci}}}, \bibinfo {author} {\bibfnamefont {E.}~\bibnamefont
  {{Keih{\"a}nen}}}, \bibinfo {author} {\bibfnamefont {R.}~\bibnamefont
  {{Keskitalo}}}, \bibinfo {author} {\bibfnamefont {K.}~\bibnamefont
  {{Kiiveri}}}, \bibinfo {author} {\bibfnamefont {J.}~\bibnamefont {{Kim}}},
  \bibinfo {author} {\bibfnamefont {T.~S.}\ \bibnamefont {{Kisner}}}, \bibinfo
  {author} {\bibfnamefont {N.}~\bibnamefont {{Krachmalnicoff}}}, \bibinfo
  {author} {\bibfnamefont {M.}~\bibnamefont {{Kunz}}}, \bibinfo {author}
  {\bibfnamefont {H.}~\bibnamefont {{Kurki-Suonio}}}, \bibinfo {author}
  {\bibfnamefont {G.}~\bibnamefont {{Lagache}}}, \bibinfo {author}
  {\bibfnamefont {J.~M.}\ \bibnamefont {{Lamarre}}}, \bibinfo {author}
  {\bibfnamefont {A.}~\bibnamefont {{Lasenby}}}, \bibinfo {author}
  {\bibfnamefont {M.}~\bibnamefont {{Lattanzi}}}, \bibinfo {author}
  {\bibfnamefont {C.~R.}\ \bibnamefont {{Lawrence}}}, \bibinfo {author}
  {\bibfnamefont {M.}~\bibnamefont {{Le Jeune}}}, \bibinfo {author}
  {\bibfnamefont {F.}~\bibnamefont {{Levrier}}}, \bibinfo {author}
  {\bibfnamefont {M.}~\bibnamefont {{Liguori}}}, \bibinfo {author}
  {\bibfnamefont {P.~B.}\ \bibnamefont {{Lilje}}}, \bibinfo {author}
  {\bibfnamefont {V.}~\bibnamefont {{Lindholm}}}, \bibinfo {author}
  {\bibfnamefont {M.}~\bibnamefont {{L{\'o}pez-Caniego}}}, \bibinfo {author}
  {\bibfnamefont {P.~M.}\ \bibnamefont {{Lubin}}}, \bibinfo {author}
  {\bibfnamefont {Y.~Z.}\ \bibnamefont {{Ma}}}, \bibinfo {author}
  {\bibfnamefont {J.~F.}\ \bibnamefont {{Mac{\'\i}as-P{\'e}rez}}}, \bibinfo
  {author} {\bibfnamefont {G.}~\bibnamefont {{Maggio}}}, \bibinfo {author}
  {\bibfnamefont {D.}~\bibnamefont {{Maino}}}, \bibinfo {author} {\bibfnamefont
  {N.}~\bibnamefont {{Mandolesi}}}, \bibinfo {author} {\bibfnamefont
  {A.}~\bibnamefont {{Mangilli}}}, \bibinfo {author} {\bibfnamefont
  {A.}~\bibnamefont {{Marcos-Caballero}}}, \bibinfo {author} {\bibfnamefont
  {M.}~\bibnamefont {{Maris}}}, \bibinfo {author} {\bibfnamefont {P.~G.}\
  \bibnamefont {{Martin}}}, \bibinfo {author} {\bibfnamefont {E.}~\bibnamefont
  {{Mart{\'\i}nez-Gonz{\'a}lez}}}, \bibinfo {author} {\bibfnamefont
  {S.}~\bibnamefont {{Matarrese}}}, \bibinfo {author} {\bibfnamefont
  {N.}~\bibnamefont {{Mauri}}}, \bibinfo {author} {\bibfnamefont {J.~D.}\
  \bibnamefont {{McEwen}}}, \bibinfo {author} {\bibfnamefont {P.~R.}\
  \bibnamefont {{Meinhold}}}, \bibinfo {author} {\bibfnamefont
  {A.}~\bibnamefont {{Melchiorri}}}, \bibinfo {author} {\bibfnamefont
  {A.}~\bibnamefont {{Mennella}}}, \bibinfo {author} {\bibfnamefont
  {M.}~\bibnamefont {{Migliaccio}}}, \bibinfo {author} {\bibfnamefont {M.~A.}\
  \bibnamefont {{Miville-Desch{\^e}nes}}}, \bibinfo {author} {\bibfnamefont
  {D.}~\bibnamefont {{Molinari}}}, \bibinfo {author} {\bibfnamefont
  {A.}~\bibnamefont {{Moneti}}}, \bibinfo {author} {\bibfnamefont
  {L.}~\bibnamefont {{Montier}}}, \bibinfo {author} {\bibfnamefont
  {G.}~\bibnamefont {{Morgante}}}, \bibinfo {author} {\bibfnamefont
  {P.}~\bibnamefont {{Natoli}}}, \bibinfo {author} {\bibfnamefont
  {F.}~\bibnamefont {{Oppizzi}}}, \bibinfo {author} {\bibfnamefont
  {L.}~\bibnamefont {{Pagano}}}, \bibinfo {author} {\bibfnamefont
  {D.}~\bibnamefont {{Paoletti}}}, \bibinfo {author} {\bibfnamefont
  {B.}~\bibnamefont {{Partridge}}}, \bibinfo {author} {\bibfnamefont
  {M.}~\bibnamefont {{Peel}}}, \bibinfo {author} {\bibfnamefont
  {V.}~\bibnamefont {{Pettorino}}}, \bibinfo {author} {\bibfnamefont
  {F.}~\bibnamefont {{Piacentini}}}, \bibinfo {author} {\bibfnamefont
  {G.}~\bibnamefont {{Polenta}}}, \bibinfo {author} {\bibfnamefont {J.~L.}\
  \bibnamefont {{Puget}}}, \bibinfo {author} {\bibfnamefont {J.~P.}\
  \bibnamefont {{Rachen}}}, \bibinfo {author} {\bibfnamefont {M.}~\bibnamefont
  {{Reinecke}}}, \bibinfo {author} {\bibfnamefont {M.}~\bibnamefont
  {{Remazeilles}}}, \bibinfo {author} {\bibfnamefont {A.}~\bibnamefont
  {{Renzi}}}, \bibinfo {author} {\bibfnamefont {G.}~\bibnamefont {{Rocha}}},
  \bibinfo {author} {\bibfnamefont {G.}~\bibnamefont {{Roudier}}}, \bibinfo
  {author} {\bibfnamefont {J.~A.}\ \bibnamefont {{Rubi{\~n}o-Mart{\'\i}n}}},
  \bibinfo {author} {\bibfnamefont {B.}~\bibnamefont {{Ruiz-Granados}}},
  \bibinfo {author} {\bibfnamefont {L.}~\bibnamefont {{Salvati}}}, \bibinfo
  {author} {\bibfnamefont {M.}~\bibnamefont {{Sandri}}}, \bibinfo {author}
  {\bibfnamefont {M.}~\bibnamefont {{Savelainen}}}, \bibinfo {author}
  {\bibfnamefont {D.}~\bibnamefont {{Scott}}}, \bibinfo {author} {\bibfnamefont
  {D.~S.}\ \bibnamefont {{Seljebotn}}}, \bibinfo {author} {\bibfnamefont
  {C.}~\bibnamefont {{Sirignano}}}, \bibinfo {author} {\bibfnamefont {L.~D.}\
  \bibnamefont {{Spencer}}}, \bibinfo {author} {\bibfnamefont {A.~S.}\
  \bibnamefont {{Suur-Uski}}}, \bibinfo {author} {\bibfnamefont {J.~A.}\
  \bibnamefont {{Tauber}}}, \bibinfo {author} {\bibfnamefont {D.}~\bibnamefont
  {{Tavagnacco}}}, \bibinfo {author} {\bibfnamefont {M.}~\bibnamefont
  {{Tenti}}}, \bibinfo {author} {\bibfnamefont {H.}~\bibnamefont
  {{Thommesen}}}, \bibinfo {author} {\bibfnamefont {L.}~\bibnamefont
  {{Toffolatti}}}, \bibinfo {author} {\bibfnamefont {M.}~\bibnamefont
  {{Tomasi}}}, \bibinfo {author} {\bibfnamefont {T.}~\bibnamefont
  {{Trombetti}}}, \bibinfo {author} {\bibfnamefont {J.}~\bibnamefont
  {{Valiviita}}}, \bibinfo {author} {\bibfnamefont {B.}~\bibnamefont {{Van
  Tent}}}, \bibinfo {author} {\bibfnamefont {P.}~\bibnamefont {{Vielva}}},
  \bibinfo {author} {\bibfnamefont {F.}~\bibnamefont {{Villa}}}, \bibinfo
  {author} {\bibfnamefont {N.}~\bibnamefont {{Vittorio}}}, \bibinfo {author}
  {\bibfnamefont {B.~D.}\ \bibnamefont {{Wandelt}}}, \bibinfo {author}
  {\bibfnamefont {I.~K.}\ \bibnamefont {{Wehus}}}, \bibinfo {author}
  {\bibfnamefont {A.}~\bibnamefont {{Zacchei}}},\ and\ \bibinfo {author}
  {\bibfnamefont {A.}~\bibnamefont {{Zonca}}},\ }\bibfield  {title} {\bibinfo
  {title} {{Planck 2018 results. IV. Diffuse component separation}},\ }\href
  {https://doi.org/10.1051/0004-6361/201833881} {\bibfield  {journal} {\bibinfo
   {journal} {\aap}\ }\textbf {\bibinfo {volume} {641}},\ \bibinfo {eid} {A4}
  (\bibinfo {year} {2020}{\natexlab{b}})},\ \Eprint
  {https://arxiv.org/abs/1807.06208} {arXiv:1807.06208 [astro-ph.CO]}
  \BibitemShut {NoStop}%
\bibitem [{\citenamefont {{Cooray}}\ and\ \citenamefont
  {{Furlanetto}}(2004)}]{2004ApJ...606L...5C}%
  \BibitemOpen
  \bibfield  {author} {\bibinfo {author} {\bibfnamefont {A.}~\bibnamefont
  {{Cooray}}}\ and\ \bibinfo {author} {\bibfnamefont {S.~R.}\ \bibnamefont
  {{Furlanetto}}},\ }\bibfield  {title} {\bibinfo {title} {{Free-Free Emission
  at Low Radio Frequencies}},\ }\href {https://doi.org/10.1086/421241}
  {\bibfield  {journal} {\bibinfo  {journal} {\apjl}\ }\textbf {\bibinfo
  {volume} {606}},\ \bibinfo {pages} {L5} (\bibinfo {year} {2004})},\ \Eprint
  {https://arxiv.org/abs/astro-ph/0402239} {arXiv:astro-ph/0402239 [astro-ph]}
  \BibitemShut {NoStop}%
\bibitem [{\citenamefont {{Ponente}}\ \emph {et~al.}(2011)\citenamefont
  {{Ponente}}, \citenamefont {{Diego}}, \citenamefont {{Sheth}}, \citenamefont
  {{Burigana}}, \citenamefont {{Knollmann}},\ and\ \citenamefont
  {{Ascasibar}}}]{2011MNRAS.410.2353P}%
  \BibitemOpen
  \bibfield  {author} {\bibinfo {author} {\bibfnamefont {P.~P.}\ \bibnamefont
  {{Ponente}}}, \bibinfo {author} {\bibfnamefont {J.~M.}\ \bibnamefont
  {{Diego}}}, \bibinfo {author} {\bibfnamefont {R.~K.}\ \bibnamefont
  {{Sheth}}}, \bibinfo {author} {\bibfnamefont {C.}~\bibnamefont {{Burigana}}},
  \bibinfo {author} {\bibfnamefont {S.~R.}\ \bibnamefont {{Knollmann}}},\ and\
  \bibinfo {author} {\bibfnamefont {Y.}~\bibnamefont {{Ascasibar}}},\
  }\bibfield  {title} {\bibinfo {title} {{The cosmological free-free signal
  from galaxy groups and clusters}},\ }\href
  {https://doi.org/10.1111/j.1365-2966.2010.17611.x} {\bibfield  {journal}
  {\bibinfo  {journal} {\mnras}\ }\textbf {\bibinfo {volume} {410}},\ \bibinfo
  {pages} {2353} (\bibinfo {year} {2011})},\ \Eprint
  {https://arxiv.org/abs/1006.2243} {arXiv:1006.2243 [astro-ph.CO]}
  \BibitemShut {NoStop}%
\bibitem [{\citenamefont {{Liu}}\ \emph {et~al.}(2019)\citenamefont {{Liu}},
  \citenamefont {{Jaacks}}, \citenamefont {{Finkelstein}},\ and\ \citenamefont
  {{Bromm}}}]{2019MNRAS.486.3617L}%
  \BibitemOpen
  \bibfield  {author} {\bibinfo {author} {\bibfnamefont {B.}~\bibnamefont
  {{Liu}}}, \bibinfo {author} {\bibfnamefont {J.}~\bibnamefont {{Jaacks}}},
  \bibinfo {author} {\bibfnamefont {S.~L.}\ \bibnamefont {{Finkelstein}}},\
  and\ \bibinfo {author} {\bibfnamefont {V.}~\bibnamefont {{Bromm}}},\
  }\bibfield  {title} {\bibinfo {title} {{Global radiation signature from early
  structure formation}},\ }\href {https://doi.org/10.1093/mnras/stz910}
  {\bibfield  {journal} {\bibinfo  {journal} {\mnras}\ }\textbf {\bibinfo
  {volume} {486}},\ \bibinfo {pages} {3617} (\bibinfo {year} {2019})},\ \Eprint
  {https://arxiv.org/abs/1901.08994} {arXiv:1901.08994 [astro-ph.GA]}
  \BibitemShut {NoStop}%
\bibitem [{\citenamefont {{Planck Collaboration}}\ and\ \citenamefont
  {{Aghanim}}(2018)}]{Planck2018_cospara}%
  \BibitemOpen
  \bibfield  {author} {\bibinfo {author} {\bibnamefont {{Planck
  Collaboration}}}\ and\ \bibinfo {author} {\bibfnamefont {N.~e.~a.}\
  \bibnamefont {{Aghanim}}},\ }\bibfield  {title} {\bibinfo {title} {{Planck
  2018 results. VI. Cosmological parameters}},\ }\href@noop {} {\bibfield
  {journal} {\bibinfo  {journal} {ArXiv e-prints}\ } (\bibinfo {year}
  {2018})},\ \Eprint {https://arxiv.org/abs/1807.06209} {arXiv:1807.06209}
  \BibitemShut {NoStop}%
\bibitem [{\citenamefont {{Navarro}}\ \emph {et~al.}(1997)\citenamefont
  {{Navarro}}, \citenamefont {{Frenk}},\ and\ \citenamefont
  {{White}}}]{NFWprofile1997}%
  \BibitemOpen
  \bibfield  {author} {\bibinfo {author} {\bibfnamefont {J.~F.}\ \bibnamefont
  {{Navarro}}}, \bibinfo {author} {\bibfnamefont {C.~S.}\ \bibnamefont
  {{Frenk}}},\ and\ \bibinfo {author} {\bibfnamefont {S.~D.~M.}\ \bibnamefont
  {{White}}},\ }\bibfield  {title} {\bibinfo {title} {{A Universal Density
  Profile from Hierarchical Clustering}},\ }\href
  {https://doi.org/10.1086/304888} {\bibfield  {journal} {\bibinfo  {journal}
  {\apj}\ }\textbf {\bibinfo {volume} {490}},\ \bibinfo {pages} {493} (\bibinfo
  {year} {1997})},\ \Eprint {https://arxiv.org/abs/astro-ph/9611107}
  {arXiv:astro-ph/9611107 [astro-ph]} \BibitemShut {NoStop}%
\bibitem [{\citenamefont {{Ishiyama}}\ \emph {et~al.}(2020)\citenamefont
  {{Ishiyama}}, \citenamefont {{Prada}}, \citenamefont {{Klypin}},
  \citenamefont {{Sinha}}, \citenamefont {{Metcalf}}, \citenamefont {{Jullo}},
  \citenamefont {{Altieri}}, \citenamefont {{Cora}}, \citenamefont {{Croton}},
  \citenamefont {{de la Torre}}, \citenamefont {{Mill{\'a}n-Calero}},
  \citenamefont {{Oogi}}, \citenamefont {{Ruedas}},\ and\ \citenamefont
  {{Vega-Mart{\'\i}nez}}}]{2020arXiv200714720I}%
  \BibitemOpen
  \bibfield  {author} {\bibinfo {author} {\bibfnamefont {T.}~\bibnamefont
  {{Ishiyama}}}, \bibinfo {author} {\bibfnamefont {F.}~\bibnamefont {{Prada}}},
  \bibinfo {author} {\bibfnamefont {A.~A.}\ \bibnamefont {{Klypin}}}, \bibinfo
  {author} {\bibfnamefont {M.}~\bibnamefont {{Sinha}}}, \bibinfo {author}
  {\bibfnamefont {R.~B.}\ \bibnamefont {{Metcalf}}}, \bibinfo {author}
  {\bibfnamefont {E.}~\bibnamefont {{Jullo}}}, \bibinfo {author} {\bibfnamefont
  {B.}~\bibnamefont {{Altieri}}}, \bibinfo {author} {\bibfnamefont {S.~A.}\
  \bibnamefont {{Cora}}}, \bibinfo {author} {\bibfnamefont {D.}~\bibnamefont
  {{Croton}}}, \bibinfo {author} {\bibfnamefont {S.}~\bibnamefont {{de la
  Torre}}}, \bibinfo {author} {\bibfnamefont {D.~E.}\ \bibnamefont
  {{Mill{\'a}n-Calero}}}, \bibinfo {author} {\bibfnamefont {T.}~\bibnamefont
  {{Oogi}}}, \bibinfo {author} {\bibfnamefont {J.}~\bibnamefont {{Ruedas}}},\
  and\ \bibinfo {author} {\bibfnamefont {C.~A.}\ \bibnamefont
  {{Vega-Mart{\'\i}nez}}},\ }\bibfield  {title} {\bibinfo {title} {{The Uchuu
  Simulations: Data Release 1 and Dark Matter Halo Concentrations}},\
  }\href@noop {} {\bibfield  {journal} {\bibinfo  {journal} {arXiv e-prints}\
  ,\ \bibinfo {eid} {arXiv:2007.14720}} (\bibinfo {year} {2020})},\ \Eprint
  {https://arxiv.org/abs/2007.14720} {arXiv:2007.14720 [astro-ph.CO]}
  \BibitemShut {NoStop}%
\bibitem [{\citenamefont {{Diemer}}\ and\ \citenamefont
  {{Joyce}}(2019)}]{2019ApJ...871..168D}%
  \BibitemOpen
  \bibfield  {author} {\bibinfo {author} {\bibfnamefont {B.}~\bibnamefont
  {{Diemer}}}\ and\ \bibinfo {author} {\bibfnamefont {M.}~\bibnamefont
  {{Joyce}}},\ }\bibfield  {title} {\bibinfo {title} {{An Accurate Physical
  Model for Halo Concentrations}},\ }\href
  {https://doi.org/10.3847/1538-4357/aafad6} {\bibfield  {journal} {\bibinfo
  {journal} {\apj}\ }\textbf {\bibinfo {volume} {871}},\ \bibinfo {eid} {168}
  (\bibinfo {year} {2019})},\ \Eprint {https://arxiv.org/abs/1809.07326}
  {arXiv:1809.07326 [astro-ph.CO]} \BibitemShut {NoStop}%
\bibitem [{\citenamefont {{Johnson}}(2013)}]{2013ASSL..396..177J}%
  \BibitemOpen
  \bibfield  {author} {\bibinfo {author} {\bibfnamefont {J.~L.}\ \bibnamefont
  {{Johnson}}},\ }\bibfield  {title} {\bibinfo {title} {{Formation of the First
  Galaxies: Theory and Simulations}},\ }in\ \href
  {https://doi.org/10.1007/978-3-642-32362-1\_4} {\emph {\bibinfo {booktitle}
  {The First Galaxies}}},\ \bibinfo {series} {Astrophysics and Space Science
  Library}, Vol.\ \bibinfo {volume} {396},\ \bibinfo {editor} {edited by\
  \bibinfo {editor} {\bibfnamefont {T.}~\bibnamefont {{Wiklind}}}, \bibinfo
  {editor} {\bibfnamefont {B.}~\bibnamefont {{Mobasher}}},\ and\ \bibinfo
  {editor} {\bibfnamefont {V.}~\bibnamefont {{Bromm}}}}\ (\bibinfo {year}
  {2013})\ p.\ \bibinfo {pages} {177},\ \Eprint
  {https://arxiv.org/abs/1105.5701} {arXiv:1105.5701 [astro-ph.CO]}
  \BibitemShut {NoStop}%
\bibitem [{\citenamefont {{Makino}}\ \emph {et~al.}(1998)\citenamefont
  {{Makino}}, \citenamefont {{Sasaki}},\ and\ \citenamefont
  {{Suto}}}]{1998ApJ...497..555M}%
  \BibitemOpen
  \bibfield  {author} {\bibinfo {author} {\bibfnamefont {N.}~\bibnamefont
  {{Makino}}}, \bibinfo {author} {\bibfnamefont {S.}~\bibnamefont {{Sasaki}}},\
  and\ \bibinfo {author} {\bibfnamefont {Y.}~\bibnamefont {{Suto}}},\
  }\bibfield  {title} {\bibinfo {title} {{X-Ray Gas Density Profile of Clusters
  of Galaxies from the Universal Dark Matter Halo}},\ }\href
  {https://doi.org/10.1086/305507} {\bibfield  {journal} {\bibinfo  {journal}
  {\apj}\ }\textbf {\bibinfo {volume} {497}},\ \bibinfo {pages} {555} (\bibinfo
  {year} {1998})},\ \Eprint {https://arxiv.org/abs/astro-ph/9710344}
  {arXiv:astro-ph/9710344 [astro-ph]} \BibitemShut {NoStop}%
\bibitem [{\citenamefont {Bell}\ \emph {et~al.}(1983)\citenamefont {Bell},
  \citenamefont {Gilbody}, \citenamefont {Hughes}, \citenamefont {Kingston},\
  and\ \citenamefont {Smith}}]{doi:10.1063/1.555700}%
  \BibitemOpen
  \bibfield  {author} {\bibinfo {author} {\bibfnamefont {K.~L.}\ \bibnamefont
  {Bell}}, \bibinfo {author} {\bibfnamefont {H.~B.}\ \bibnamefont {Gilbody}},
  \bibinfo {author} {\bibfnamefont {J.~G.}\ \bibnamefont {Hughes}}, \bibinfo
  {author} {\bibfnamefont {A.~E.}\ \bibnamefont {Kingston}},\ and\ \bibinfo
  {author} {\bibfnamefont {F.~J.}\ \bibnamefont {Smith}},\ }\bibfield  {title}
  {\bibinfo {title} {Recommended data on the electron impact ionization of
  light atoms and ions},\ }\href {https://doi.org/10.1063/1.555700} {\bibfield
  {journal} {\bibinfo  {journal} {Journal of Physical and Chemical Reference
  Data}\ }\textbf {\bibinfo {volume} {12}},\ \bibinfo {pages} {891} (\bibinfo
  {year} {1983})},\ \Eprint
  {https://arxiv.org/abs/https://doi.org/10.1063/1.555700}
  {https://doi.org/10.1063/1.555700} \BibitemShut {NoStop}%
\bibitem [{\citenamefont {{Pequignot}}\ \emph {et~al.}(1991)\citenamefont
  {{Pequignot}}, \citenamefont {{Petitjean}},\ and\ \citenamefont
  {{Boisson}}}]{1991A&A...251..680P}%
  \BibitemOpen
  \bibfield  {author} {\bibinfo {author} {\bibfnamefont {D.}~\bibnamefont
  {{Pequignot}}}, \bibinfo {author} {\bibfnamefont {P.}~\bibnamefont
  {{Petitjean}}},\ and\ \bibinfo {author} {\bibfnamefont {C.}~\bibnamefont
  {{Boisson}}},\ }\bibfield  {title} {\bibinfo {title} {{Total and effective
  radiative recombination coefficients.}},\ }\href@noop {} {\bibfield
  {journal} {\bibinfo  {journal} {\aap}\ }\textbf {\bibinfo {volume} {251}},\
  \bibinfo {pages} {680} (\bibinfo {year} {1991})}\BibitemShut {NoStop}%
\bibitem [{\citenamefont {{Rybicki}}\ and\ \citenamefont
  {{Lightman}}(1986)}]{Radiative_process_in_Astrophysics}%
  \BibitemOpen
  \bibfield  {author} {\bibinfo {author} {\bibfnamefont {G.~B.}\ \bibnamefont
  {{Rybicki}}}\ and\ \bibinfo {author} {\bibfnamefont {A.~P.}\ \bibnamefont
  {{Lightman}}},\ }\href@noop {} {\emph {\bibinfo {title} {{Radiative Processes
  in Astrophysics}}}}\ (\bibinfo {year} {1986})\BibitemShut {NoStop}%
\bibitem [{\citenamefont {{Draine}}(2011)}]{2011piim.book.....D}%
  \BibitemOpen
  \bibfield  {author} {\bibinfo {author} {\bibfnamefont {B.~T.}\ \bibnamefont
  {{Draine}}},\ }\href@noop {} {\emph {\bibinfo {title} {{Physics of the
  Interstellar and Intergalactic Medium}}}}\ (\bibinfo {year}
  {2011})\BibitemShut {NoStop}%
\bibitem [{\citenamefont {{Weinberg}}(2008)}]{2008cosm.book.....W}%
  \BibitemOpen
  \bibfield  {author} {\bibinfo {author} {\bibfnamefont {S.}~\bibnamefont
  {{Weinberg}}},\ }\href@noop {} {\emph {\bibinfo {title} {{Cosmology}}}}\
  (\bibinfo  {publisher} {Oxford University Press},\ \bibinfo {year}
  {2008})\BibitemShut {NoStop}%
\bibitem [{\citenamefont {{Komatsu}}\ and\ \citenamefont
  {{Seljak}}(2002)}]{2002MNRAS.336.1256K}%
  \BibitemOpen
  \bibfield  {author} {\bibinfo {author} {\bibfnamefont {E.}~\bibnamefont
  {{Komatsu}}}\ and\ \bibinfo {author} {\bibfnamefont {U.}~\bibnamefont
  {{Seljak}}},\ }\bibfield  {title} {\bibinfo {title} {{The Sunyaev-Zel'dovich
  angular power spectrum as a probe of cosmological parameters}},\ }\href
  {https://doi.org/10.1046/j.1365-8711.2002.05889.x} {\bibfield  {journal}
  {\bibinfo  {journal} {\mnras}\ }\textbf {\bibinfo {volume} {336}},\ \bibinfo
  {pages} {1256} (\bibinfo {year} {2002})},\ \Eprint
  {https://arxiv.org/abs/astro-ph/0205468} {arXiv:astro-ph/0205468 [astro-ph]}
  \BibitemShut {NoStop}%
\bibitem [{\citenamefont {{Nicholson}}\ and\ \citenamefont
  {{Contaldi}}(2009)}]{2009JCAP...07..011N}%
  \BibitemOpen
  \bibfield  {author} {\bibinfo {author} {\bibfnamefont {G.}~\bibnamefont
  {{Nicholson}}}\ and\ \bibinfo {author} {\bibfnamefont {C.~R.}\ \bibnamefont
  {{Contaldi}}},\ }\bibfield  {title} {\bibinfo {title} {{Reconstruction of the
  primordial power spectrum using temperature and polarisation data from
  multiple experiments}},\ }\href
  {https://doi.org/10.1088/1475-7516/2009/07/011} {\bibfield  {journal}
  {\bibinfo  {journal} {\jcap}\ }\textbf {\bibinfo {volume} {2009}},\ \bibinfo
  {eid} {011} (\bibinfo {year} {2009})},\ \Eprint
  {https://arxiv.org/abs/0903.1106} {arXiv:0903.1106 [astro-ph.CO]}
  \BibitemShut {NoStop}%
\bibitem [{\citenamefont {{Nicholson}}\ \emph {et~al.}(2010)\citenamefont
  {{Nicholson}}, \citenamefont {{Contaldi}},\ and\ \citenamefont
  {{Paykari}}}]{2010JCAP...01..016N}%
  \BibitemOpen
  \bibfield  {author} {\bibinfo {author} {\bibfnamefont {G.}~\bibnamefont
  {{Nicholson}}}, \bibinfo {author} {\bibfnamefont {C.~R.}\ \bibnamefont
  {{Contaldi}}},\ and\ \bibinfo {author} {\bibfnamefont {P.}~\bibnamefont
  {{Paykari}}},\ }\bibfield  {title} {\bibinfo {title} {{Reconstruction of the
  primordial power spectrum by direct inversion}},\ }\href
  {https://doi.org/10.1088/1475-7516/2010/01/016} {\bibfield  {journal}
  {\bibinfo  {journal} {\jcap}\ }\textbf {\bibinfo {volume} {2010}},\ \bibinfo
  {eid} {016} (\bibinfo {year} {2010})},\ \Eprint
  {https://arxiv.org/abs/0909.5092} {arXiv:0909.5092 [astro-ph.CO]}
  \BibitemShut {NoStop}%
\bibitem [{\citenamefont {{Bird}}\ \emph {et~al.}(2011)\citenamefont {{Bird}},
  \citenamefont {{Peiris}}, \citenamefont {{Viel}},\ and\ \citenamefont
  {{Verde}}}]{2011MNRAS.413.1717B}%
  \BibitemOpen
  \bibfield  {author} {\bibinfo {author} {\bibfnamefont {S.}~\bibnamefont
  {{Bird}}}, \bibinfo {author} {\bibfnamefont {H.~V.}\ \bibnamefont
  {{Peiris}}}, \bibinfo {author} {\bibfnamefont {M.}~\bibnamefont {{Viel}}},\
  and\ \bibinfo {author} {\bibfnamefont {L.}~\bibnamefont {{Verde}}},\
  }\bibfield  {title} {\bibinfo {title} {{Minimally parametric power spectrum
  reconstruction from the Lyman {\ensuremath{\alpha}} forest}},\ }\href
  {https://doi.org/10.1111/j.1365-2966.2011.18245.x} {\bibfield  {journal}
  {\bibinfo  {journal} {\mnras}\ }\textbf {\bibinfo {volume} {413}},\ \bibinfo
  {pages} {1717} (\bibinfo {year} {2011})},\ \Eprint
  {https://arxiv.org/abs/1010.1519} {arXiv:1010.1519 [astro-ph.CO]}
  \BibitemShut {NoStop}%
\end{thebibliography}%


%

\end{document}